\documentclass[apj]{emulateapj}
\usepackage{morefloats}
\usepackage{color}
\usepackage{amsmath}
\usepackage{multirow}
\usepackage{graphicx}
\usepackage{xspace}

\newcommand{\hi}{H\,{\sc i}\xspace}
\newcommand{\cii}{[C\,{\sc ii}]\xspace}
\newcommand{\HI}{H\,{\sc i}\xspace}
\newcommand{\kms}{km s$^{-1}$\xspace}

\shorttitle{\cii\ dynamics}
\shortauthors{de Blok et al.}
\submitted{To be submitted to the Astronomical Journal}
\begin{document}
\title{Comparing \cii, \hi, and CO dynamics of nearby galaxies} 
\author{W.J.G. de Blok\altaffilmark{1,2,3},
  F. Walter\altaffilmark{4},
  J.-D.T Smith\altaffilmark{5},
  R. Herrera-Camus\altaffilmark{6,7},
  A.D. Bolatto\altaffilmark{6,7},
  M. A. Requena-Torres\altaffilmark{8} 
  A.F. Crocker\altaffilmark{5,9},  
  K.V. Croxall\altaffilmark{10},
  R.C. Kennicutt\altaffilmark{11},
  J. Koda\altaffilmark{12},
  L. Armus\altaffilmark{13},
  M. Boquien\altaffilmark{14},
  D. Dale\altaffilmark{15},
  K. Kreckel\altaffilmark{4},
  S. Meidt\altaffilmark{4}
}
\altaffiltext{1}{Netherlands Institute for Radio Astronomy (ASTRON), Postbus 2, 7990 AA Dwingeloo, the Netherlands}
\altaffiltext{2}{Astrophysics, Cosmology and Gravity Centre,Univ.\ of Cape Town, Private Bag X3, Rondebosch 7701, South Africa}
\altaffiltext{3}{Kapteyn Astronomical Institute, University ofGroningen, PO Box 800, 9700 AV Groningen, The Netherlands}
\altaffiltext{4}{Max-Planck Institut f\"ur Astronomie, K\"onigstuhl 17, 69117, Heidelberg, Germany}
\altaffiltext{5}{Department of Physics and Astronomy, University of Toledo, Toledo, OH 43606, USA}
\altaffiltext{6}{Department of Astronomy, University of Maryland, College Park, MD 20742, USA}
\altaffiltext{7}{Laboratory for Millimeter-Wave Astronomy, University of Maryland, College Park, MD 20742, USA}
\altaffiltext{8}{Space Telescope Science Institute, 3700 San Martin Dr., Baltimore, 21218 MD, USA}
\altaffiltext{9}{Department of Physics, Reed College, Portland, OR 97202, USA}
\altaffiltext{10}{Department of Astronomy, The Ohio State University, 4051 McPherson Laboratory, 140 West 18th Avenue, Columbus, OH 43210, USA}
\altaffiltext{11}{Institute of Astronomy, University of Cambridge, Madingley Road, Cambridge CB3 0HA, UK}
\altaffiltext{12}{Department of Physics and Astronomy, Stony Brook University, Stony Brook, NY 11794-3800, USA}
\altaffiltext{13}{Spitzer Science Center, California Institute of Technology, Pasadena, CA 91125, USA}
\altaffiltext{14}{Universidad de Antofagasta, Unidad de Astronom\'\i{}a, Avenida Angamos 601, 02800 Antofagasta, Chile}
\altaffiltext{15}{Department of Physics \& Astronomy, University of Wyoming, WY 82071, USA}



\begin{abstract}
The \hi and CO components of the interstellar medium (ISM) are usually
used to derive the dynamical mass $M_{\rm dyn}$ of nearby
galaxies. Both components become too faint to be used as a tracer in
observations of high-redshift galaxies.  In those cases, the 158
$\mu$m line of atomic carbon (\cii) may be the only way to derive
$M_{\rm dyn}$. As the distribution and kinematics of the ISM tracer
affects the determination of $M_{\rm dyn}$, it is important to
quantify the relative distributions of \hi, CO and \cii. \HI and CO
are well-characterised observationally, however, for \cii only very
few measurements exist.  Here we compare observations of CO, \hi, and
\cii emission of a sample of nearby galaxies, drawn from the HERACLES,
THINGS and KINGFISH surveys. We find that within $R_{25}$, the average
\cii exponential radial profile is slightly shallower than that of the
CO, but much steeper than the \hi distribution. This is also reflected
in the integrated spectrum (``global profile''), where the \cii
spectrum looks more like that of the CO than that of the \hi. For one
galaxy, a spectrally resolved comparison of integrated spectra was
possible; other comparisons were limited by the intrinsic line-widths
of the galaxies and the coarse velocity resolution of the \cii
data. Using high-spectral-resolution SOFIA \cii data of a number of
star forming regions in two nearby galaxies, we find that their \cii
linewidths agree better with those of the CO than the \HI.  As the
radial extent of a given ISM tracer is a key input in deriving $M_{\rm
  dyn}$ from spatially unresolved data, we conclude that the relevant
length-scale to use in determining $M_{\rm dyn}$ based on \cii data,
is that of the well-characterised CO distribution. This length scale
is similar to that of the optical disk.
\end{abstract}


\section{Introduction}

Observations of the gas component of galaxies yield important
information on their dynamics. For example, through the use of
rotation curves, detailed knowledge of the distribution of dark matter
in galaxies has become available. Where rotating galaxies are not
spatially resolved, their global profile (or integrated spectrum,
i.e., the distribution of spatially integrated flux per velocity
interval as a function of apparent radial Doppler velocity) still
provides information on dynamical masses $M_{\rm dyn} \propto
  V^2R/G$, where $V$ is the rotation velocity at radius $R$, and $G$
  is the gravitational constant.

The dynamical mass therefore measures the total mass content of a
galaxy within a certain radius $R$, including that of the dark
matter. The dark matter distribution does, however, extend
significantly beyond that of the visible matter, and the \emph{total}
dynamical mass (i.e, as measured at the virial radius) cannot be
determined in the absence of visible tracers at these large radii.

Dynamical mass measurements using the visible galaxy are therefore
always lower limits and depend on the radius at which the measurement
is made. A consistent choice for the radius $R$ at which $M_{\rm dyn}$
is measured is critically important for comparisons between different
galaxies, but also for measurements using different tracers in the
same galaxy. For example, in local disk galaxies the \HI disk
typically extends significantly beyond the main stellar disk. Using the
outer radii of either the stellar or the \HI disk will result in two
different values of $M_{\rm dyn}$. These can only be consistently
compared once the typical distribution of each of the tracers is taken
into account.

For low-redshift galaxies, both atomic neutral hydrogen (\hi) and
H$\alpha$ are routinely used to study galaxy dynamics. H$\alpha$ has
the advantage that it can be observed at high spatial resolution, but
it is generally confined to the optical disks of galaxies. It can also
be more prone to the effects of extinction as well as random motions
due to, e.g., massive star formation.

\HI is easily observable in the 21-cm radio line, constituting the main
gas component in most disk galaxies. It usually extends beyond the
optical disk, enabling studies of the dynamics out to large radii. The
\HI tends to have a high disk surface covering factor, and has a
fairly constant surface density over most of its radial extent.  Many
studies of the kinematics of disk galaxies thus rely heavily on \HI
observations.

The \HI line has a disadvantage, though, in that the nature of the
transition that gives rise to it makes it intrinsically faint. This
makes it still challenging to observe \HI at redshifts $z \gtrsim 0.2$
(e.g., \citealt{verheijen07}).  Future telescopes, such as the Square
Kilometre Array (SKA), should make these observations more feasible,
but even with such large facilities, observations of \HI in galaxies
at $z \sim 1$ and beyond will not be trivial (see, e.g.,
\citealt{blyth15}).

Observations of the kinematics of galaxies at high redshift are also
being obtained by observing redshifted optical emission lines such as
H$\alpha$ using Integral Field Unit instruments. Compared to
low-redshift observations an added uncertainty in the interpretation
of these kinematics is the importance of non-circular motions (due to
turbulence, winds, and/or streaming motions in non-axisymmetric
potentials) in very actively star-forming galaxies. It is generally
expected that at higher redshifts these are more important. Some early
observations of the kinematics of disk galaxies at high redshift are
presented in, e.g., \citet{vogt06}, \citet{forster06} and
\citet{genzel06}.

At higher redshifts, CO has also been used as a tracer for galaxy
dynamics.  CO lines are intrinsically much brighter than the \HI line,
but the CO distribution is more compact and, at least at low redshift,
tied to that of the stellar component (e.g., \citealt{regan06}, \citealt{leroy08,
  leroy13} and \citealt{schruba11}). Despite the more compact extent
it is still a useful tool to study galaxy dynamics.

However, even CO detections are challenging at very high ($z>5$)
redshifts. This is due to a combination of different factors. First,
low-$J$ CO transitions, that trace the bulk of the molecular gas, are
shifted to long wavelengths that are difficult to access with current
facilities (e.g., \citealt{carilli13}). Second, the cosmic microwave
background (which has a temperature proportional to $1+z$) sets the
effective bottom temperature of the interstellar medium (ISM) of any
galaxy which in turn makes it difficult to detect gas at intrinsically
low temperatures \citep{dacunha13}. Lastly, if the metallicities are
sufficiently low, studies at low redshift have shown that the CO
emission is decreased for a given H$_2$ mass (e.g.,
\citealt{wilson95}, \citealt{schruba12}).  This will also affect
measurements in main sequence galaxies at very high redshift, where
metallicities are depressed (e.g., \citealt{genzel12}).

An alternative gas tracer is the 158 $\mu$m line of atomic carbon,
C$^+$.  This is one of the main cooling lines of the ISM, and
is typically the brightest line (for exceptions see, e.g.,
\citealt{cormier15, madden11}).

\cii is also well-known to undergo relatively large changes in its
fractional cooling power relative to bolometric infrared luminosity (TIR) --
the so-called ``cooling line deficit'' (e.g., \citealt{luhman03}, Smith et al.,
in prep).  At high star formation rate (SFR) surface density, as is found in the local
universe at the highest infrared luminosities, \cii/TIR can drop
significantly.  At higher redshift, however, luminous systems are
often extended and can have Milky-Way like or even higher \cii/TIR
ratios \citep{brisbin15}.

On the other hand, \citet{cormier15} and \citet{madden11} show that in
low-metallicity environments C$^+$ remains bright with respect to CO
and FIR emission and increases in relative power as metallicity
declines in resolved nearby galaxies (Smith et al. in prep). C$^+$ is
distributed throughout the ISM and as it can be excited by collisions
with electrons, \HI and H$_2$, it is expected to trace most of the
phases of the ionized, atomic and molecular ISM (e.g.,
\citealt{pineda13}).

The downside of \cii is that its rest-frequency is very high: at $\sim
1900$\,GHz (158\,$\mu$m) observations of low-redshift galaxies from
the ground are impossible. However, moving to higher redshifts means
the frequency of the line shifts into regions of the spectrum that are
observable from the ground.  With the advent of ALMA (particularly in
Bands 8, 9, 10) it is therefore expected that observations of C$^+$
can be routinely done for galaxies at $z>1$ (see e.g.,
\citealt{maiolino05, walter12, venemans12}), and in principle out to
extreme redshifts ($z \sim 20$).

C$^+$ observations are therefore likely to become an important tool in
determining dynamical masses and other kinematical properties of
galaxies at high redshifts. A disadvantage of the high redshifts is
that detailed spatial information will not be available since galaxies
will be hard to resolve. As an example, an $0.5''$ beam will
  measure $\sim 3-4$ kpc beyond $z\sim 1$. Surface brightness
  sensitivity considerations make it unlikely that significantly
  higher resolutions can be achieved with current instruments unless
  significant amounts of observing time (e.g., $>10^h$ with ALMA) are
  invested.

Galaxies at these redshifts will therefore typically only be resolved
by a small number of beams along their semi-major axis.  This means
that determinations of quantities such as the dynamical mass of
systems will have to be based on global velocity profiles (integrated
spectra).  Determining a dynamical mass using this method requires a
velocity width and a representative radius. The former is readily
determined from the profile, but the latter is generally unknown. In
practice, a multiple of the optical radius or some other measure of
the expected extent of the tracer is used. This choice is important
in, e.g., proper comparisons between the baryonic and the dynamical
mass. Choosing a radius that is not appropriate for the tracer used
can lead to severe over- or underestimates of the dynamical mass and
therefore skewed comparisons.  See \citet{deblok14} for further
discussion on this.

A spatially integrated spectrum, or `global profile' as it is usually
called in radio-astronomy, is a general observable of galaxies,
independent of their dynamical state. For the case of a rotating
galaxy, with gas on circular orbits, the global profile is defined by
the rotation curve and the gas tracer distribution. The rotation
  curve depends on the total mass distribution within the galaxy
  (including stars and dark matter), and the rotation velocity as a
  function of radius therefore does not depend on the type of tracer
  used to measure the velocities. The radial mass surface density
  distributions of the tracers can, however, differ.
\citet{deblok14} demonstrate how the shape of the radial distribution
of the tracer has an impact on the shape and velocity width of the
global profile. In other words, determining the integrated spectrum of
a galaxy in either \HI or CO will likely not lead to identical
profiles (see \citealt{frank15} for examples).  In interpreting any
measurement from a global profile it is therefore important to know
what tracer was used and what its likely radial distribution is. This
directly affects any interpretation of the Tully-Fisher relation
\citep{tf}. Several papers compare the observed linewidths as measured
for different tracers within this context.  For example,
\citet{mathewson92} compare H$\alpha$ and \HI linewidths. Observed CO
and \HI linewidths are compared in, e.g., \citet{sofue92},
\citet{dickey92}, \citet{schoniger94} and
\citet{frank15}. \citet{deblok14} use simple models to quantify the
effects of the radial distributions of various tracers on linewidths
and the Tully-Fisher relation.

Here we discuss and compare radial profiles and spatially integrated
spectra as derived from \cii observations.  For this we will use
spatially resolved \cii data of nearby galaxies and make a direct
comparison between the overall distributions and kinematics of the
\cii, CO and \HI. The goal is to test whether global profiles derived
from \cii measurements resemble global profiles derived from CO data
(with an inherently compact tracer distribution) or from \hi data
(where the tracer is much more extended). We use observations of
  individual star-forming regions to also extend the comparison to
  velocity dispersions.

In Sect.\ 2 we present the sample and the data products that we use
in our analysis. Section 3 compares the radial profiles of the gas
tracers.  In Sect.\ 4 we discuss the kinematics of the three gas
tracers and construct spatially integrated spectra. Section 5 presents
measurements of the spectra of individual star forming regions.
We summarize our
findings in Sect.\ 6.

\section{Sample and data}

The only set of galaxies where detailed, resolved information is
available for all three tracers, CO, \HI and \cii, is the overlap of
the THINGS (\HI; \citealt{walter08}), HERACLES (CO $J = 2 \rightarrow
1$; \citealt{leroy09}), and KINGFISH (\cii 158 $\mu$m;
\citealt{kennicutt11}) samples.  Galaxies in these samples were
selected to mostly be disk galaxies dominated by rotation and without strong
bulge or bar features.  By design, a significant number of galaxies is
in common to these three surveys and can thus be used for a direct
comparison of the detailed \HI, CO and \cii distribution and
kinematics. For many of the galaxies in these samples detailed studies
of the kinematics in the \HI and CO are already available
(\citealt{deblok08, frank15}).

The main criterion for our sample selection was the availability of
high signal-to-noise and extended \HI, CO and \cii data for each of
the galaxies.  This resulted in a final sample of 10 galaxies.  Table
1 lists this sample, along with some basic properties\footnote{NGC
  7331 and NGC 3198 would have qualified as well. However, for these
  two galaxies the KINGFISH data were taken along the minor axis,
  making them less suited for an analysis of the dynamics. NGC 925 and
  NGC 2841 also have detections in all three tracers; however, the CO
  emission is too faint to provide useful constraints. Finally, NGC
  3077 was not added to the sample as it is not a regularly rotating
  disk galaxy.}.

A few of the galaxies have very low inclinations which would make them
less suitable for dynamical mass measurements (due to the $\sin i$
correction that needs to be applied to correct the line-of-sight
velocities); however, we include them here as the emphasis of this
paper is on a comparison of observed radial mass distributions and
integrated spectra and we do not try to calculate the actual dynamical
masses.
  
In the end, the presence of a clear CO detection
turned out to be the most stringent criterion. For this reason our
sample mostly consists of spiral galaxies. As described below, all
three data sets have a similar angular resolution of $\sim 12''$. For
a galaxy at the typical distance in our sample ($D \sim 8$ Mpc), this
corresponds to a linear resolution of $\sim 0.5$ kpc.

We show the integrated emission (zeroth moment) maps and the velocity
fields (first moment) for each galaxy in Fig.\ \ref{fig:mom_01} (and
Figs.\ \ref{fig:mom_02}--\ref{fig:mom_10} in the Appendix). In all
panels, the \hi data are shown in greyscale in the background, with
\hi, CO and \cii data overlaid as contours. Both the \hi and CO
observations cover the entire galaxy out to at least $R_{25}$, but the
\cii observations are typically restricted to a smaller region, as
indicated in the Figures.

\begin{figure*}
\includegraphics[width=0.9\hsize]{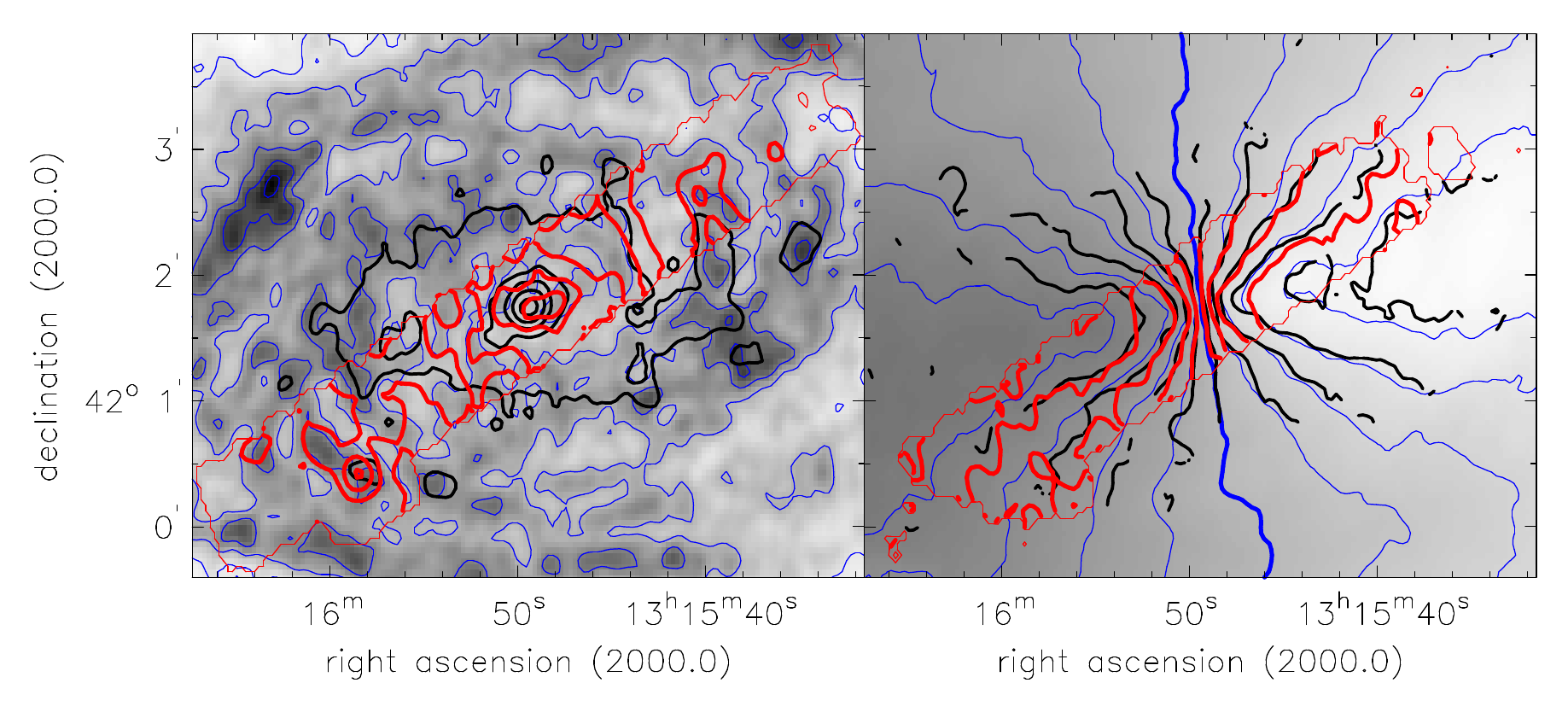}
\caption{\HI, CO and \cii moment maps of NGC 5055. \emph{Left panel:}
  Integrated intensity of zeroth moment map. Grayscale and thin blue
  contours shown the \HI distribution. Thick black contours show the
  CO distribution. Thick red contours show the \cii distribution
  within the area mapped by Herschel/PACS. The blue, thick black and
  thick red contours start at 10 percent of the respective tracer
  maximum value and increase in steps of 20 percent. \emph{Right
    panel:} Velocity field or first-moment map. Grayscale and blue
  contours show the \HI velocity field. Light-gray indicates the
  approaching side, dark-gray the receding side. The thick blue
  contour indicates the systemic velocity of 496.8 \kms. The blue
  contours are spaced by 40 \kms. The thick black and thick red contours
  show the CO and \cii velocities, respectively. Contour levels and
  spacings are identical to those used for the \HI data. In both
  panels the thin red contour shows the extent of the \cii
  observations and indicates the masks used for the CO and \HI
  observations.
\label{fig:mom_01}}
\end{figure*}

\begin{deluxetable}{crcccc}
\tabletypesize{\scriptsize} \tablewidth{0pt} \tablecaption{Properties
  of sample galaxies.\label{tab:template}}
\tablehead{\colhead{Name} & \colhead{$D$} & \colhead{$M_B$} & \colhead{$i$} & \colhead{$\log{D_{25}}$} & \colhead{Metal.} \\
\colhead{} & \colhead{(Mpc)} & \colhead{(mag)} & \colhead{($^{\circ}$)} & \colhead{($\log{0.1'}$)} & \colhead{(12+log(O/H))}\\
 \colhead{(1)} & \colhead{(2)} & \colhead{(3)} & \colhead{(4)} & \colhead{(5)} & \colhead{(6)} }
\startdata 
NGC 0628 & 7.3  & $-19.97$ & 7  & 1.99 & 8.33\\
NGC 2976 & 3.6  & $-17.78$ & 65 & 1.86 & 8.30\\
NGC 3184 & 11.1 & $-19.92$ & 16 & 1.87 & 8.48\\
NGC 3351 & 10.1 & $-19.88$ & 41 & 1.86 & 8.60\\
NGC 3521 & 10.7 & $-20.94$ & 73 & 1.92 & 8.36\\
NGC 3627 & 9.3  & $-20.74$ & 62 & 2.01 & 8.43\\
NGC 4736 & 4.7  & $-19.80$ & 41 & 1.89 & 8.31\\
NGC 5055 & 10.1 & $-21.12$ & 59 & 2.07 & 8.42\\
NGC 5457 & 7.4  & $-21.05$ & 22\tablenotemark{a} & 2.38 & 8.52\\
NGC 6946 & 5.9  & $-20.61$ & 33 & 2.06 & 8.40
\enddata 
\tablecomments{(1) Name of galaxy (2) Distance (Mpc). (3) Absolute
  $B$-band magnitude (4) inclination (degrees) (5) $\log{D_{25}}$
  ($\log{0.1'}$) (6) Metallicity. Quantities in columns (2)--(6) are from
  \citet{walter08} and \citet{kennicutt03}.} \tablenotetext{a}{Inclination from \citet{colombo14}.}
\end{deluxetable} 

\subsection{\cii data}

The \cii 158 $\mu$m data were obtained as part of the KINGFISH survey
\citep{kennicutt11}.  The observations were done with the PACS
instrument \citep{poglitsch10} on board \textit{Herschel}
\citep{pilbratt10}.  For a detailed description of the data reduction
we refer to \citet{croxall13} and Croxall et al.\ (in prep).

Most of the data consists of strip maps, usually (though not always)
along the major axis of the galaxy. In our sample selection we have
selected only galaxies where the orientation of the strip is close to
or along the major axis. The orientation is defined by that of the
SINGS observations with the \emph{Spitzer} Infra-Red Spectrograph
(IRS) instrument's Long-Low module of strips covering the central
regions of the galaxies (see \citealt{kennicutt03,smith07}).  For a
number of galaxies extra-nuclear pointings of approximately 1
arcmin$^2$ in size were also obtained, which are combined with the
corresponding strip map.

The angular resolution of the \cii data is $12''$. A pixel size of
$2.6''$ was adopted. The spectral resolution is $\sim$239\,\kms. Here
we use the \cii data cubes, integrated intensity maps (zeroth moment)
and the velocity fields (first moment) that are part of the KINGFISH
DR3 data products\footnote{\tt
  ftp://hsa.esac.esa.int/LEGACY\_PRODUCTS/UPDP/KINGFISH-DR3/}. Derivation
of these maps is described in Croxall et al.\ (in prep).

For two of our galaxies (NGC 5457 and NGC 6946), high
spectral-resolution SOFIA observations of the \cii in individual
star-forming regions are available which are discussed in Sect.~\ref{sec:sofia}.

\subsection{CO data}

We use the CO ($J=2\rightarrow 1$) data obtained as part of the
HERACLES survey \citep{leroy09}.  The angular resolution of the
HERACLES data is $13''$ by construction, the velocity spacing is
5.2\,\kms.  We use the data cubes, the integrated column density
(zeroth moment) maps and the intensity-weighted mean velocity fields
(first moment) that are part of the HERACLES data release\footnote{\tt
  http://www.mpia.de/HERACLES}.  We refer to \citet{leroy09} for a
further description of the properties and derivations of these maps.
The velocities in the HERACLES data were corrected from LSR to
heliocentric velocities.

\subsection{\HI data}

For the \HI data we use the datacubes, the integrated column density
(zeroth moment) maps and the intensity-weighted mean velocity fields
(first moment) from the THINGS
survey\footnote{\tt http://www.mpia.de/THINGS}, where we choose the maps
derived from the natural-weighted, residual-scaled data cubes.

The THINGS observations of the galaxies in this paper have an angular
resolution of $\sim 10''$, the velocity spacing is 2.6 or 5.2\,\kms.
We refer to \citet{walter08} for a description of the properties and
derivation of the individual maps and data cubes.

\subsection{Resolutions and spatial coverage}

As can be seen from the brief descriptions above, the angular
resolutions of the CO, \HI and \cii data are similar.  We will
therefore use and analyse these data sets at their native angular
resolutions, without any additional spatial smoothing. The situation
is different with the velocity resolutions. The CO and \hi data have
similar resolutions, in contrast with the \cii data. The effects of
the differences in spectral resolutions will be discussed in
Sect.\ \ref{sec:vel}. The \cii data consist mainly of strip maps with
restricted areal coverage. In our further analysis we therefore only
consider those areas in the CO and \HI data that are also covered by
the \cii observations.

\subsection{Comparison of velocity fields}

We compare the \cii, \hi and CO velocities on a pixel-by-pixel
basis. The CO and \HI velocity maps were regridded to have the same
grid spacings as the \cii data.  The \cii velocity fields contain a
small number of noisy pixels, mainly at the edges of the observed
emission where the \cii intensity is low. These were removed by
applying an intensity cut: velocity field pixels where the
corresponding intensity in the integrated \cii map was below $8 \cdot
10^{-9}$ W m$^{-2}$ sr$^{-1}$ were blanked, as were the corresponding
pixels in the CO and \HI velocity fields.

\begin{figure*}
\includegraphics[width=0.9\hsize]{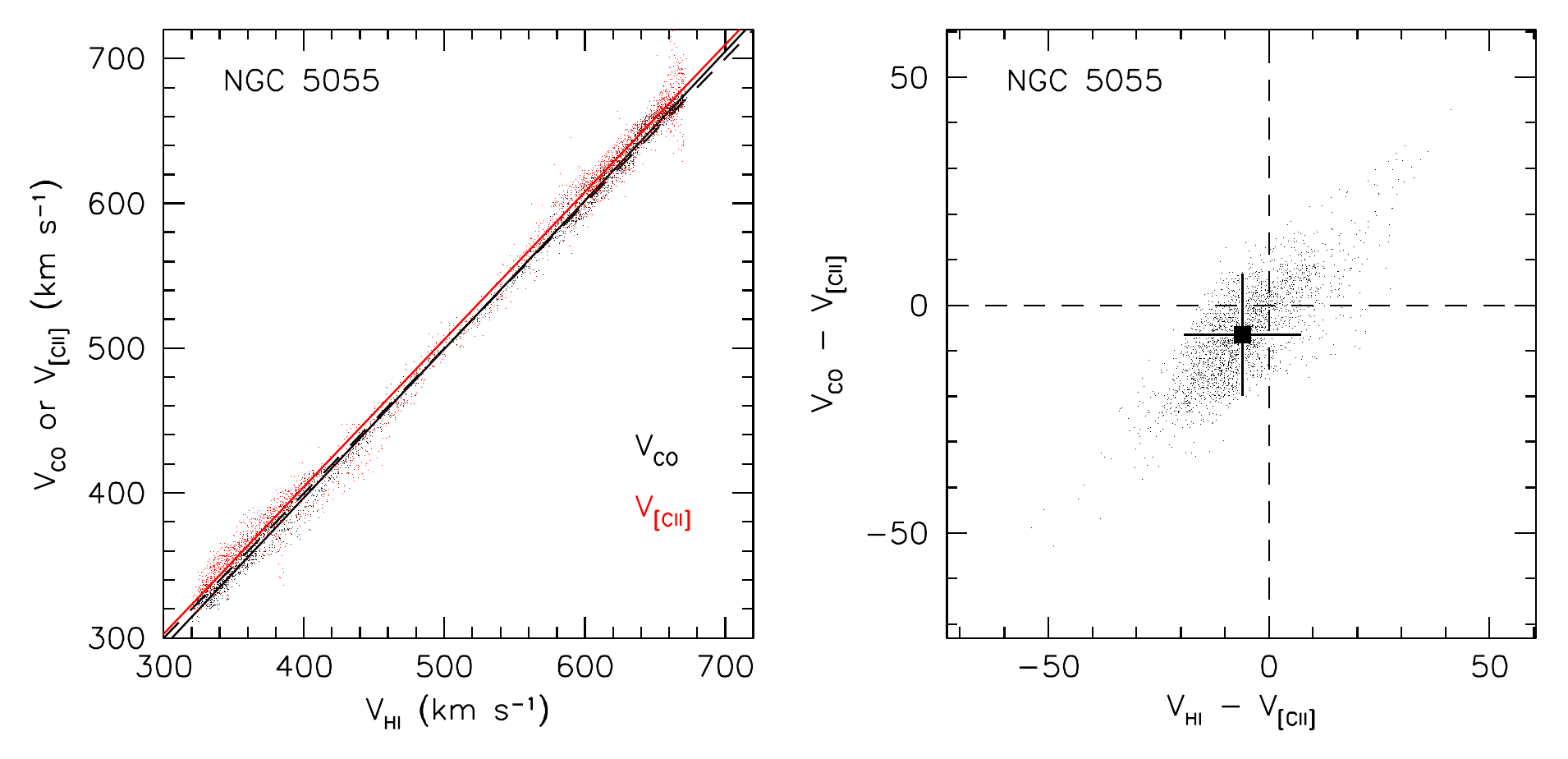}
\caption{Pixel-by-pixel comparison of the \cii, \hi and CO velocity
  field values for NGC 5055. {\bf Left:} CO (black) and \cii (red)
  velocities plotted against the \HI velocities. The black dashed line
  is the line of equality. The red and black full lines are linear
  least-squares fits of the CO (black) and the \cii (red)
  velocities. Similar plots for the other galaxies are given in the
  Appendix. {\bf Right:} Differences between \HI and \cii velocities
  plotted against the differences between CO and \cii velocities.  The
  filled square indicates the mean value, the errorbars indicate the
  $1\sigma$ spread in differences.  There are small offsets which we
  attribute to systematics in the determination of the \cii velocities
  (see text).\label{fig:vels}}
\end{figure*}

In the left panel of Fig.\ \ref{fig:vels} we plot the measured
velocities (vertical axis: CO and \cii, horizontal axis: \HI) for each
pixel. Overall there is excellent agreement between the velocities, as
expected for rotating gas disks in which all components are embedded
in the same gravitational potential. However, in some cases, small
offsets are seen.

Therefore, in the right panel of Fig.\ \ref{fig:vels} we plot the
difference of the CO and \cii velocities (vertical axis) as a function
of the difference of the \HI and \cii velocities (horizontal axis) for
each pixel.  The small offsets between the \cii velocities on the one hand, and the
CO and \hi velocities on the other hand, are also seen clearly
here. These systematic offsets of the \cii velocities, which are up to
$\sim 20$ \kms on average (and usually much less), are small compared
to the native velocity resolution of $\sim 239$ \kms of the \cii
data. We consider it unlikely that these offsets represent true,
physical velocity differences between gas components.  Rather, they
are likely the result of systematics in the \cii data reduction
process combined with the low velocity resolution of the \cii
profiles.

\subsection{SOFIA \cii data}\label{sec:sofia}

\begin{figure*}
\centering \includegraphics[width=0.9\hsize]{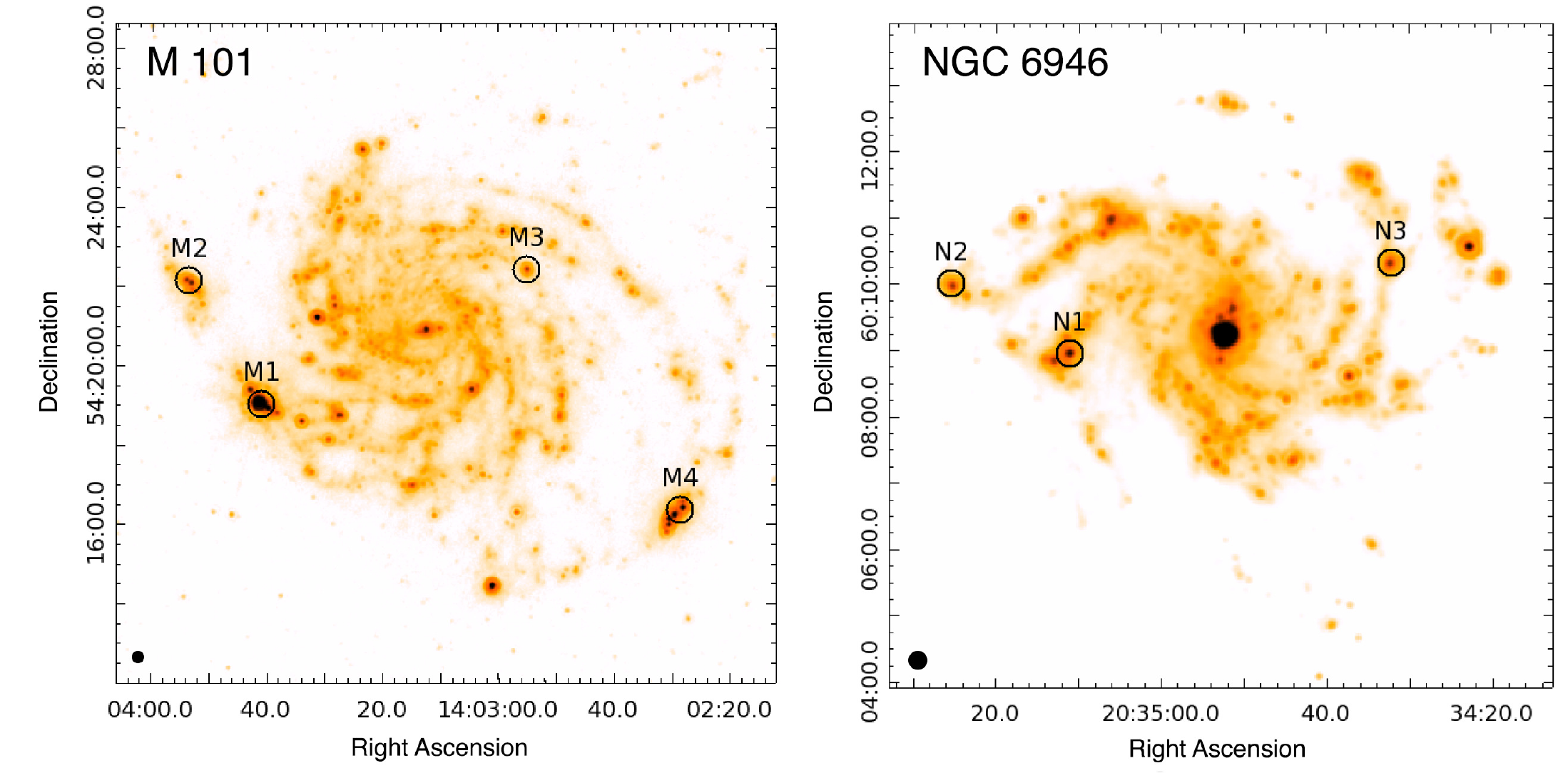}
\caption{Positions of the four star-forming regions in M~101 (left
  panel) and the three star-forming regions in NGC~6946 (right panel)
  that were observed with SOFIA overplotted on MIPS 24 $\mu$m
  images.\label{fig:MIPS}}
\end{figure*}

Observations of a number of star-forming regions in M101 (NGC 5457)
and NGC 6946 (PI Herrera-Camus, 02\_0098) were carried out using the
GREAT\footnote{GREAT is a development by the MPI f\"ur Radioastronomie
  and the KOSMA/Universit\"at zu K\"oln, in cooperation with the MPI
  f\"ur Sonnensystemforschung and the DLR Institut f\"ur
  Planetenforschung.} heterodyne receiver \citep{heyminck2012} on
board SOFIA\footnote{SOFIA is jointly operated by the Universities
  Space Research Association, Inc.\ (USRA), under NASA contract
  NAS2-97001, and the Deutsches SOFIA Institut (DSI) under DLR
  contract 50 OK 0901 to the University of Stuttgart.}
\citep{young12}.  Figure \ref{fig:MIPS} shows the positions of the
star-forming regions observed overplotted on {\it Spitzer} MIPS 24
$\mu$m maps.

The four star-forming regions in M 101 (M1, M2, M3 and M4) were
observed on May 20, 2014 on a $\sim 1.2$ hr leg at an altitude of
45,000 ft. The measured system temperature was $\sim 3100$ K for \cii
158 $\mu$m and on-source the integration times on M1, M2, M3 and M4
were 5.6, 6.3, 3.2 and 6.3 min, respectively. These data were obtained
in beam switching mode, with a chopper throw of $\lesssim 50''$.

The three star-forming regions in NGC 6946 (N1, N2 and N3) were
observed on May 21, 2014 on a $\sim 1.2$ hr leg at an altitude of
39,000 ft. The measured system temperature was $\sim 3200$ K for \cii
158 $\mu$m and the on-source integration times on N1, N2 and N3 were
6.8, 6.3 and 6.3 min, respectively. These observations used beam
switching with amplitudes of $120''$ and $135''$.

GREAT was set to its L1:L2 configuration, with L1 tuned to the
[N\,{\sc ii}] 205 $\mu$m transition (1461.13 GHz) in the lower
sideband and the L2 channel tuned to the \cii 158 $\mu$m transition
(1900.53 GHz) in the upper sideband. The half power beam width (HPBW)
determined during the flight was 14.1$''$ at 1.89 THz and 19.9$''$ at
1.34 THz.  The spectra were analyzed in CLASS\footnote{\tt
  http://www.iram.fr/IRAMFR/GILDAS/} and calibrated for a main beam
coupling efficiency 0.70 for the L1 channel and 0.65 for the L2
channel (numbers provided by the GREAT project).  The GREAT \cii spectra
were smoothed to a velocity resolution of 2.88 km s$^{-1}$.

\section{Radial surface density profiles}\label{sec:radial}

We investigate the radial distribution of the \cii from the
\emph{Herschel} observations, and compare to that of \hi and CO. We
use the integrated intensity (zeroth moment) maps for this. As noted
above, we have masked the \HI and CO data so that only regions where
\cii observations exist are considered in this comparison. An example
of these integrated intensity maps is shown in Fig.\ \ref{fig:mom_01}
(left panel). The full collection of maps is shown in the Appendix
(Figs.\ \ref{fig:mom_02}--\ref{fig:mom_10})

To derive the radial profiles, we adopt the orientation parameters
listed in \citet{walter08} (see also Tab.~1).  We adopt a fixed
(deprojected) ring width of $15''$ for all galaxies and tracers,
implying that all values presented here are independent.  The
profiles are corrected for inclination, and we use the conversion
factors listed in \citet{leroy08} to convert the \HI and CO fluxes to
column densities. We convert the inclination-corrected \cii values
from their original units of W m$^{-2}$ sr$^{-1}$ into surface
brightness units of $L_{\odot}$ pc$^{-2}$.

\begin{figure}
\includegraphics[width=0.9\hsize]{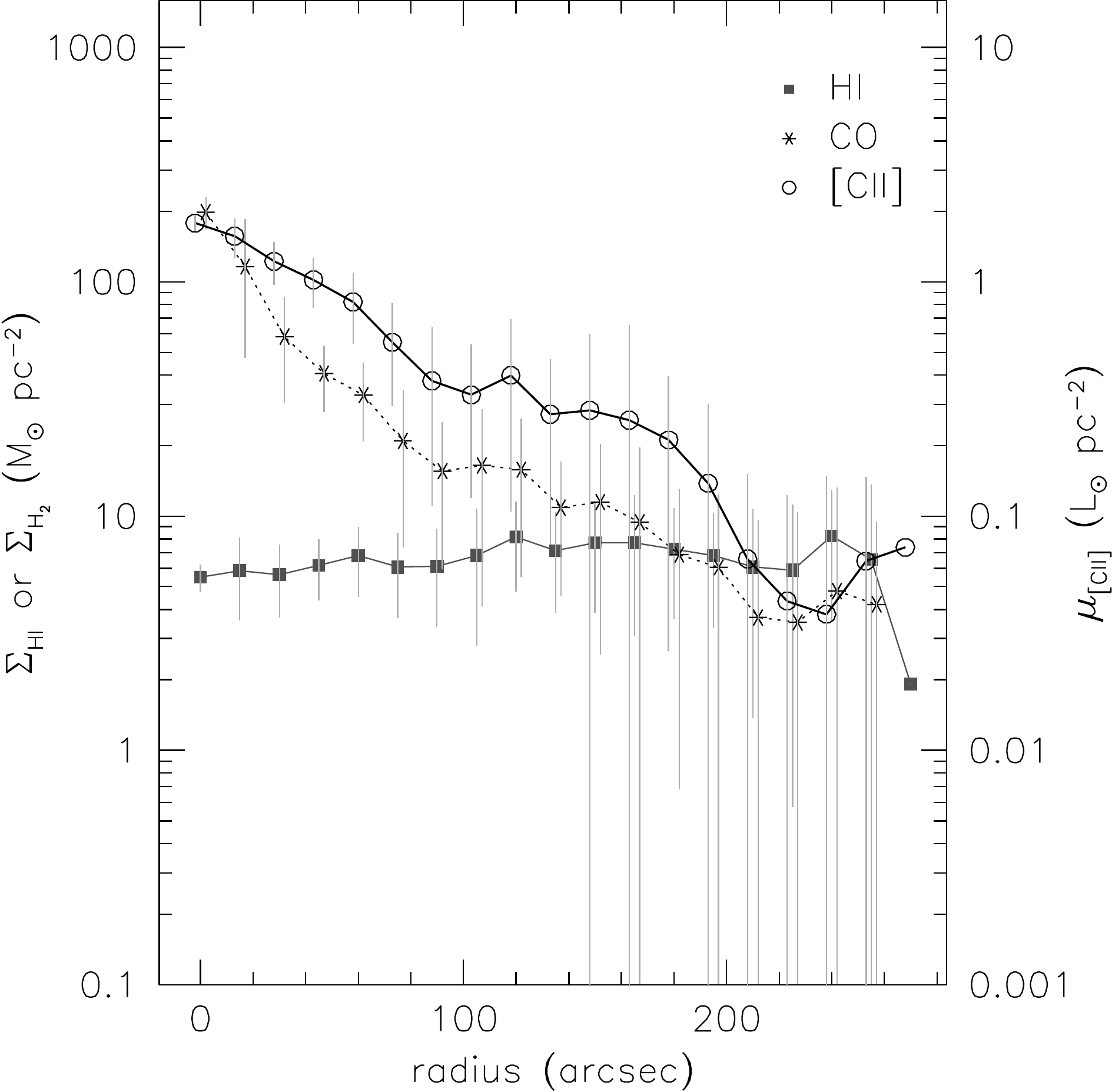}
\caption{CO (H$_2$), \HI and \cii radial profiles for NGC 5055 over
  the area covered by the \cii observations. Filled squares show the
  \HI surface density in $M_{\odot}$ pc$^{-2}$. The H$_2$ surface
  density in $M_{\odot}$ pc$^{-2}$ is shown with the star symbols. The
  \cii surface brightness in $L_{\odot}$ pc$^{-2}$ is shown with open
  circles. For clarity, the CO and \cii profiles have been offset by
  $2''$ and $-2''$ with respect to the \HI points, respectively. The
  errorbars indicate the RMS spread in values found along each
  ring. All surface densities and brightnesses have been corrected for
  inclination.
\label{fig:radialprofs}}
\end{figure}

\begin{figure*}
\includegraphics[width=0.49\hsize]{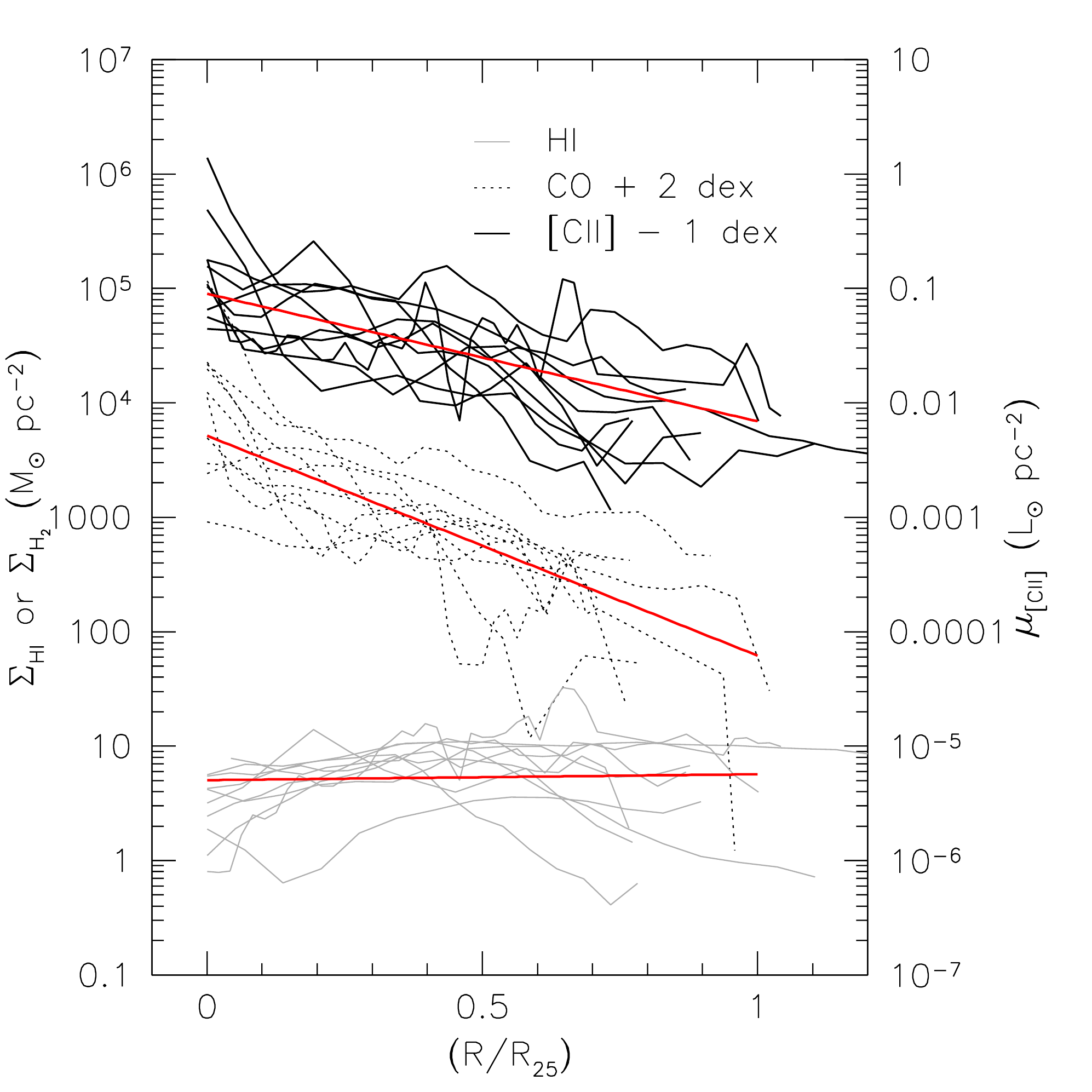}
\includegraphics[width=0.49\hsize]{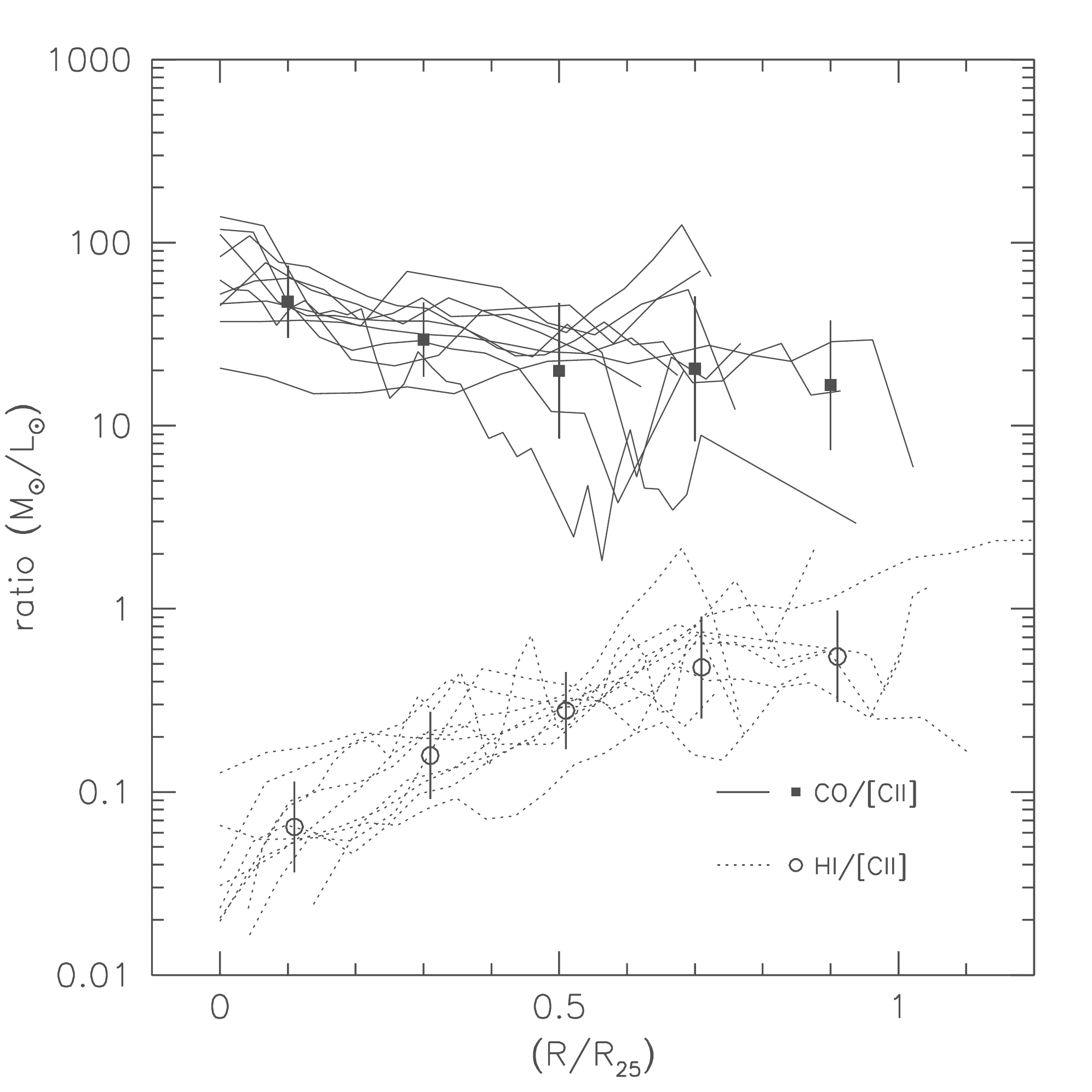}
\caption{Left: Radial profiles of the \hi (thin grey lines), CO
  (dotted black lines), and \cii (thick black lines) as shown in
  Figs.\ \ref{fig:radialprofs} and
  \ref{fig:radialprofs_app1}-\ref{fig:radialprofs_app4}. For clarity
  the CO profiles have been offset vertically by $+2$ dex, the \cii profiles by
  $-1$ dex. The red lines show the unweighted exponential fits to the
  combined profiles of each tracer. Right: Ratios of the CO, \HI and
  \cii radial profiles shown in Figs.\ \ref{fig:radialprofs} and
  \ref{fig:radialprofs_app1}-\ref{fig:radialprofs_app4}. Full curves
  show the CO/\cii surface density ratio as a function of radius for
  the individual galaxies. The dotted curves show the same for the
  \HI/\cii surface density ratios. The radius is expressed in terms of
  the optical radius $R_{25}$.  Also shown are the average values and
  RMS spread in bins of $0.2\,R_{25}$. Filled squares show the CO/\cii
  averages, open circles the \HI/\cii values. The ratios on the
  vertical axis are derived from the original units in
  Figs.\ \ref{fig:radialprofs} and
  \ref{fig:radialprofs_app1}-\ref{fig:radialprofs_app4}. The \HI/\cii
  data have been shifted down by 2 dex for clarity.
\label{fig:radialratio}}
\end{figure*}

Figures \ref{fig:radialprofs} and
\ref{fig:radialprofs_app1}--\ref{fig:radialprofs_app4} in the Appendix
show the azimuthally averaged profiles for our sample galaxies. The
errorbars show the RMS spread in intensity values along each ring.
These figures show that for the majority of cases the \hi radial
profile remains approximately constant with radius over the radial
range covered here, while those of the CO and \cii decline. The \cii
radial distribution therefore follows that of the CO more closely than
that of the \HI.  We illustrate this further by overplotting the
radial profiles for each tracer (expressed in terms of $R_{25}$;
\citealt{walter08} and Table 1), as shown in
Fig.~\ref{fig:radialratio} (left panel). To quantify the slope for
each tracer we make (unweighted) exponential fits of the form
$e^{-r/h}$ to each of the three sets of combined profiles. As
expected, the slope of the average \hi profile is consistent with
being flat, with an exponential scale length of $h = -8.4R_{25}$ with
large uncertainties in the $h$-value. The scale length values of the
CO and \cii tracer are better defined. For the CO we find $h = (0.22
^{+0.03}_{-0.01}) R_{25}$, which is consistent with the value found by
\citet{schruba11} of $h = 0.2 R_{25}$ for a partially different
subsample of the HERACLES galaxies.  For the \cii we find
$h=(0.38^{+0.03}_{-0.02})R_{25}$. For comparison, a purely exponential
stellar Freeman disk \citep{freeman70} has a scale length $h = 0.31
R_{25}$. The \cii distribution is therefore slightly less compact than
the CO one, and comparable with that of the optical component.

To confirm that this behaviour applies to the entire disk, and is not
dominated by, e.g., the bright inner parts, we also
determine the exponential scale lengths for the radial ranges $R < 0.5
R_{25}$ and $R>0.5 R_{25}$ separately. For $R<0.5 R_{25}$ we find that
the scale lengths for \HI, CO and \cii are $h =
(-0.41^{+0.06}_{-0.08},\ 0.19^{+0.03}_{-0.02},\ 0.35^{+0.08}_{-0.05})$,
respectively. For the radii $R>0.5 R_{25}$, the respective values are $h=
(1.24^{+0.98}_{-0.38},\ 0.28^{+0.17}_{-0.07},\ 0.45^{+0.09}_{-0.06})$. In
these two radial ranges we therefore see the same trends as over the
entire disk, showing that the behaviour seen is systematic.

We quantify this further by plotting the ratios of the surface densities,
\HI/\cii and CO/\cii, in Fig.\ \ref{fig:radialratio} (right panel), where we again express
the radius in terms of the optical radius, $R_{25}$. We also
show the average value and RMS spread for all galaxies combined and
determined in bins of width $0.2\, R_{25}$.

The CO/\cii ratio declines by a factor $\sim 2.8$ between the center
and $R_{25}$. This implies again that the \cii\ is somewhat less
concentrated than the CO emission. On the other hand, the \hi/\cii
ratio increases by a factor of $\sim 9.8$, consistent with a much
`flatter' radial \hi\ distribution.  We conclude that the
\cii\ follows the CO more closely than the \hi and note that this is
consistent with the finding that both CO and \cii are approximately
linear tracers of the SFR surface density, whereas \HI is not
(e.g., \citealt{bigiel08, leroy08, leroy13, herrera15}).

\section{Velocities}\label{sec:vel}

While pixel-by-pixel comparisons are a good way to show that
velocities agree with each other locally, these kind of data will not
be available for the majority of galaxies studied at high
redshift. For many galaxies only an integrated spectrum (global
profile) will be available.

\begin{figure*}
\begin{center}
\includegraphics[width=0.4\hsize]{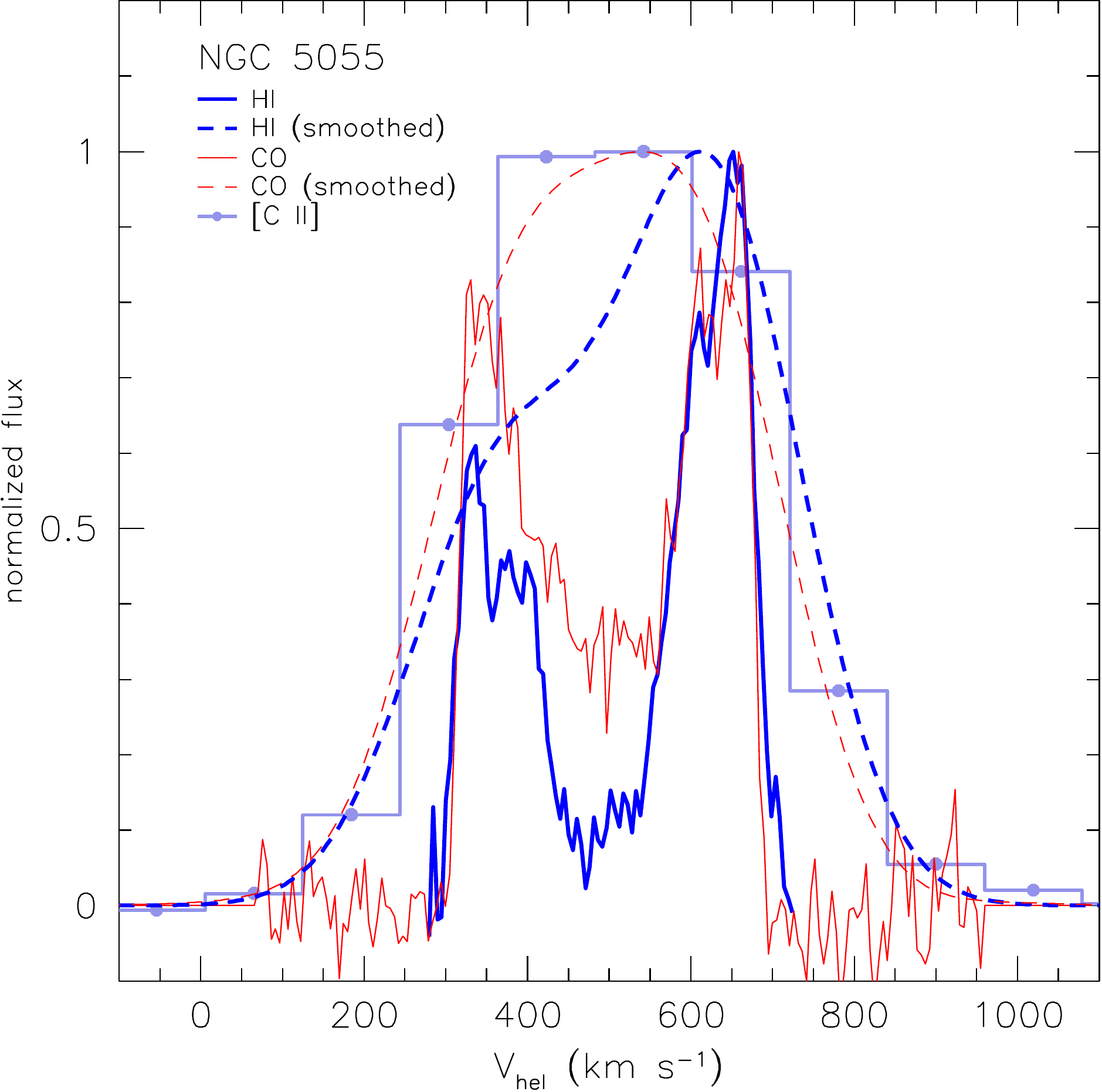}\hspace{1cm}\includegraphics[width=0.4\hsize]{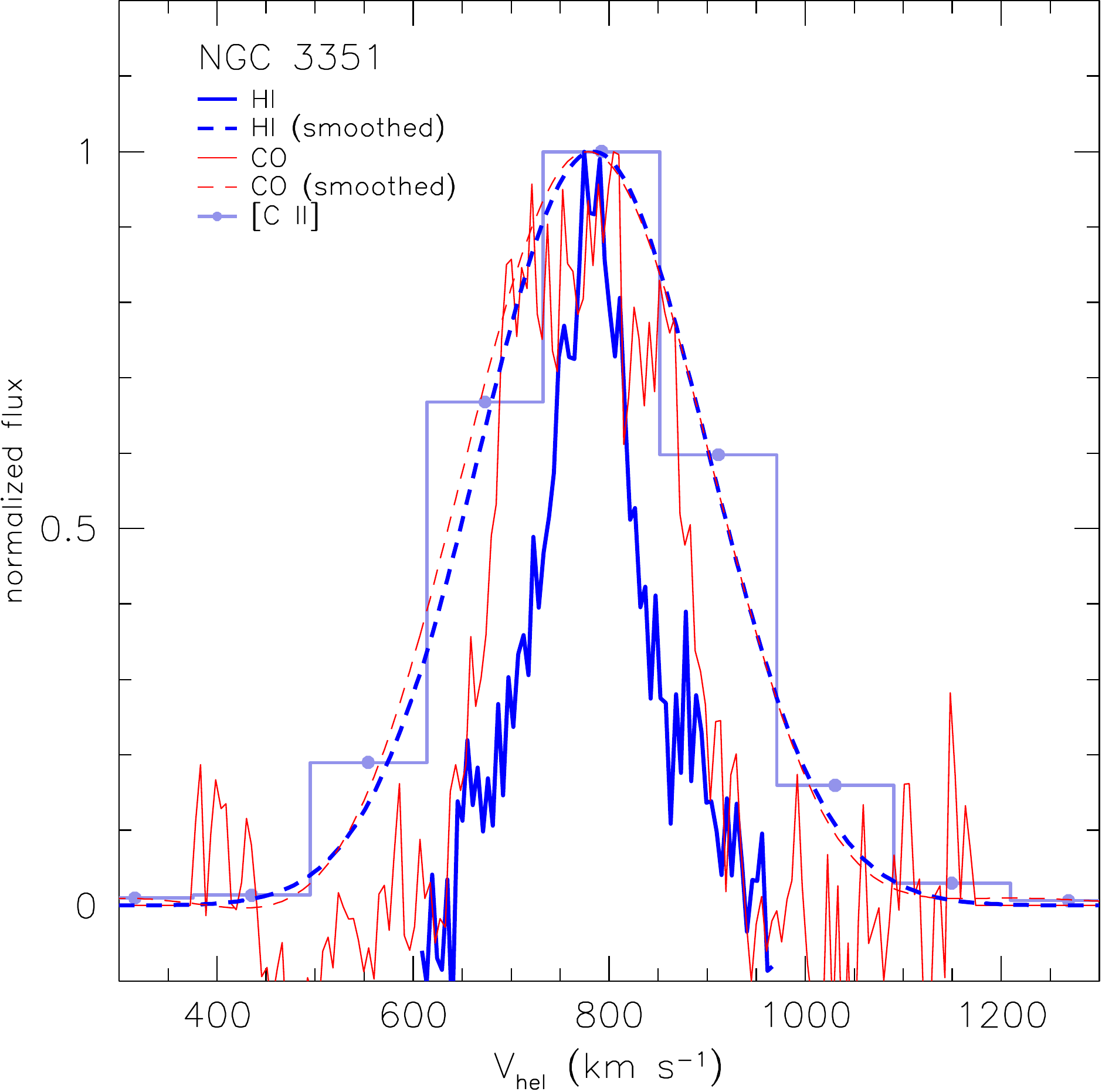}
\caption{Global profiles as measured in CO, \cii and \HI. Global
  profiles are all derived from the masked ``strip'' cubes. The thick,
  blue full line shows the \HI profile, the thin red full line the CO
  profile. The light-blue histogram with superimposed filled circles
  shows the \cii global profile. The thick blue and thin red dashed
  lines show the \HI and CO global profiles, respectively, both
  convolved to the velocity resolution of 239 \kms of the \cii data. Left panel
  shows NGC 5055, right panel shows NGC 3351.
\label{fig:globprofs}}
\end{center}
\end{figure*}

Using the \HI, CO and \cii data cubes we determine integrated spectra
of our sample galaxies by adding the total flux in each velocity
channel, using only the areas where \cii observations were done.
As these regions cover the major axes of the galaxies, the global
profiles should contain most of the rotation signal of the sample
galaxies.  The profiles are shown in Figs.\ \ref{fig:globprofs} and
\ref{fig:globprofs_apps1}--\ref{fig:globprofs_apps4}.  Due to the
large difference in velocity resolution between the \cii data (239
\kms) on the one hand and \HI and CO data on the other hand (2.6 or
5.2 \kms), we show the \HI and CO profiles both at their native
velocity resolution (full curves) and also convolved in velocity
(dashed curves, using a Gaussian convolution function) to the \cii
resolution.

The coarse velocity resolution of the \cii data in combination with
the velocity width of the galaxies means that for many of the galaxies
the \cii data do not allow us to resolve the double horn profile that
is typically seen in \HI and CO (see \citealt{walter08,
  frank15}). Figure \ref{fig:globprofs} shows one galaxy (NGC\,5055,
left panel) where the velocity width of the galaxy as measured in \cii
is significantly (a factor of $\sim$ 2) larger than the velocity
resolution of the \cii data.  Despite the coarse velocity resolution,
it is still possible to directly compare features in the \cii spectrum
with those in the convolved CO and \HI spectra.  We see that for this
galaxy, the \cii spectrum follows the convolved CO spectrum more
closely than the convolved \HI one.  Also shown in that Figure is a
more typical case (NGC 3351, right) where the \cii velocity width is
comparable to the \cii velocity resolution.  Here, no
distinctive features can be made out in the spectra.  This lack of
features also applies to the remaining galaxies.

A global profile, or integrated spectrum, is determined by both the
galaxy rotation curve and the surface density profile of the tracer
(e.g., CO, \HI, \cii) used to measure the velocities.  As all three
gas tracers are embedded in the same mass potential, the measured
rotation curve will to a very large degree only depend on the \emph{total}
mass density as a function of radius. Any difference between
integrated spectra of the same galaxy, can therefore be directly
attributed to differences in the radial distribution of the tracer
used (see \citealt{deblok14} for an extensive discussion).  With none
of the galaxies having a \cii radial distribution significantly
different from the CO distribution, we expect spectrally resolved \cii
global profiles to look more like the CO global profiles than their
\hi counterparts.

\section{Small-scale comparisons}

In the previous sections we established that the \cii and CO
radial distributions are more similar to each other than to the \HI
distribution. This raises the question whether the CO and \cii
emission also originates from similar volumes in the galaxies.  One
possible way of studying this is by looking at the velocity
dispersions of the components, which are an indication of the extent
of the distribution along the line-of-sight.

\begin{figure}
\centering \includegraphics[width=\hsize]{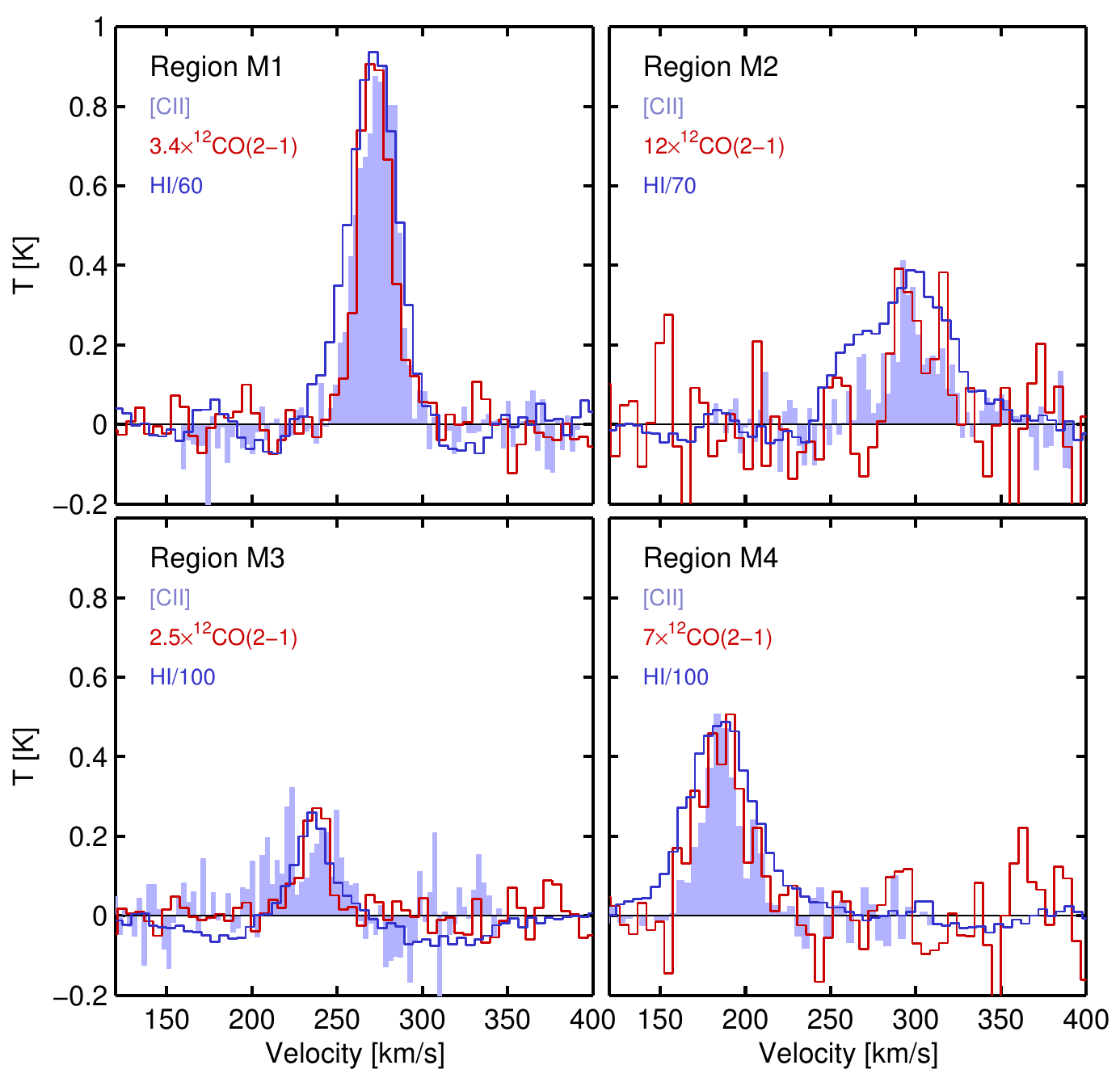}
\caption{\cii, \hi and CO spectra of the four star-forming
  regions in M101. The \hi and CO peak intensities are scaled to
  approximately match the \cii intensity. Positions of the regions are indicated in Fig.\ \ref{fig:MIPS}.
\label{fig:SOF_M101}}
\end{figure}

\begin{figure}
\includegraphics[width=\hsize]{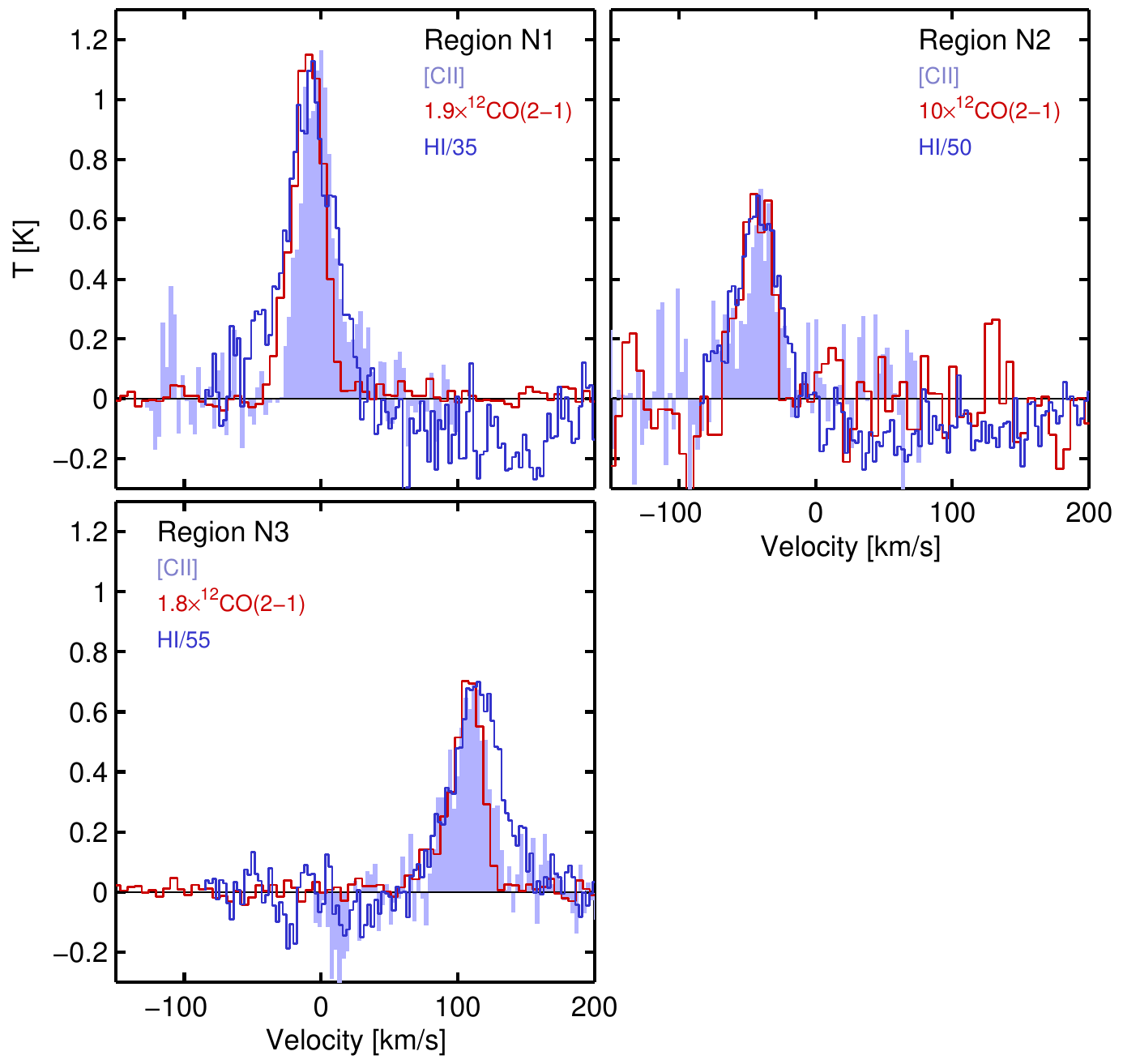}
\centering \caption{\cii, \hi and CO spectra of the three star-forming
  regions in NGC 6946. The \hi and CO peak intensities are scaled to
  approximately match the \cii intensity.  Positions of the regions are indicated in Fig.\ \ref{fig:MIPS}.
\label{fig:SOF_6946}}
\end{figure}

We use the highly resolved SOFIA data obtained for star-forming
regions in M101 (NGC 5457) and NGC 6946.  Figures \ref{fig:SOF_M101}
and \ref{fig:SOF_6946} compare the \cii, CO and \hi spectra at the
positions of these star-forming regions.  It is clear that \cii is
only found at the velocities where \hi and/or CO also occur.  Note
that the spatial resolution of the SOFIA observations is $\sim 0.5$
kpc, so the individual pointings are probing a significant volume.

\begin{deluxetable*}{crrrrrrrrr}
  \tabletypesize{\scriptsize} \tablewidth{0pt}
  \tablecaption{Gaussian fit parameters.\label{tab:gau}}
  \tablehead{\colhead{Region} & \multicolumn{3}{c}{\cii} & \multicolumn{3}{c}{CO} & \multicolumn{3}{c}{\hi}\\
    \noalign{\smallskip}\cline{2-4} \cline{5-7} \cline{8-10}\noalign{\smallskip}
    \colhead{} & \colhead{peak} & \colhead{vel.} & \colhead{disp.} & \colhead{peak} & \colhead{vel.} & \colhead{disp.} & \colhead{peak} & \colhead{vel.} & \colhead{disp.} \\
\colhead{} & \colhead{[K]} & \colhead{[km s$^{-1}$]} & \colhead{[km s$^{-1}$]} & \colhead{[K]} & \colhead{[km s$^{-1}$]} & \colhead{[km s$^{-1}$]} & \colhead{[K]} & \colhead{[km s$^{-1}$]} & \colhead{[km s$^{-1}$]}}
\startdata
         M1 & $0.90\pm0.48$ & $273.8\pm6.9$ & $16.2\pm9.8$ & $0.28\pm0.02$ & $274.7\pm0.8$ & $13.3\pm1.1$ & $56.38\pm2.42$ & $273.6\pm0.7$ & $19.9\pm1.0$ \\
        M2 & $0.26\pm0.14$ & $297.0\pm9.5$ & $22.7\pm13.4$ & $0.03\pm0.02$ & $302.9\pm6.8$ & $18.8\pm9.7$ & $24.81\pm1.65$ & $297.5\pm2.1$ & $37.5\pm2.9$ \\         
        M3 & $0.19\pm0.06$ & $229.7\pm7.3$ & $27.4\pm10.4$ & $0.11\pm0.02$ & $240.1\pm1.7$ & $10.3\pm2.4$ & $24.16\pm4.40$ & $238.2\pm2.1$ & $14.4\pm3.0$ \\         
        M4 & $0.44\pm0.32$ & $185.5\pm9.6$ & $16.3\pm13.7$ & $0.06\pm0.01$ & $188.7\pm3.4$ & $19.6\pm4.7$ & $48.05\pm1.80$ & $188.1\pm0.9$ & $28.1\pm1.2$ \\ 
         \hline
         N1 & $1.09\pm0.64$ & $-3.1\pm8.2$ &$17.1\pm11.6$ &$0.61\pm0.02$ &$-6.8\pm0.4$ &$16.1\pm0.6$ &$34.04\pm3.18$ & $-7.5\pm2.0$ & $25.6\pm2.8$ \\
         N2 & $0.55\pm0.21$& $-41.0\pm5.9$ & $19.0\pm8.3$ & $0.07\pm0.02$ & $-41.1\pm2.8$ &  $15.4\pm4.0$ & $31.27\pm3.95$ & $-41.5\pm2.2$ & $21.4\pm3.1$ \\
         N3 &$0.63\pm0.09$ &$109.0\pm2.3$ &$20.2\pm3.3$ &$0.39\pm0.02$ &$109.9\pm0.7$ &$15.5\pm1.0$ &$36.62\pm2.48$ &$115.4\pm1.4$ &$25.9\pm2.0$ 
\enddata 
\tablecomments{Gaussian fit parameters of the profiles shown in Figs.\ \ref{fig:SOF_M101} and \ref{fig:SOF_6946}. ``peak'', ``vel.'', and ``disp.'' indicate the peak flux, central velocity and velocity dispersion, respectively.
}
\end{deluxetable*} 

\begin{figure}
\includegraphics[width=0.9\hsize]{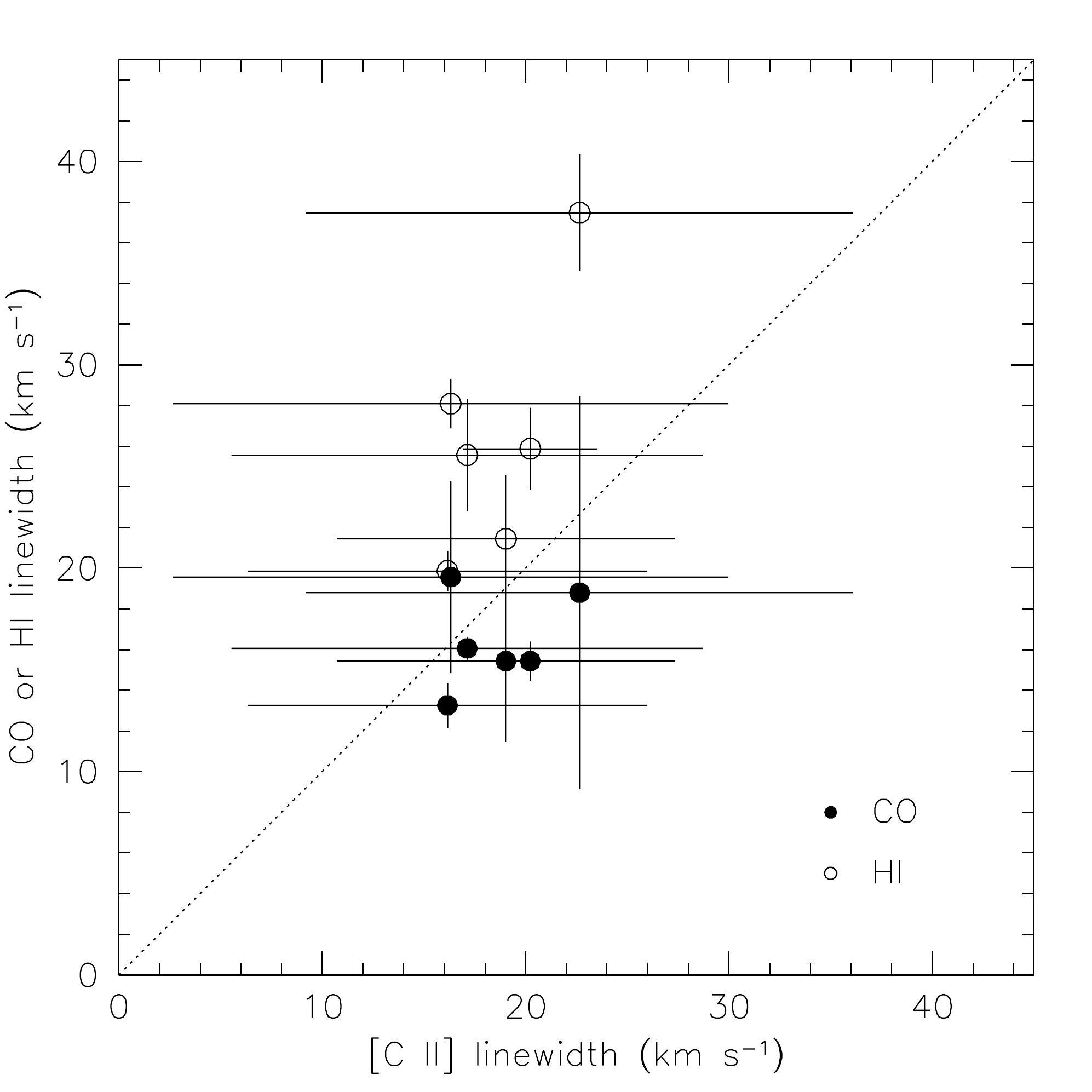}
\centering \caption{Comparison of the \cii, \hi and CO linewidths
  (Gaussian dispersions) of the spectra shown in
  Figs.\ \ref{fig:SOF_M101} and \ref{fig:SOF_6946}. Filled circles
  compare the CO and \cii linewidths, open circles compare the \hi and
  the \cii. Errorbars indicate 95\% confidence intervals. 
  Region M3 is not shown due to the low signal-to-noise of the \cii spectrum.
  Note that the CO values linewidths agree better with the \cii values than the \hi linewidths do.
\label{fig:SOF_disp}}
\end{figure}

The majority of the profiles are approximately Gaussian, and we
characterize them as such. The parameters of the fits are given in
Table~\ref{tab:gau}. In Fig.\ \ref{fig:SOF_disp} we compare the fitted
velocity widths (Gaussian dispersions) of the various profiles. We
have excluded region M3 due to the low signal-to-noise of the
\cii. The \cii linewidths are intermediate between those of CO and
\hi, but always closer to those of the CO. For the regions probed
here, the agreement in CO and \cii velocity dispersion values suggests
that the distribution of CO and \cii along the line-of-sight is
similar.

It also suggests that the contribution from photo-dissociation regions
dominates the emission, at least in these regions.  This implies that
at least the bright \cii emission is associated with CO.  This is a
very similar situation to what we found for the integrated spectra
above (see also \citealt{braine, okada15, mookerjea, mookerjea15}).

\section{Discussion and summary.}

We have shown that for our sample of nearby disk galaxies, the \cii
radial surface density distribution follows that of the CO more
closely than that of the \HI.  For one galaxy (NGC 5055) we find that,
at common spectral resolution, the \cii integrated spectrum resembles
that of the CO more than that of the \hi.

We note that it is not a surprise that the \cii follows the CO more
closely than the \hi: \citet{herrera15} show that for the KINGFISH
galaxies the \cii surface brightness is a good tracer of the SFR
surface density. With the CO surface density also being a good tracer
\citep{leroy08}, a tight relation between CO and \cii is therefore
expected (on the scales considered here).  This correspondence is also
found on the scales of individual star-forming regions, as shown by
the SOFIA \cii data presented here.  We note that these conclusions
may not be applicable for very low metallicity objects, where CO can
be deficient, or ultra-luminous infra-red galaxies, where a \cii
deficit has been observed, as already noted in the Introduction.

The physical origins of the tight relation between CO and \cii are not
entirely clear. \cii emission can originate from both the ionized and
the neutral phase of the ISM.  While naively one would expect that the
good correlation with SFR suggests that the majority of the \cii
emission arises from the ionized phase, there is no evidence for the
presence of a very significant ionized component in most galaxies.
The ionized gas contribution of \cii can be traced reliably by the
[N{\sc\, ii}] 205$\mu$m line, which has a very similar critical
density. This fraction has been found to be relatively small in nearby
galaxies ($\sim 30\%$), Croxall et al.\ in prep).  Therefore, at least
in nearby galaxies, it is more likely that the \cii emission is
dominated by the contribution of the neutral gas, in which case the
observed relation is due to the heating of the neutral gas by 6--13 eV
photons which are a reasonable tracer of the instantaneous SFR.  An
extensive and more complete discussion of the origins of the \cii
emission is given in Croxall et al.\ (in preparation).

We conclude that if spatially unresolved \cii observations are used to
determine dynamical masses then the well-characterised length scale of
the CO is preferred over that of the \HI (see
\citealt{schruba11,deblok14}) and approximates the dynamical mass that
would have been derived from the \cii itself to within a factor of
two.  The CO scale length is known to be similar to that of the
stellar disk in nearby galaxies \citep{young95, regan01, schruba11}
and maybe even in high-$z$ main sequence galaxies \citep{bolatto15}.

Our conclusions can be applied to galaxies observed at high redshift,
provided they are dominated by rotation and assuming that the locally
observed relations between \cii, CO and SFR hold there as well.

\acknowledgements WJGdB was supported by the European Commission
(grant FP7-PEOPLE-2012-CIG \#333939).  Beyond the Peak research has
been supported by a NASA/JPL grant (RSA 1427378). JDS gratefully
acknowledges visiting support from the Alexander von Humboldt
Foundation and the Max Planck Institute f\"ur Astronomie.  ADB and RHC
acknowledge support from grants USRA-SOF020098, NSF-AST0955836, and
NSF-AST1412419, as well as visiting support from the Humboldt
Foundation.  This work is based in part on observations made with the
NASA/DLR Stratospheric Observatory for Infrared Astronomy
(SOFIA). Financial support for this work was provided by NASA.

\clearpage

\appendix

\begin{figure*}
\includegraphics[width=0.9\hsize]{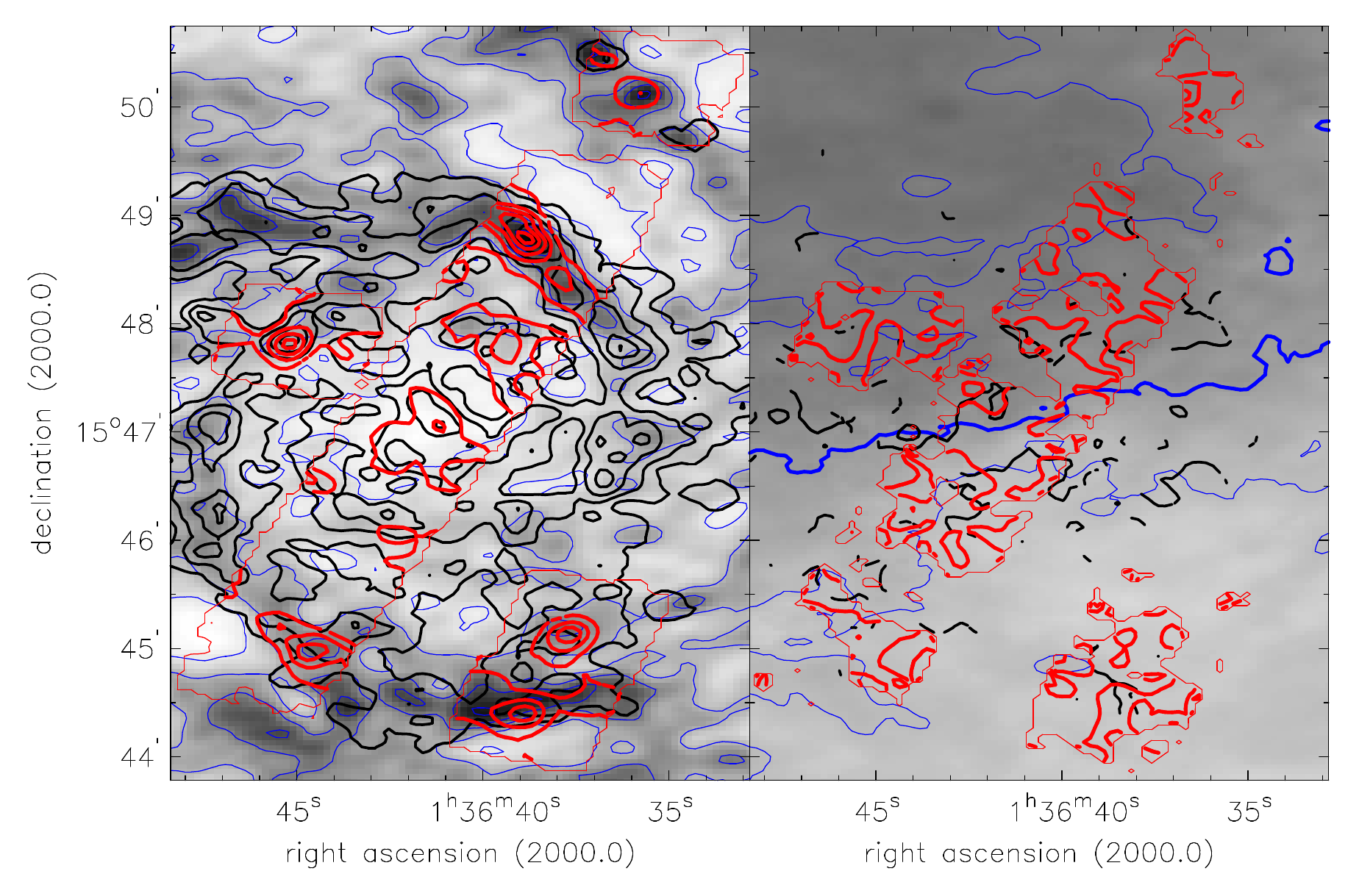}
\caption{\HI, CO and \cii moment maps of NGC 0628.  \emph{Left panel:}
  Integrated intensity of zeroth moment map.  \emph{Right panel:}
  Velocity field or first-moment map. All grayscales and contours are
  as in Fig.\ \ref{fig:mom_01}, except for the velocity contours, which indicate
a systemic velocity of 659.1 \kms (thick blue contour) and have a spacing of 20 \kms.
\label{fig:mom_02}}
\end{figure*}

\begin{figure*}
\includegraphics[width=0.9\hsize]{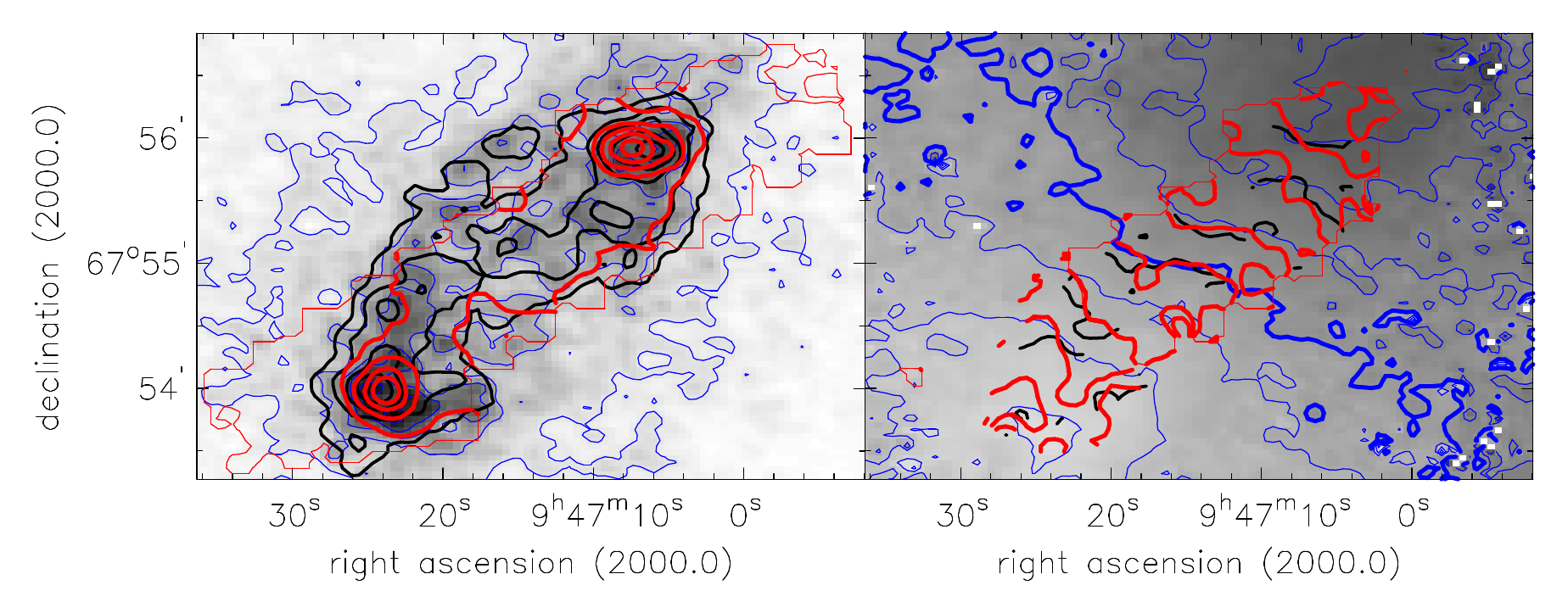}
\caption{\HI, CO and \cii moment maps of NGC 2976.  \emph{Left panel:}
  Integrated intensity of zeroth moment map.  \emph{Right panel:}
  Velocity field or first-moment map. All grayscales and contours are
  as in Fig.\ \ref{fig:mom_01}, except for the velocity contours, which indicate
a systemic velocity of 1.1 \kms (thick blue contour) and have a spacing of 20 \kms.
\label{fig:mom_03}}
\end{figure*}

\begin{figure*}
\includegraphics[width=0.9\hsize]{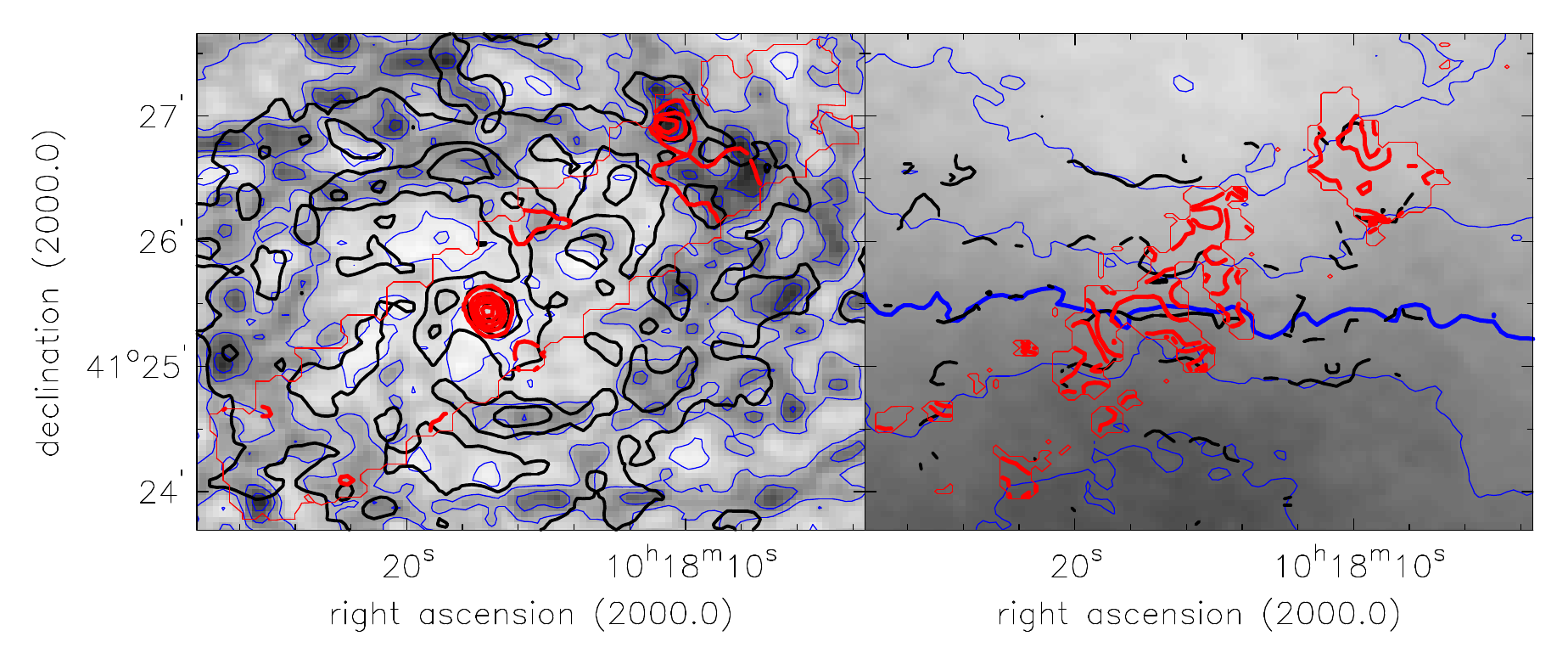}
\caption{\HI, CO and \cii moment maps of NGC 3184.  \emph{Left panel:}
  Integrated intensity of zeroth moment map.  \emph{Right panel:}
  Velocity field or first-moment map. All grayscales and contours are
  as in Fig.\ \ref{fig:mom_01}, except for the velocity contours, which indicate
a systemic velocity of 593.3 \kms (thick blue contour) and have a spacing of 20 \kms.
\label{fig:mom_04}}
\end{figure*}

\begin{figure*}
\includegraphics[width=0.9\hsize]{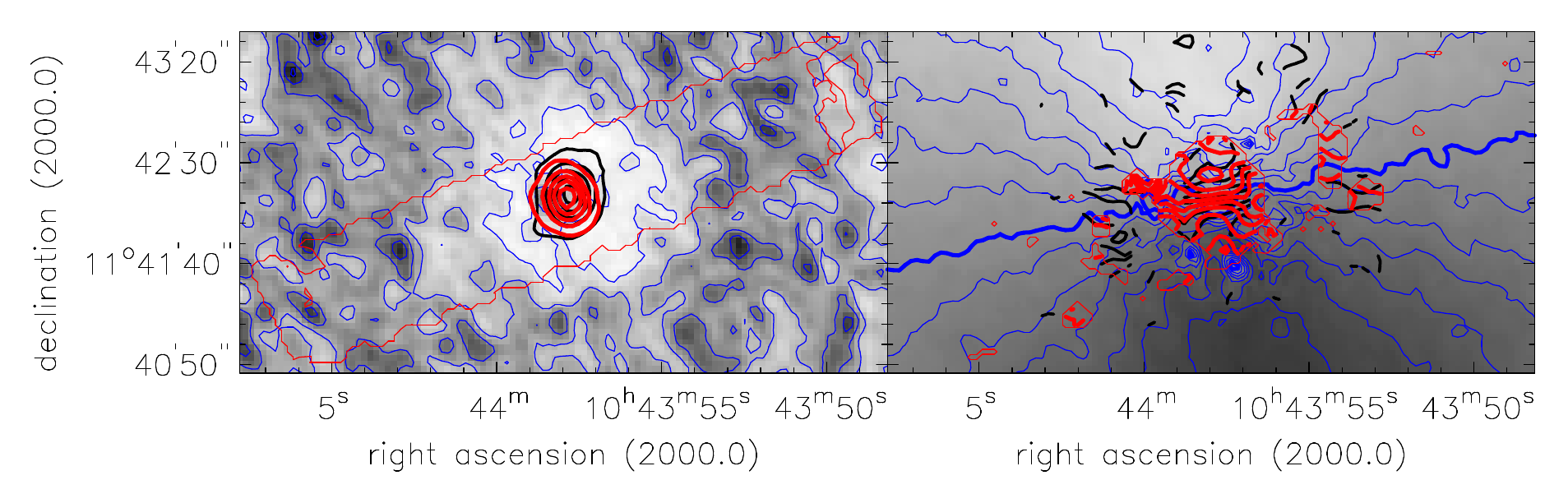}
\caption{\HI, CO and \cii moment maps of NGC 3351.  \emph{Left panel:}
  Integrated intensity of zeroth moment map.  \emph{Right panel:}
  Velocity field or first-moment map. All grayscales and contours are
  as in Fig.\ \ref{fig:mom_01}, except for the velocity contours, which indicate
a systemic velocity of 779.0 \kms (thick blue contour) and have a spacing of 20 \kms.
\label{fig:mom_05}}
\end{figure*}

\begin{figure*}
\includegraphics[width=0.9\hsize]{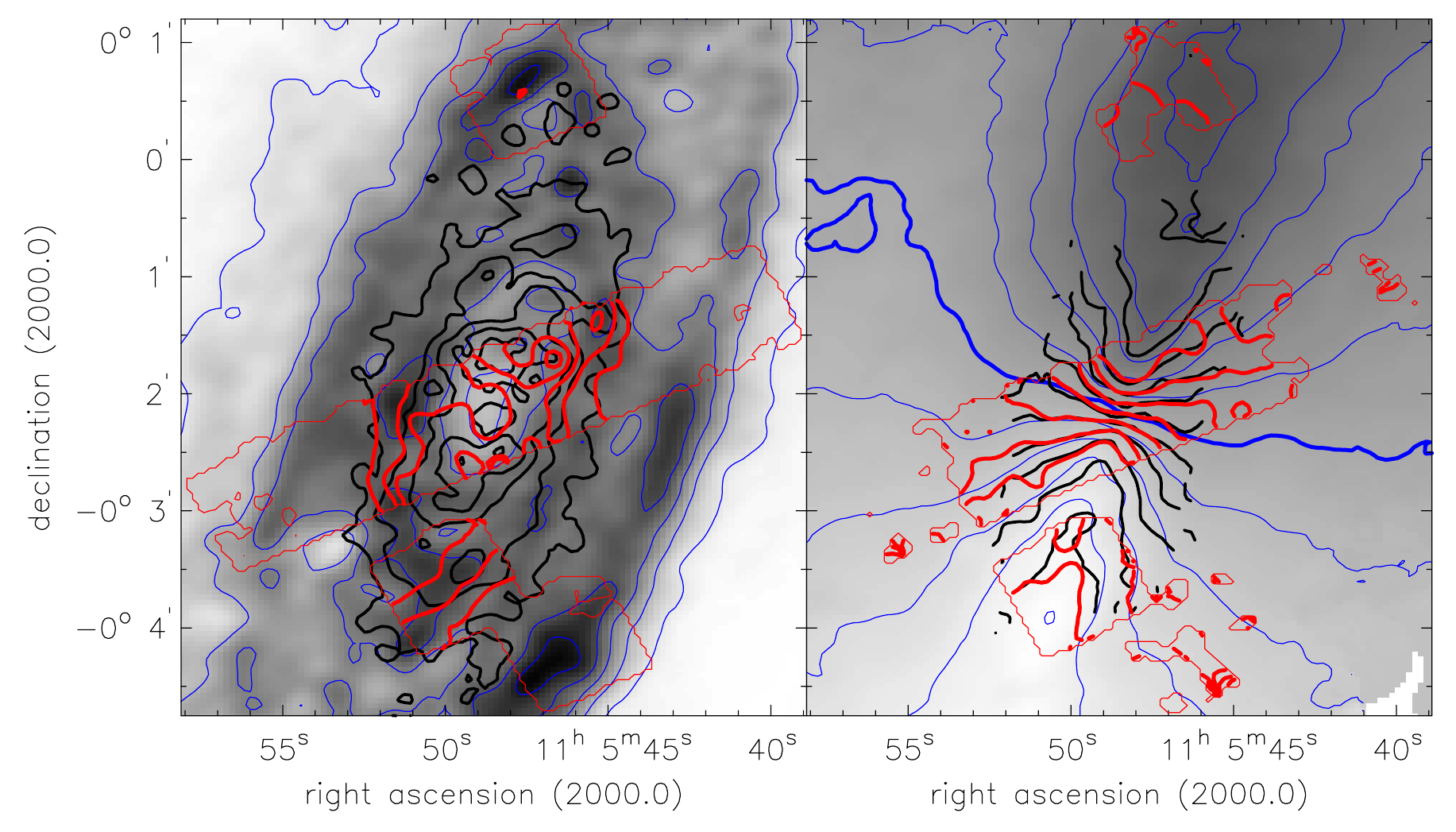}
\caption{\HI, CO and \cii moment maps of NGC 3521.  \emph{Left panel:}
  Integrated intensity of zeroth moment map.  \emph{Right panel:}
  Velocity field or first-moment map. All grayscales and contours are
  as in Fig.\ \ref{fig:mom_01}, except for the velocity contours, which indicate
a systemic velocity of 803.5 \kms (thick blue contour) and have a spacing of 40 \kms.
\label{fig:mom_06}}
\end{figure*}

\begin{figure*}
\includegraphics[width=0.9\hsize]{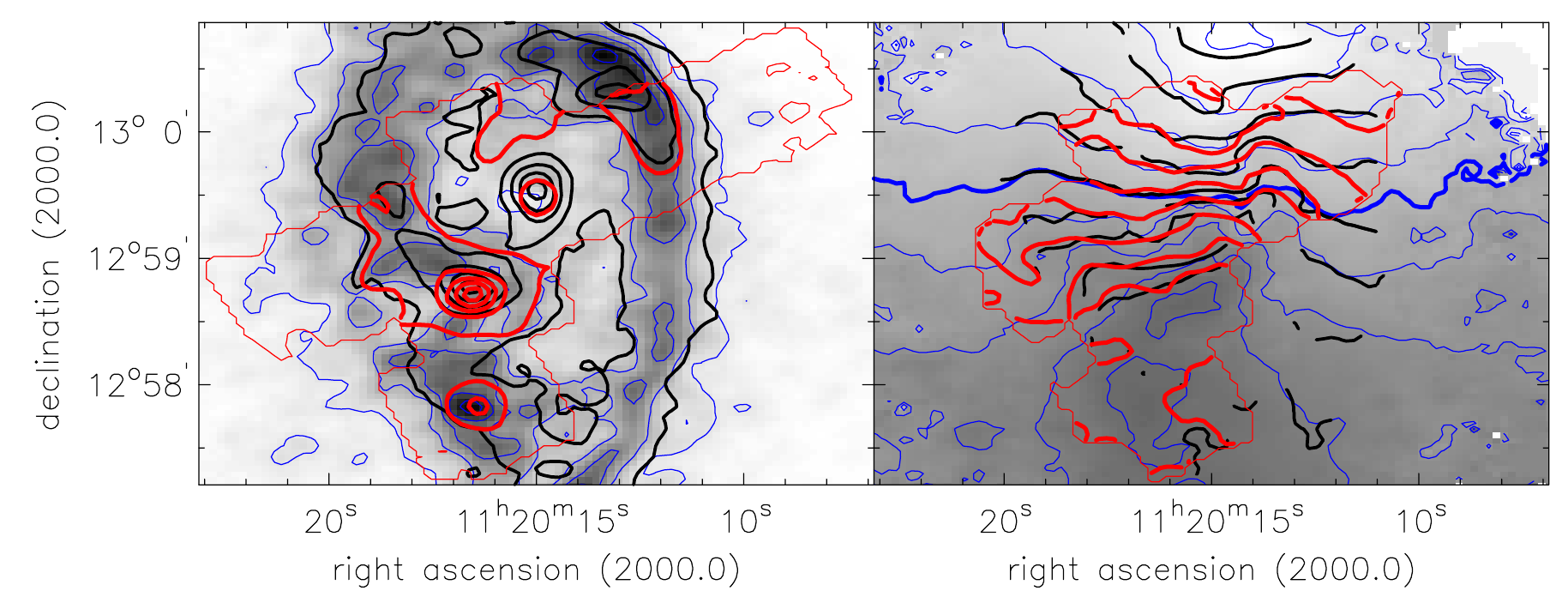}
\caption{\HI, CO and \cii moment maps of NGC 3627.  \emph{Left panel:}
  Integrated intensity of zeroth moment map.  \emph{Right panel:}
  Velocity field or first-moment map. All grayscales and contours are
  as in Fig.\ \ref{fig:mom_01}, except for the velocity contours, which indicate
a systemic velocity of 708.2 \kms (thick blue contour) and have a spacing of 40 \kms.
\label{fig:mom_07}}
\end{figure*}

\begin{figure*}
\includegraphics[width=0.9\hsize]{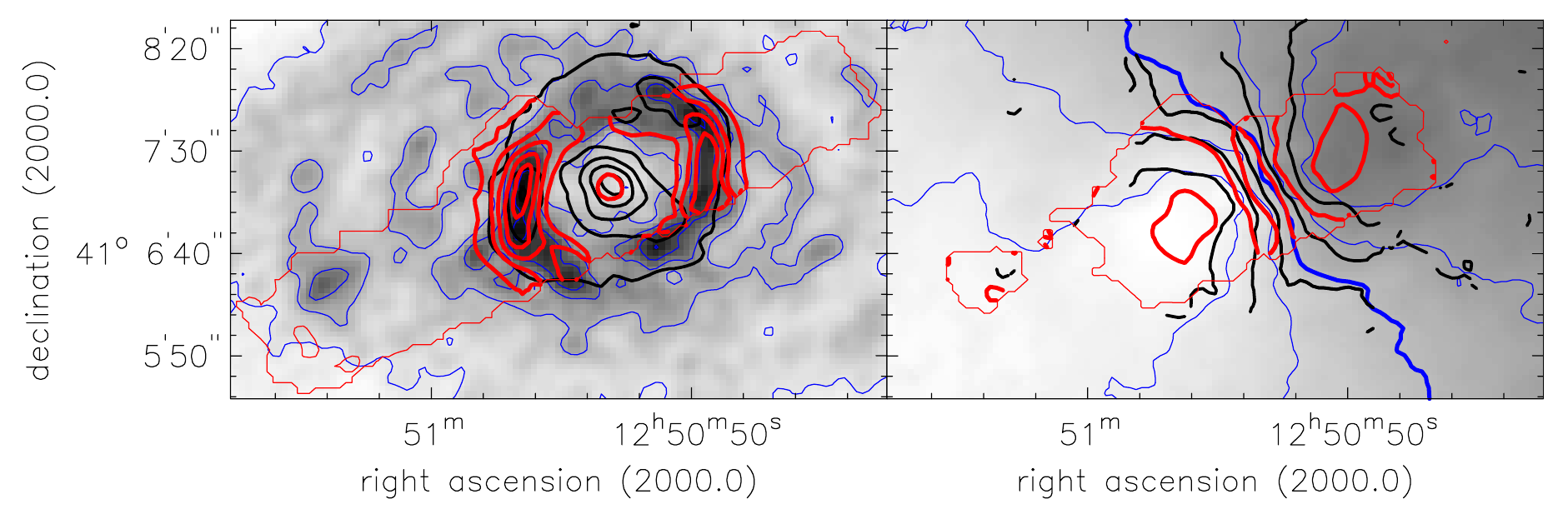}
\caption{\HI, CO and \cii moment maps of NGC 4736.  \emph{Left panel:}
  Integrated intensity of zeroth moment map.  \emph{Right panel:}
  Velocity field or first-moment map. All grayscales and contours are
  as in Fig.\ \ref{fig:mom_01}, except for the velocity contours, which indicate
a systemic velocity of 306.7 \kms (thick blue contour) and have a spacing of 40 \kms.
\label{fig:mom_08}}
\end{figure*}

\begin{figure*}
\includegraphics[width=0.9\hsize]{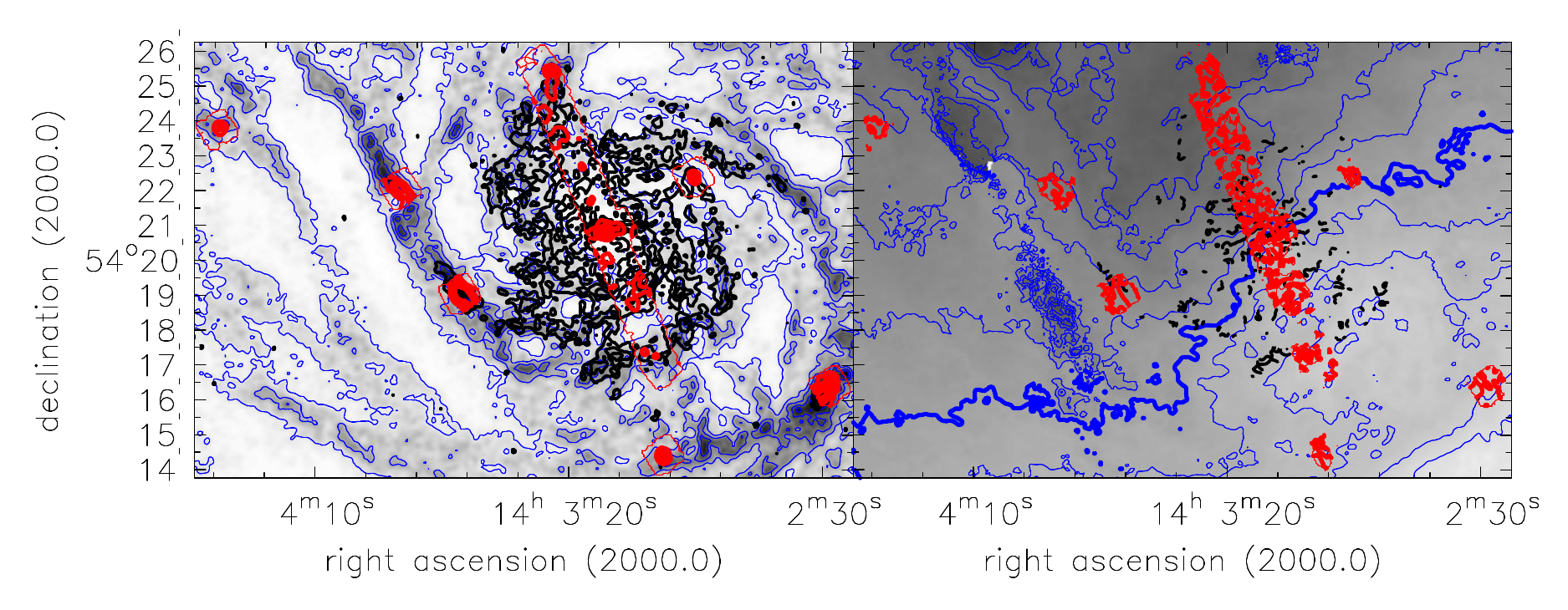}
\caption{\HI, CO and \cii moment maps of NGC 5457.  \emph{Left panel:}
  Integrated intensity of zeroth moment map.  \emph{Right panel:}
  Velocity field or first-moment map. All grayscales and contours are
  as in Fig.\ \ref{fig:mom_01}, except for CO intensity values which
  start at 30 percent of the maximum and the velocity contours, which
  indicate a systemic velocity of 226.5 \kms (thick blue contour) and
  have a spacing of 20 \kms.
\label{fig:mom_09}}
\end{figure*}

\begin{figure*}
\includegraphics[width=0.9\hsize]{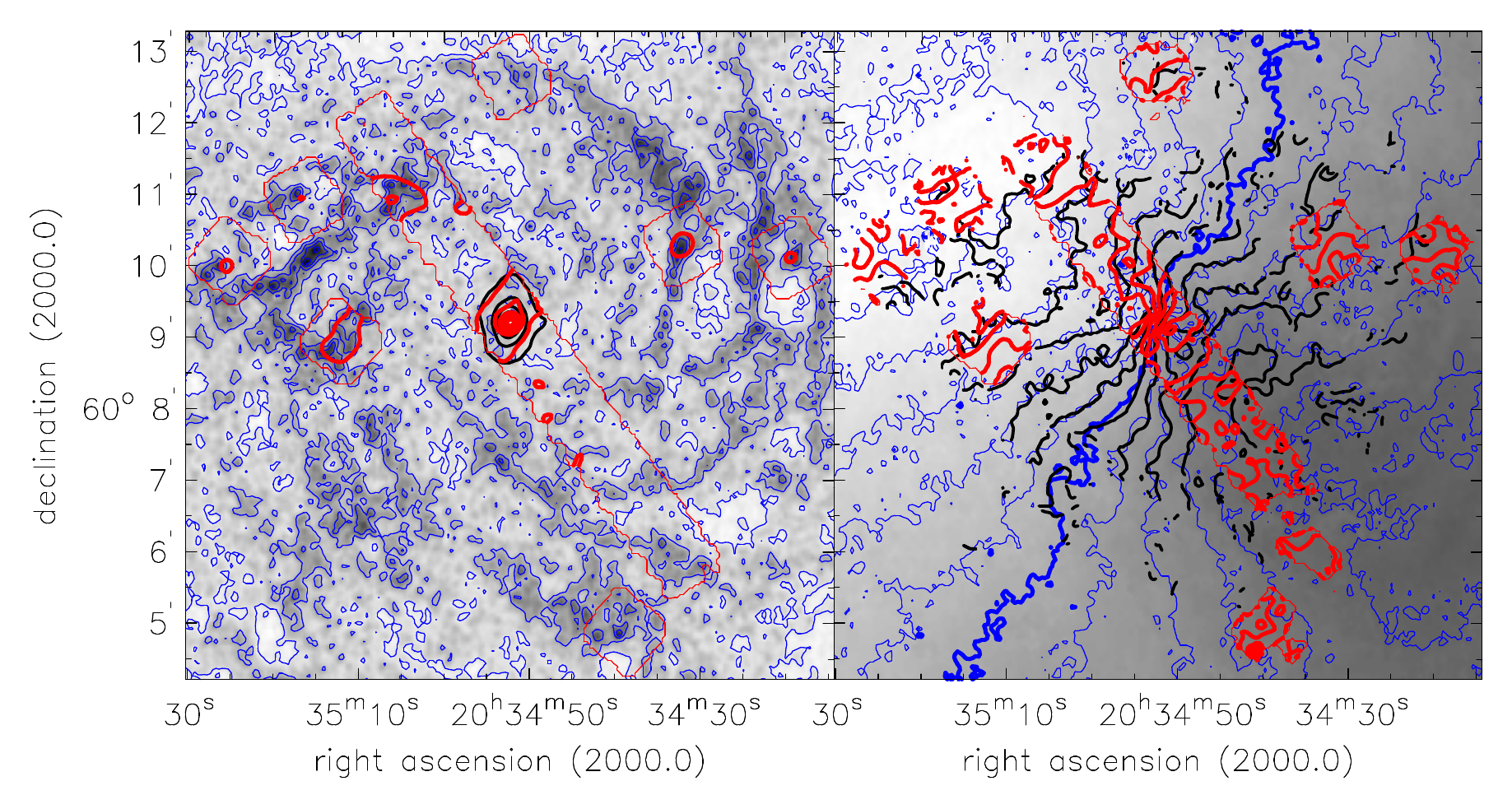}
\caption{\HI, CO and \cii moment maps of NGC 6946.  \emph{Left panel:}
  Integrated intensity of zeroth moment map.  \emph{Right panel:}
  Velocity field or first-moment map. All grayscales and contours are
  as in Fig.\ \ref{fig:mom_01}, except for CO intensity values which
  start at 30 percent of the maximum and the velocity contours, which
  indicate a systemic velocity of 43.3 \kms (thick blue contour) and
  have a spacing of 20 \kms.
\label{fig:mom_10}}
\end{figure*}

{\bf Moment maps:} Figures \ref{fig:mom_02}--\ref{fig:mom_10} show the CO, \HI and \cii
integrated intensity maps and velocity field (zeroth and first moment
maps, respectively). The maps for NGC 5055 are shown in
Fig.\ \ref{fig:mom_01} in the main text.

\begin{figure*}
\begin{center}
\includegraphics[width=0.4\hsize]{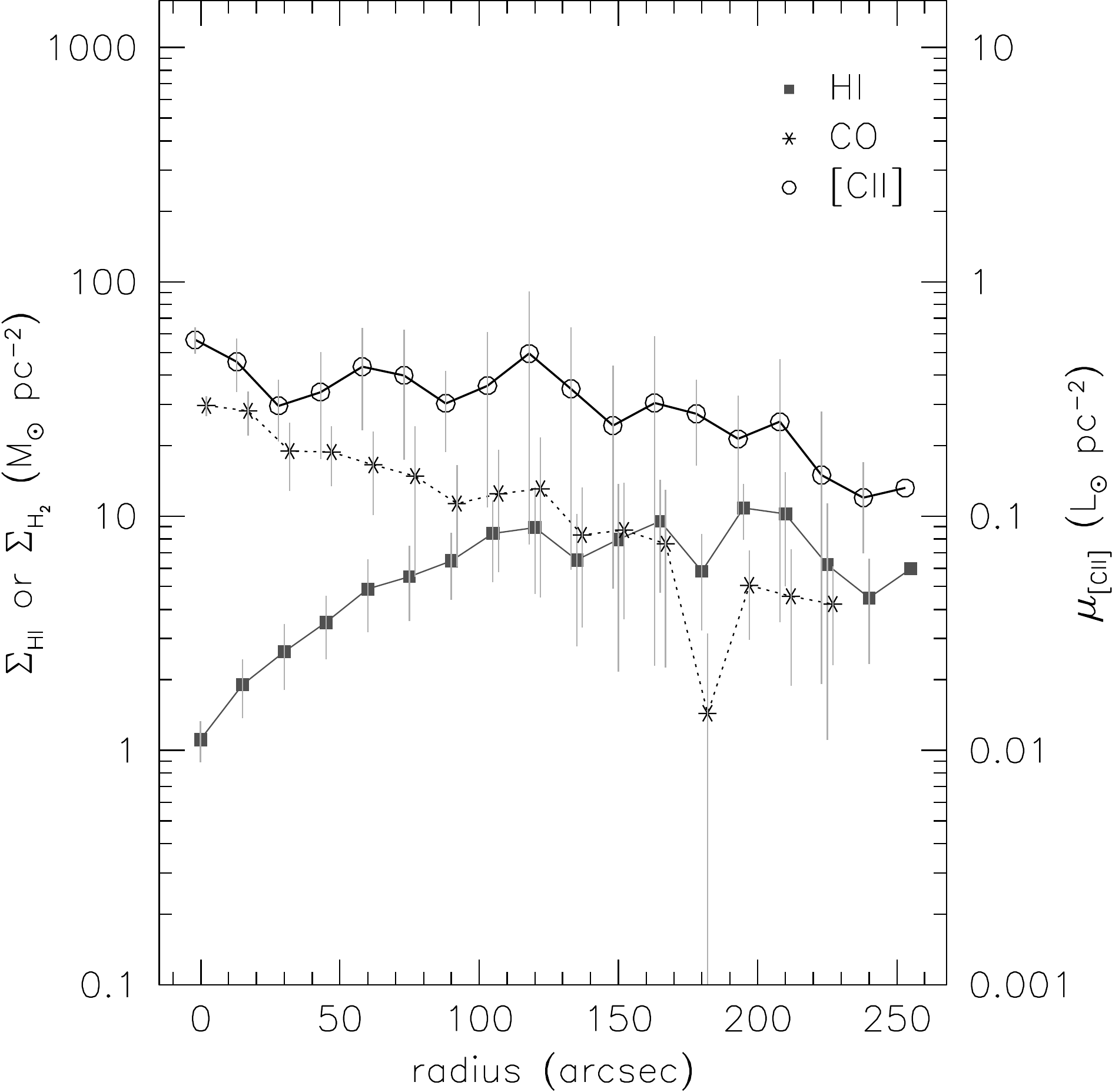}\hspace{1cm}\includegraphics[width=0.4\hsize]{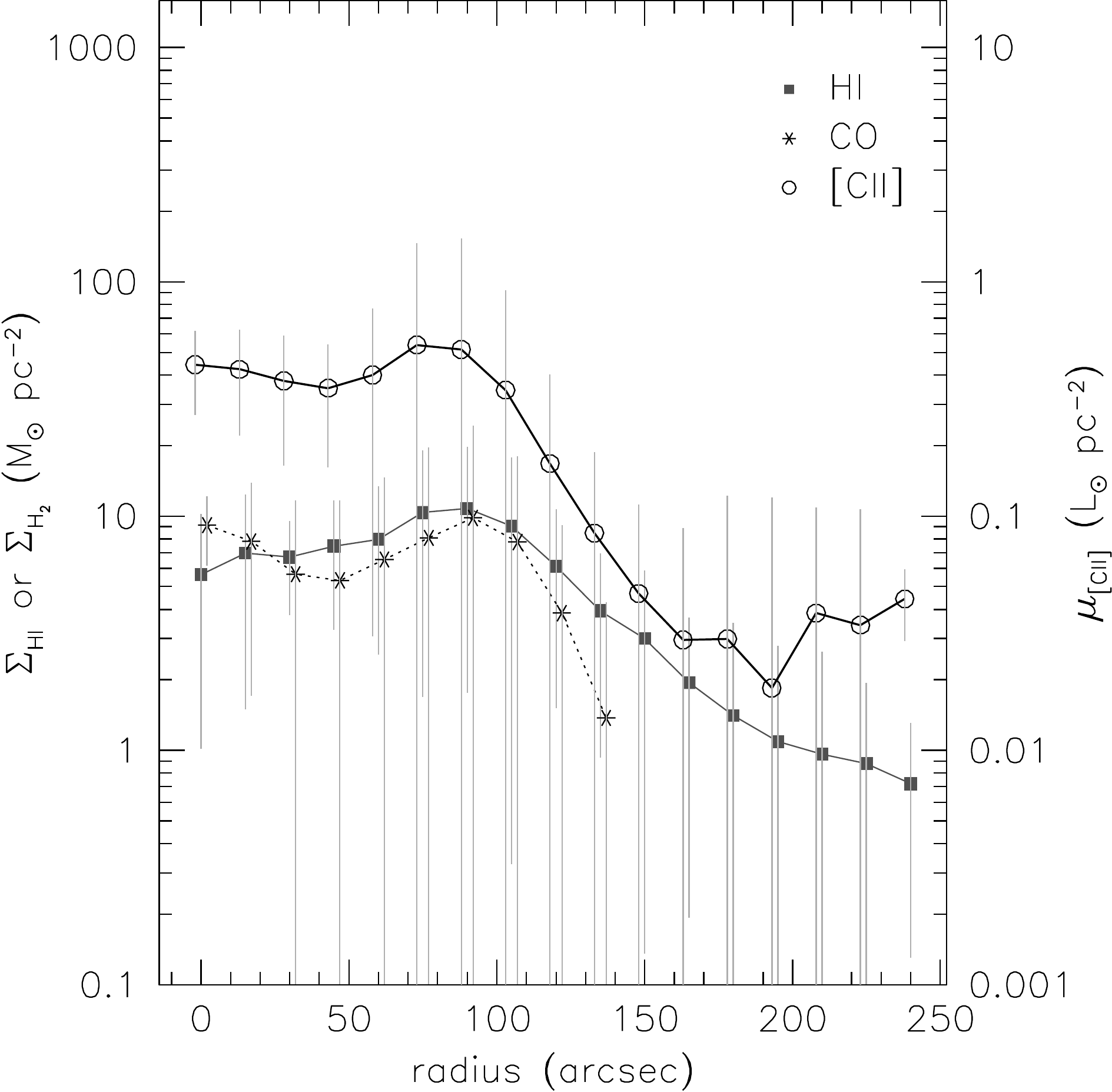}
\caption{CO, \HI and \cii radial profiles for NGC 628 (left panel) and
  NGC 2976 (right panel) over the area covered by the \cii
  observations. Filled squares show the inclination-corrected \HI
  surface density in $M_{\odot}$ pc$^{-2}$. The H$_2$ surface density
  in $M_{\odot}$ pc$^{-2}$ is shown with the star symbols. The \cii
  surface brightness in $L_{\odot}$ pc$^{-2}$ is shown with open
  circles. For clarity, the CO and \cii profiles have been offset by
  $2''$ and $-2''$ with respect to the \HI points. The errorbars
  indicate the RMS spread in values found along each ring.
\label{fig:radialprofs_app1}}
\end{center}
\end{figure*}

\begin{figure*}
\begin{center}
\includegraphics[width=0.4\hsize]{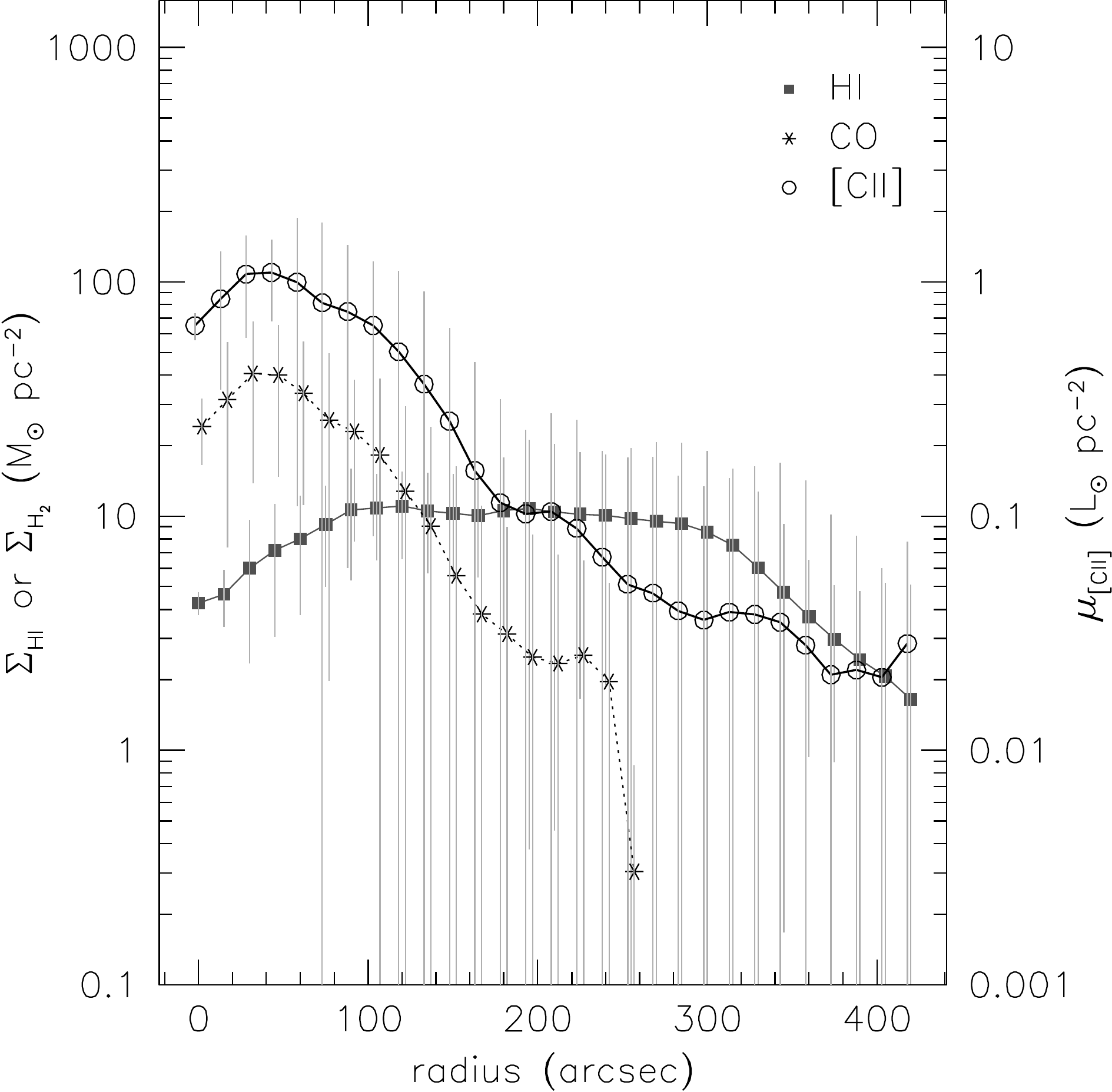}\hspace{1cm}\includegraphics[width=0.4\hsize]{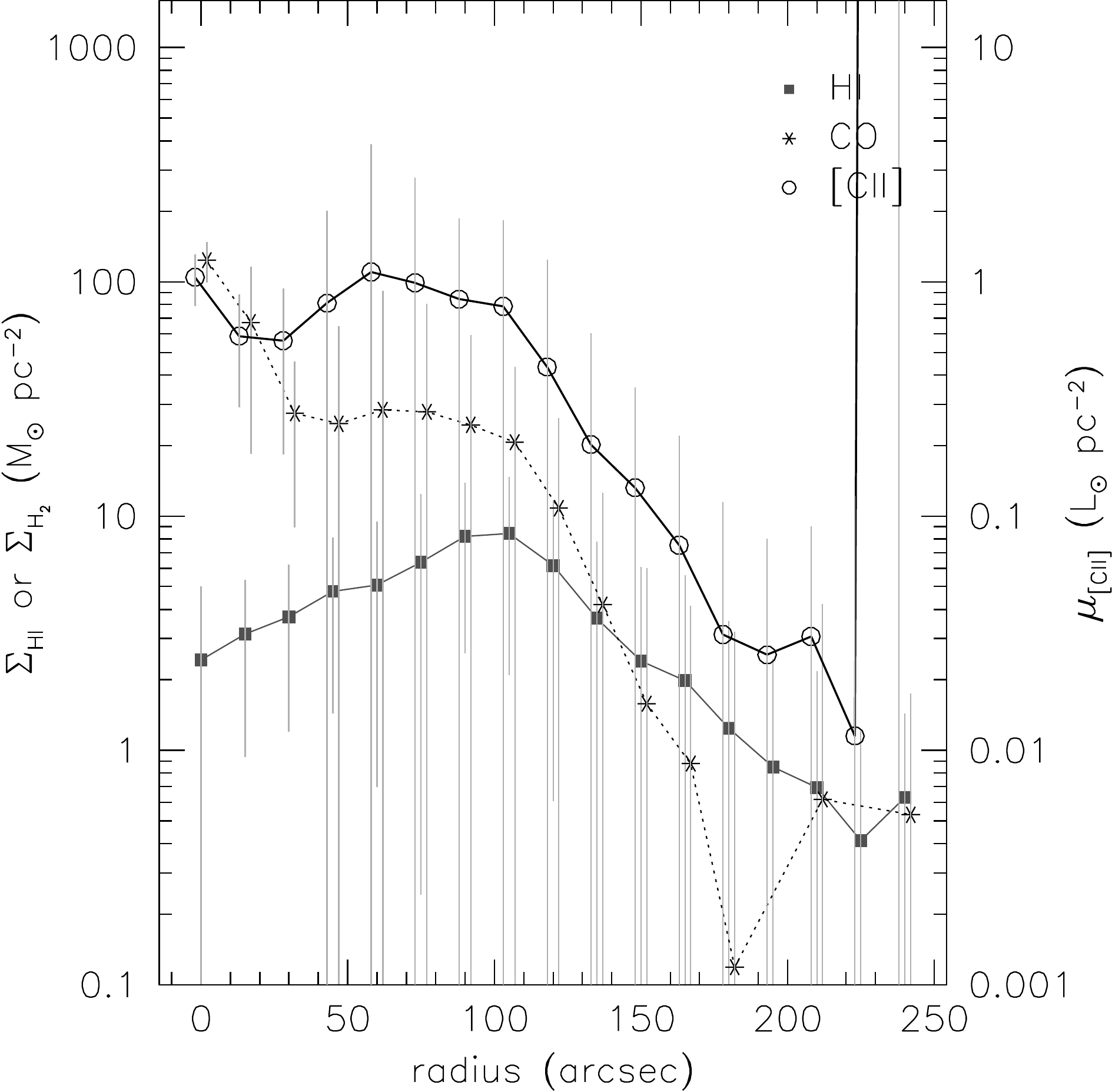}
\caption{CO, \HI and \cii radial profiles for NGC 3521 (left panel) and NGC 3627 (right panel). Symbols and lines as in Fig.\ \ref{fig:radialprofs_app1}.
\label{fig:radialprofs_app2}}
\end{center}
\end{figure*}

\begin{figure*}
\begin{center}
\includegraphics[width=0.4\hsize]{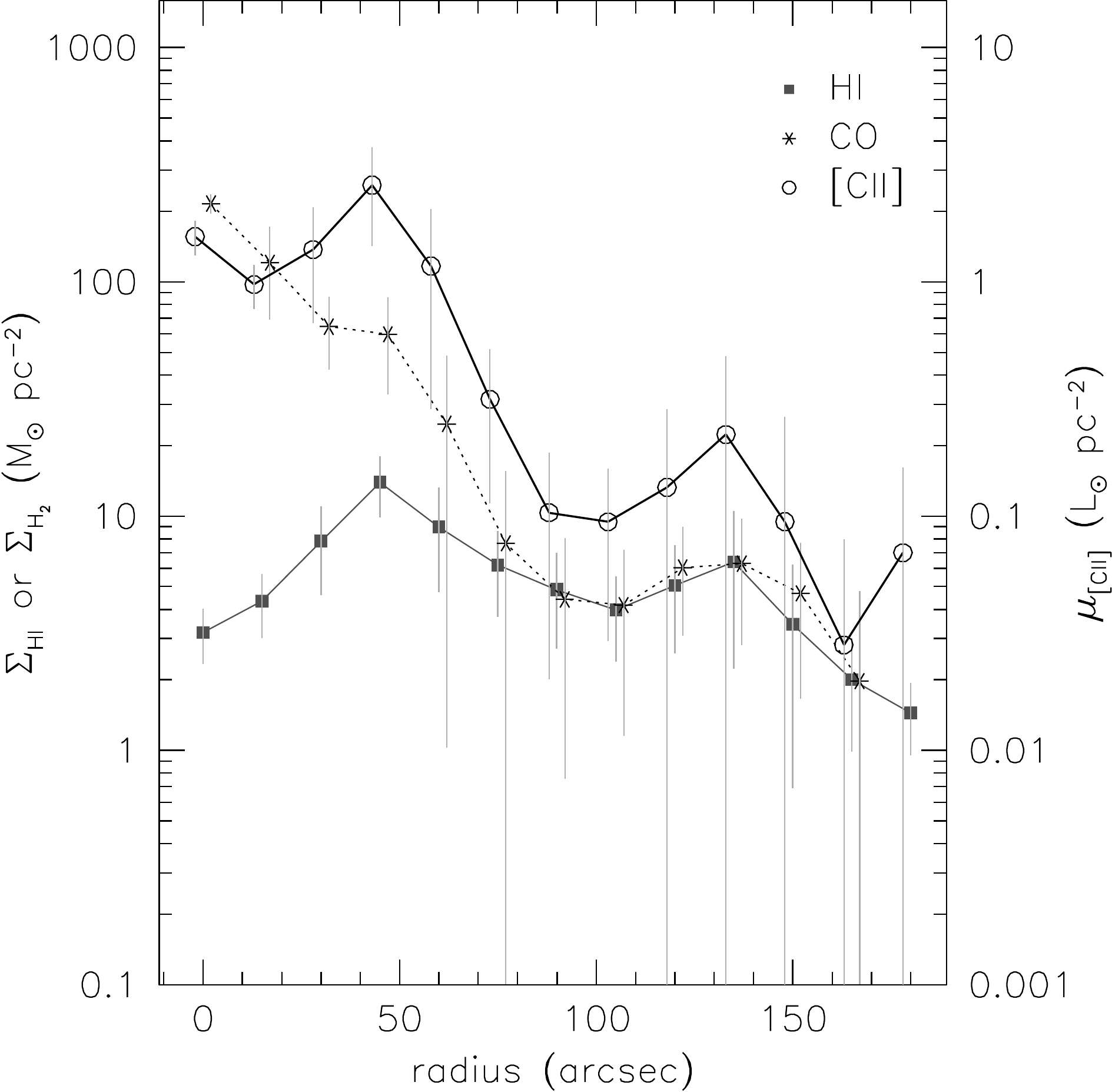}\hspace{1cm}\includegraphics[width=0.4\hsize]{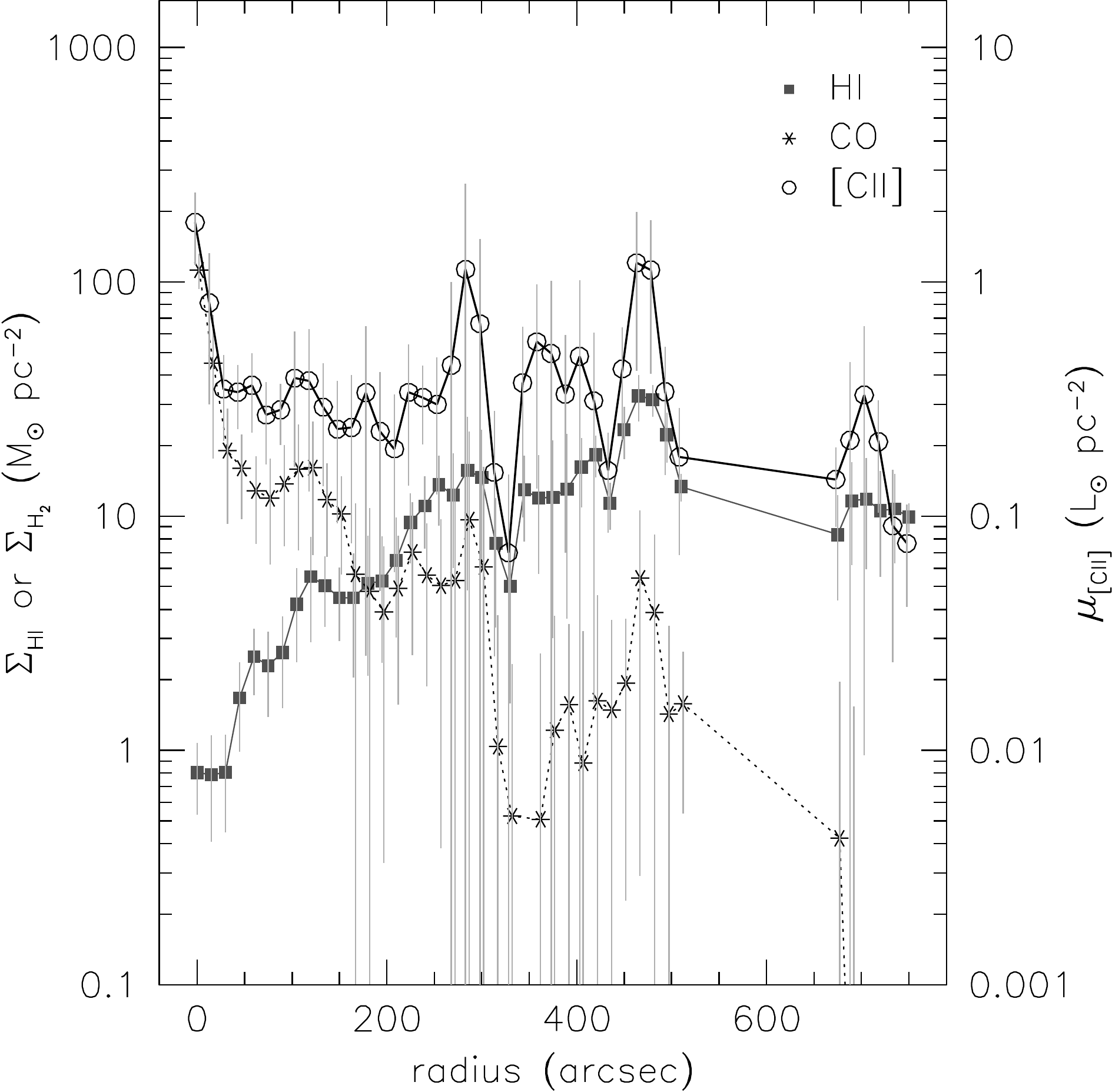}
\caption{CO, \HI and \cii radial profiles for NGC 4736 (left panel) and NGC 5457 (right panel). Symbols and lines as in Fig.\ \ref{fig:radialprofs_app1}.
\label{fig:radialprofs_app3}}
\end{center}
\end{figure*}

\begin{figure*}
\begin{center}
\includegraphics[width=0.4\hsize]{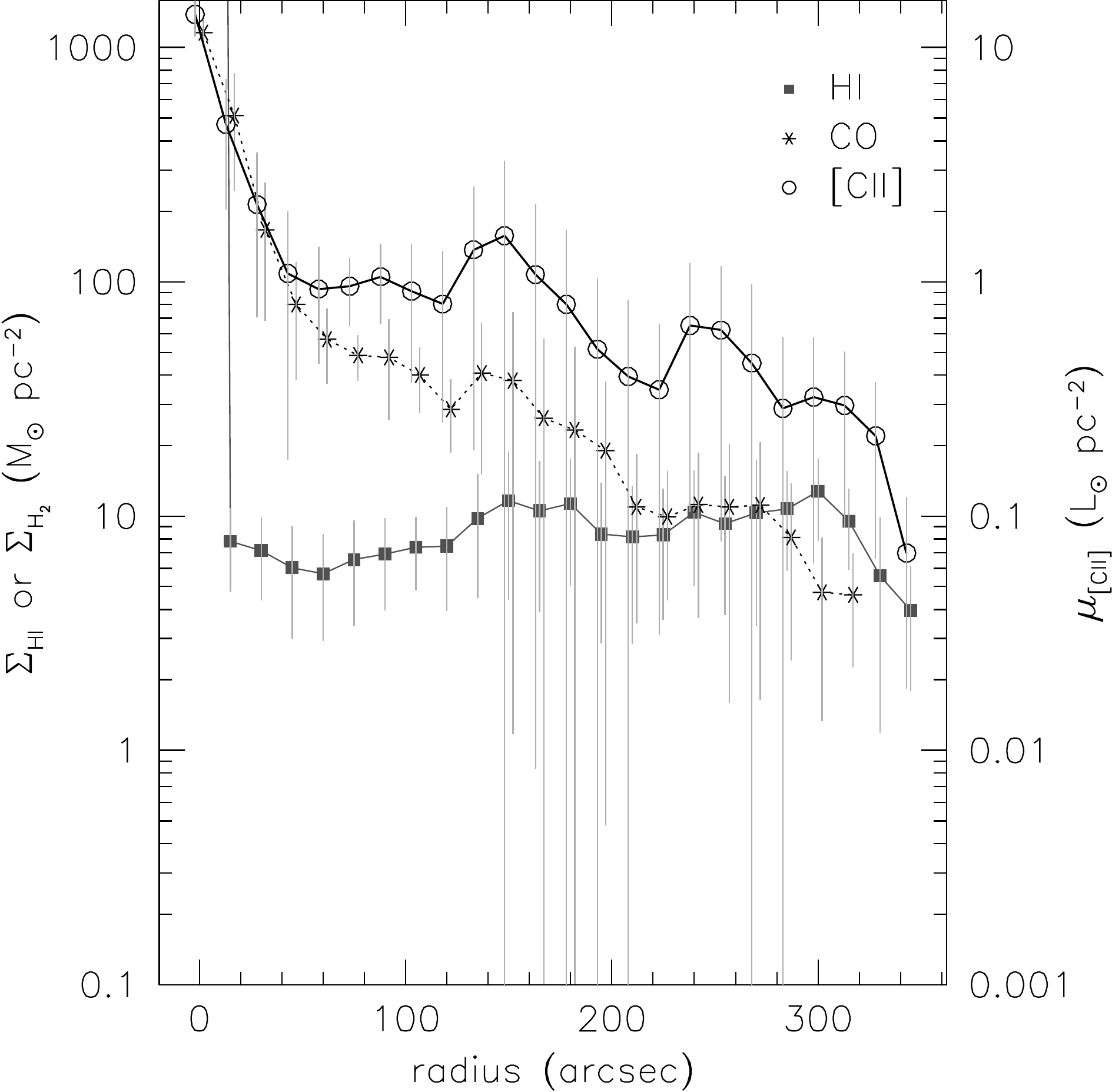}
\caption{CO, \HI and \cii radial profiles for NGC 6946. Symbols and lines as in Fig.\ \ref{fig:radialprofs_app1}.
\label{fig:radialprofs_app4}}
\end{center}
\end{figure*}

{\bf Radial profiles:} We show the \cii, CO and \HI radial surface
density profiles in Figs.\ \ref{fig:radialprofs_app1} to
\ref{fig:radialprofs_app4}. The results for NGC 5055 are shown in
Fig.\ \ref{fig:radialprofs} in the main text.

\begin{figure*}
\includegraphics[width=0.9\hsize]{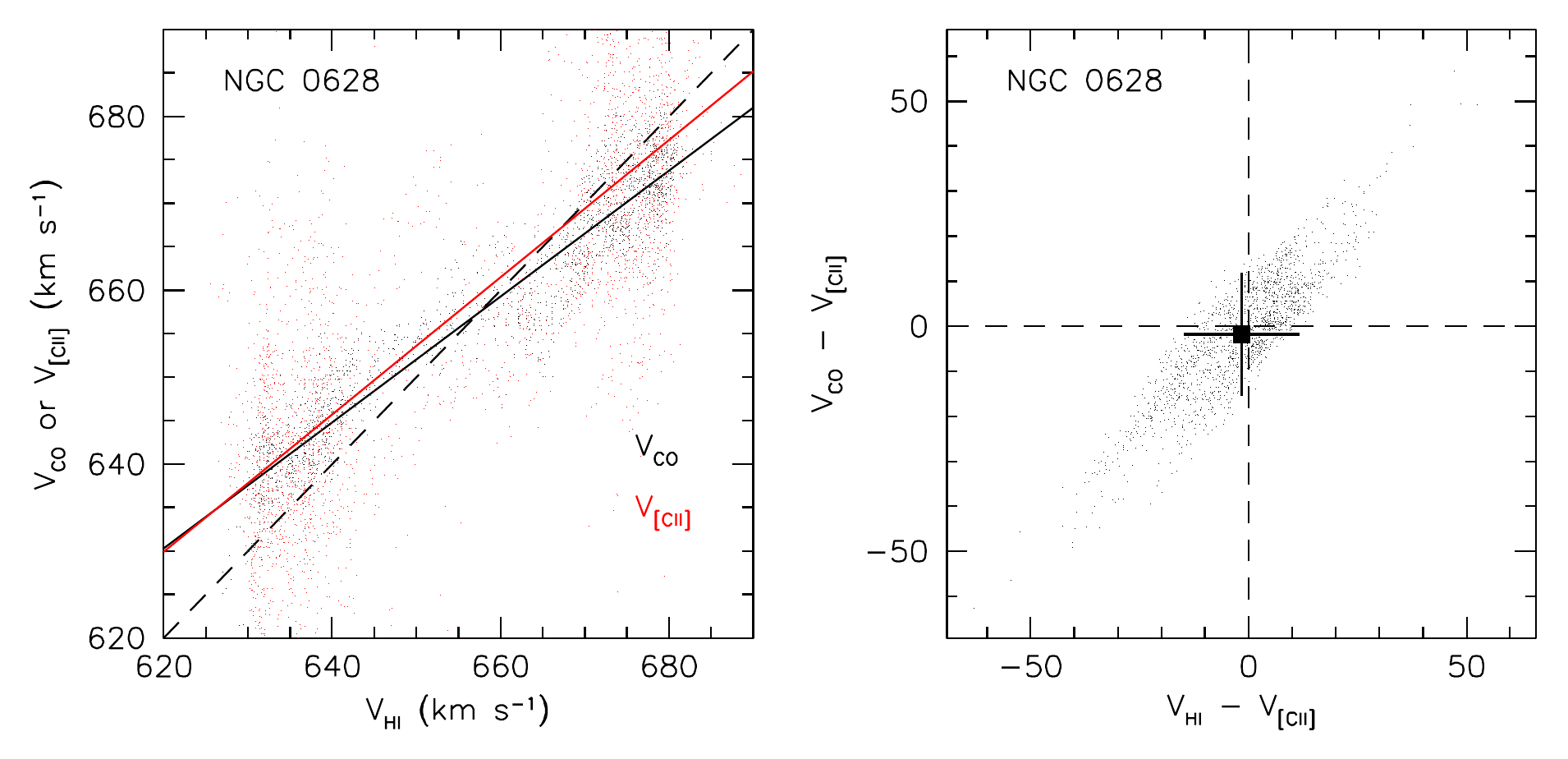}
\caption{ Comparison of the \cii, \hi and CO velocity field values for
  NGC 628. {\bf Left:} CO (black) and \cii (red) velocities plotted
  against the \HI velocities at each position in the velocity
  field. The black dashed line is the line of equality. The red and
  black full lines are linear least-squares fits of the CO (black) and
  the \cii (red) velocities.  {\bf Right:} Differences between \HI and
  \cii velocities plotted against the differences between CO and \cii
  velocities.  The filled square indicates the mean value, the
  errorbars indicate the $1\sigma$ spread in differences.  The small
  systematic offset is due to the uncertainty in determining the \cii
  velocity field given the large velocity resolution of 239 \kms of
  the \cii data.
  \label{fig:vels_app1}}
\end{figure*}

\begin{figure*}
\includegraphics[width=0.9\hsize]{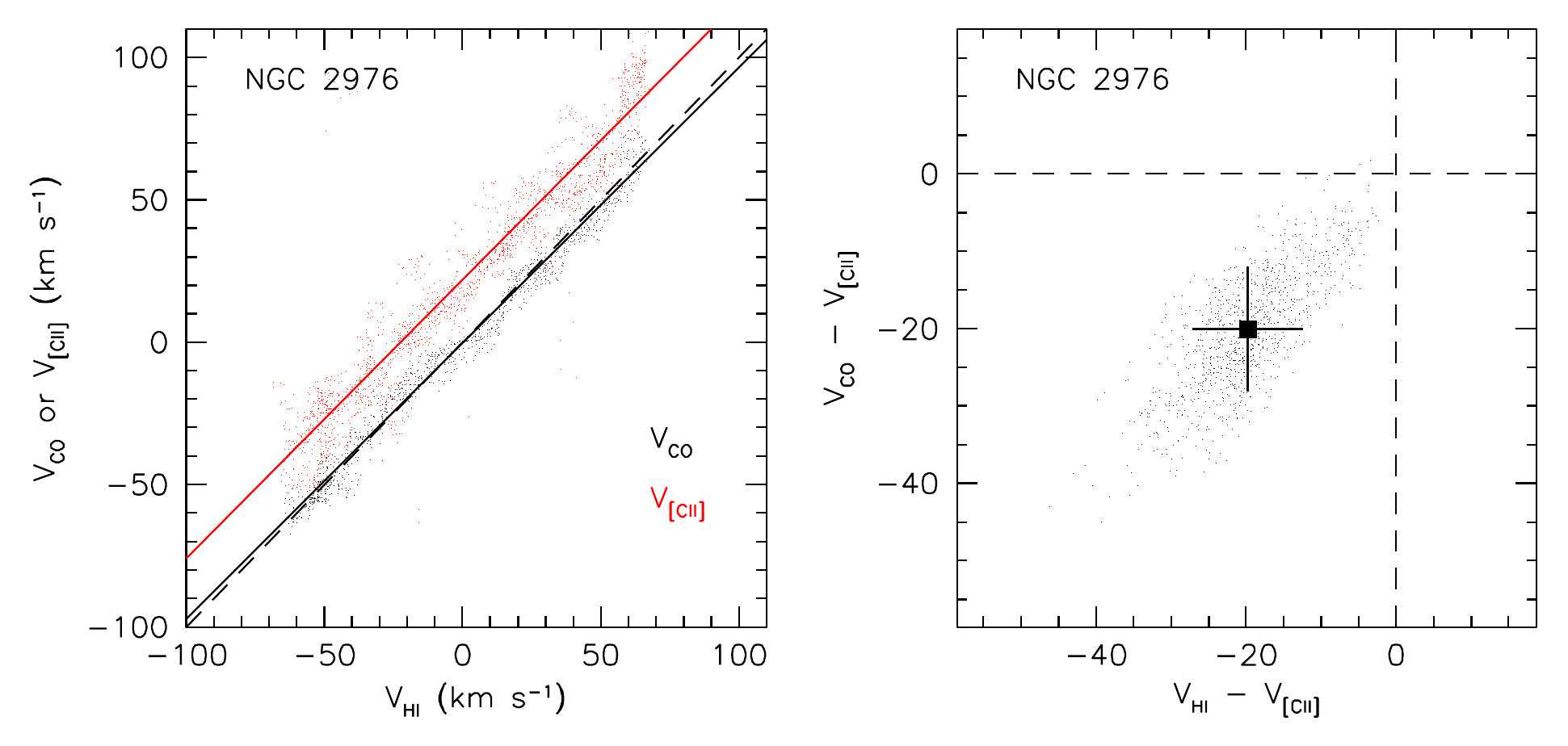}
\caption{As Fig.\ \ref{fig:vels_app1} but for NGC 2976.\label{fig:vels_app2}}
\end{figure*}

\begin{figure*}
\includegraphics[width=0.9\hsize]{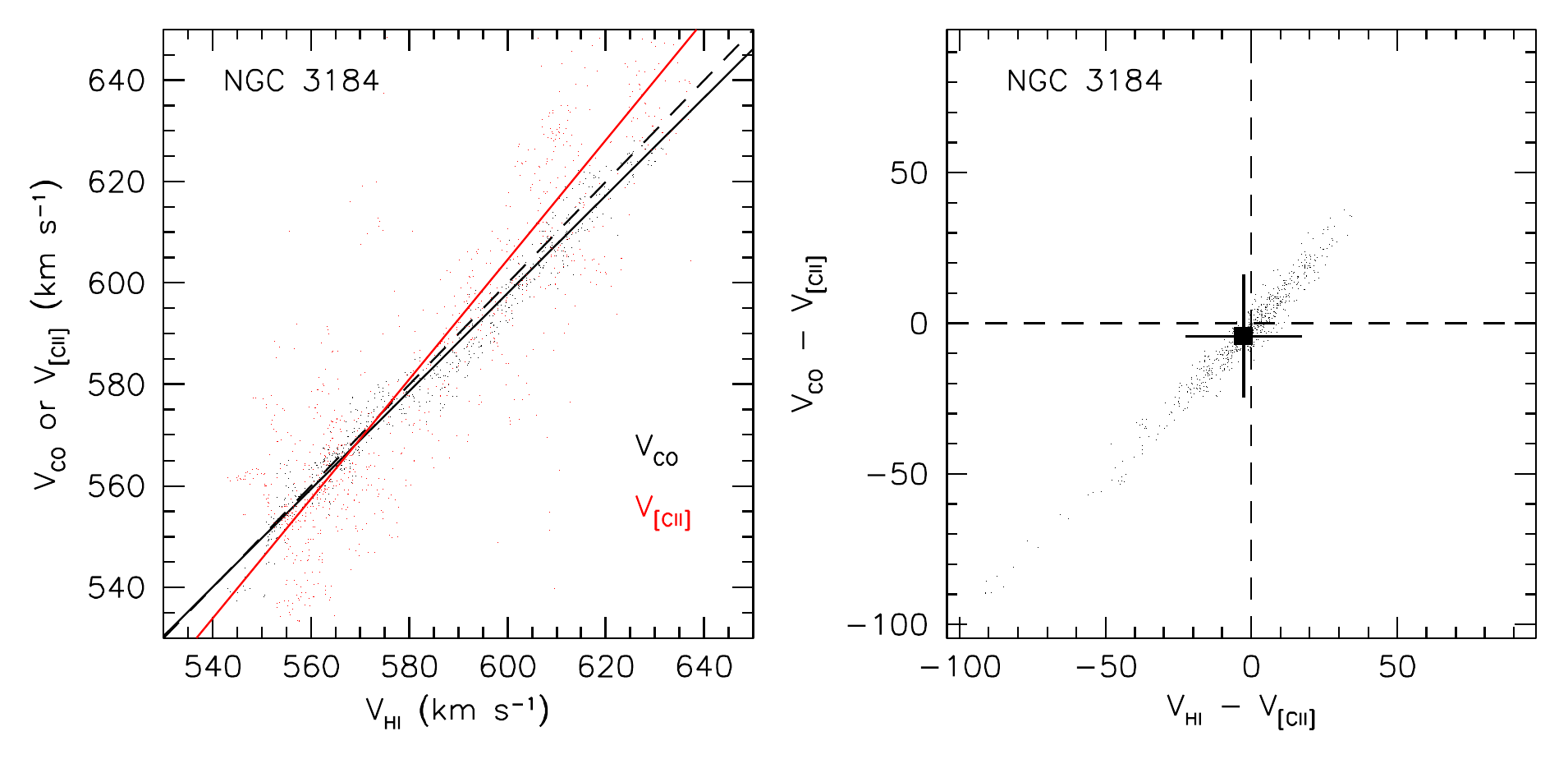}
\caption{As Fig.\ \ref{fig:vels_app1} but for NGC 3184.\label{fig:vels_app3}}
\end{figure*}

\begin{figure*}
\includegraphics[width=0.9\hsize]{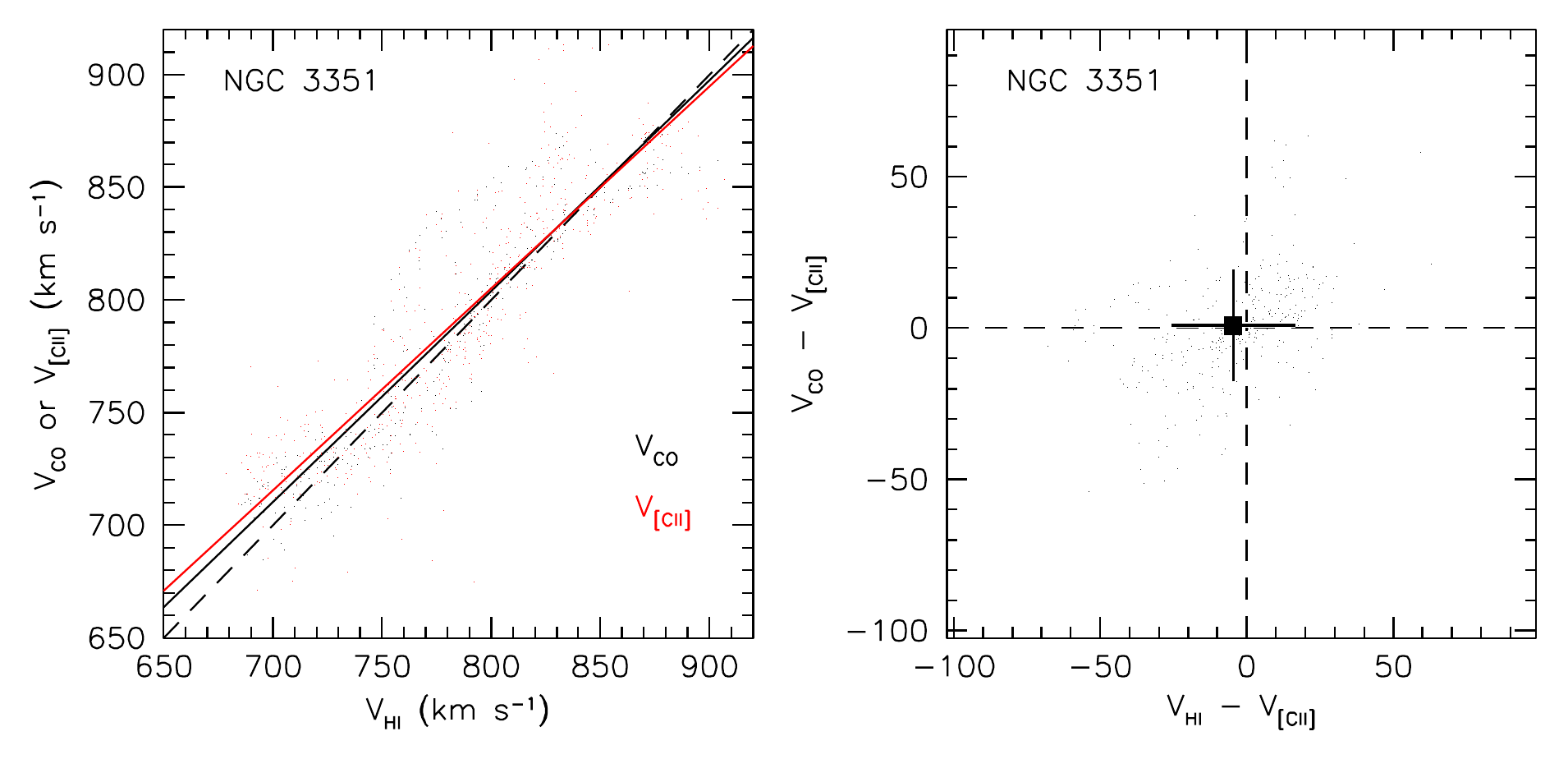}
\caption{As Fig.\ \ref{fig:vels_app1} but for NGC 3351.\label{fig:vels_app4}}
\end{figure*}

\begin{figure*}
\includegraphics[width=0.9\hsize]{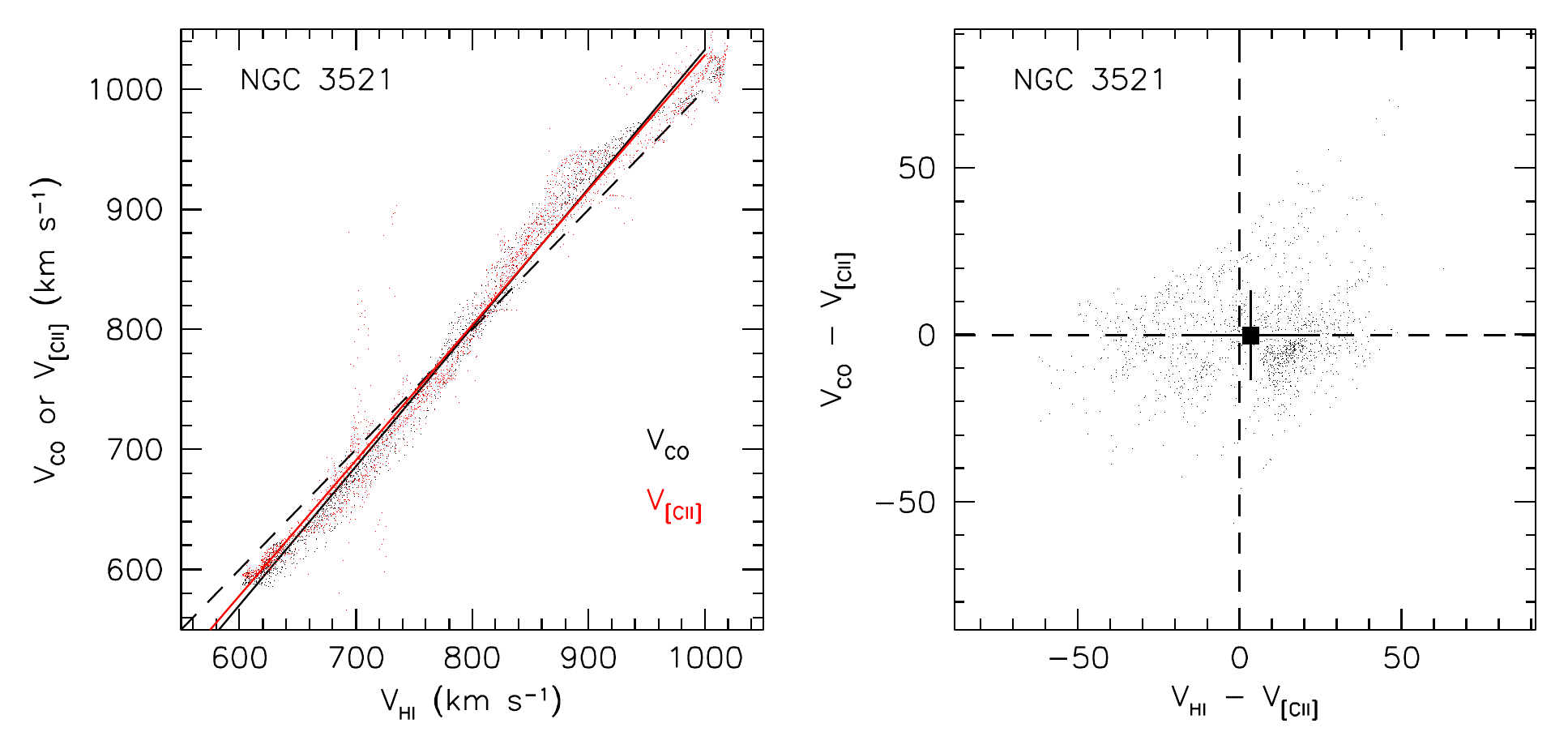}
\caption{As Fig.\ \ref{fig:vels_app1} but for NGC 3521.
\label{fig:vels_app5}}
\end{figure*}

\begin{figure*}
\includegraphics[width=0.9\hsize]{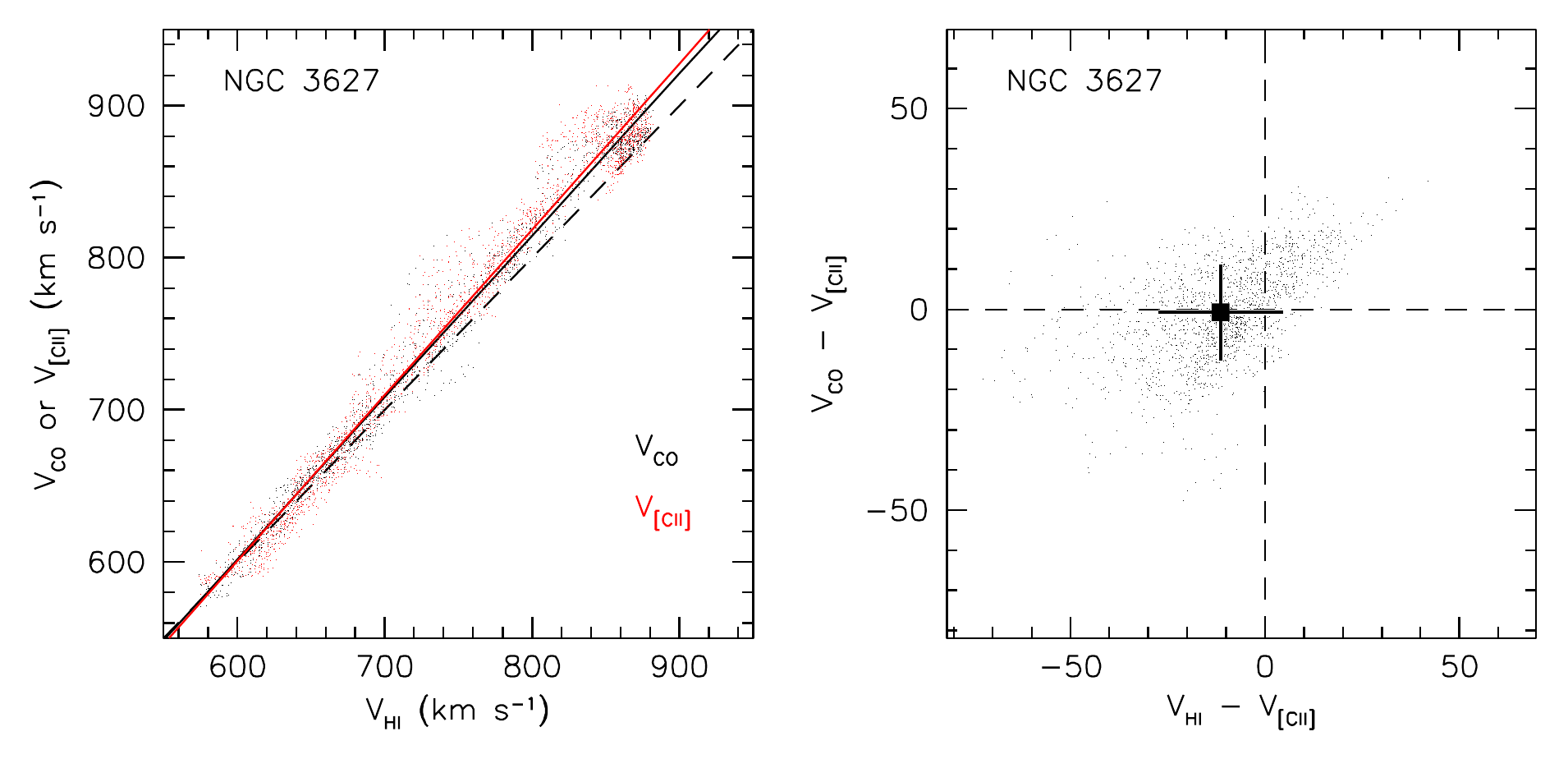}
\caption{As Fig.\ \ref{fig:vels_app1} but for NGC 3627.\label{fig:vels_app6}}
\end{figure*}

\begin{figure*}
\includegraphics[width=0.9\hsize]{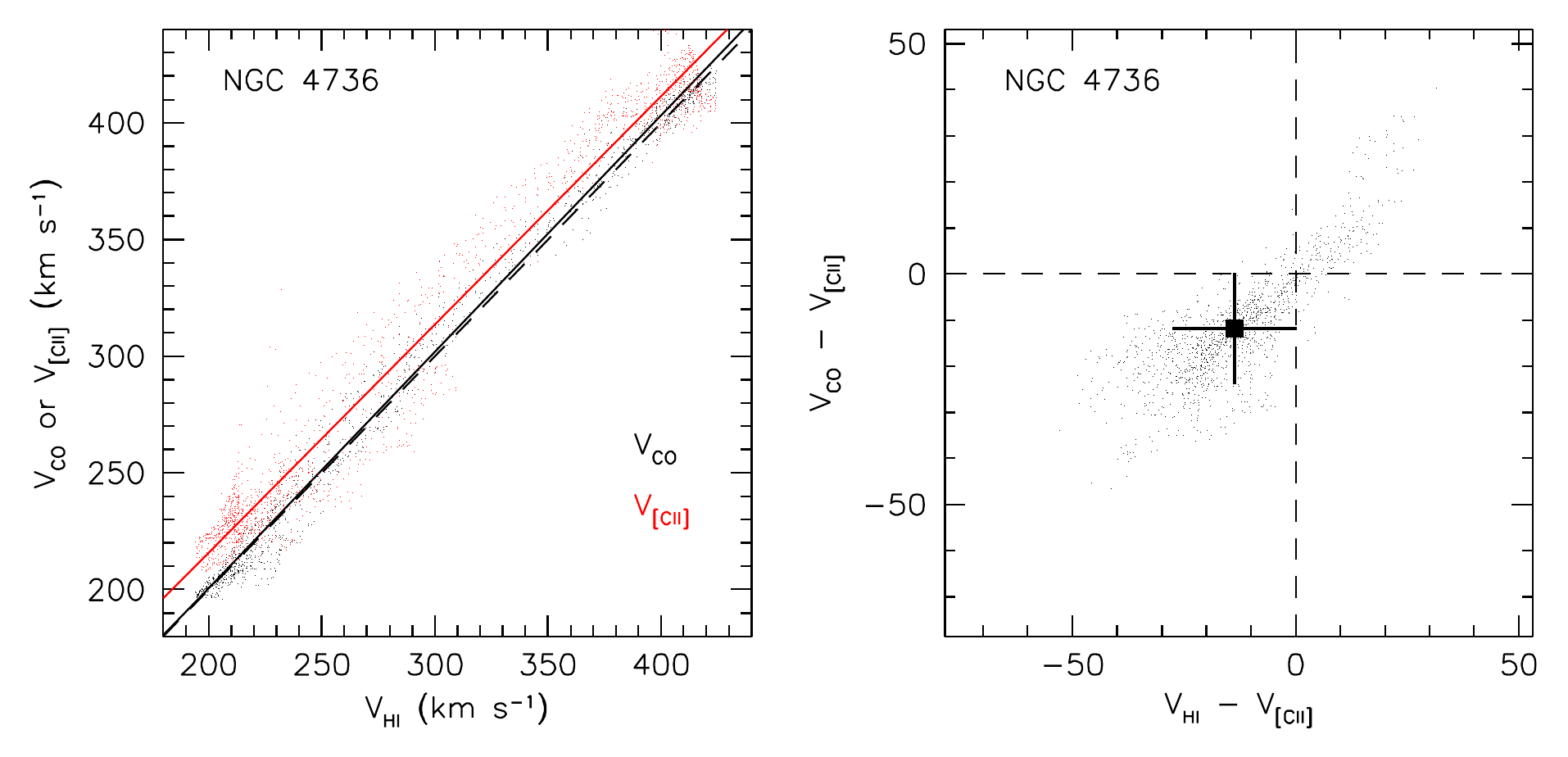}
\caption{As Fig.\ \ref{fig:vels_app1} but for NGC 4736.\label{fig:vels_app7}}
\end{figure*}

\begin{figure*}
\includegraphics[width=0.9\hsize]{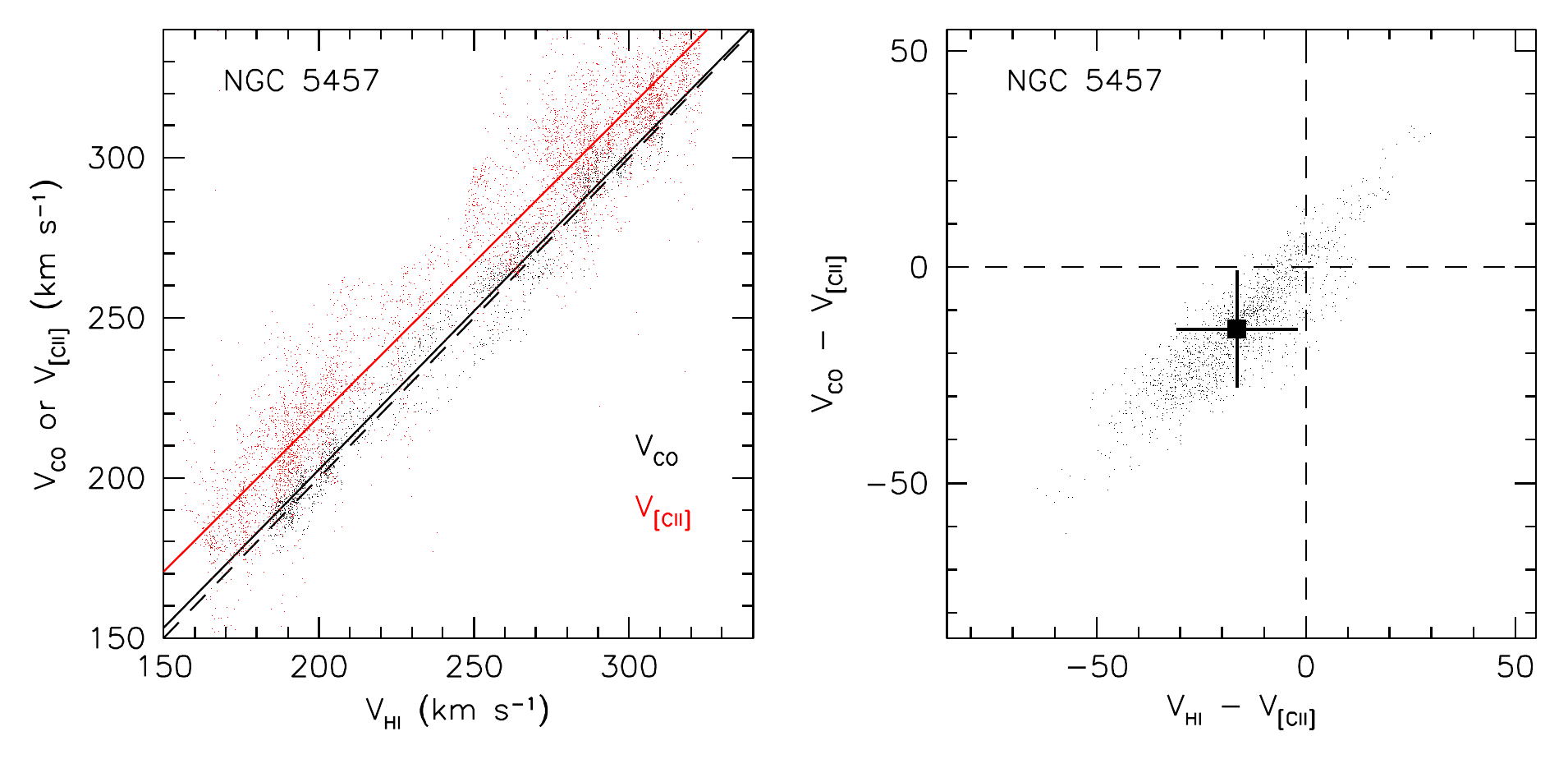}
\caption{As Fig.\ \ref{fig:vels_app1} but for NGC 5457.\label{fig:vels_app8}}
\end{figure*}

\begin{figure*}
\includegraphics[width=0.9\hsize]{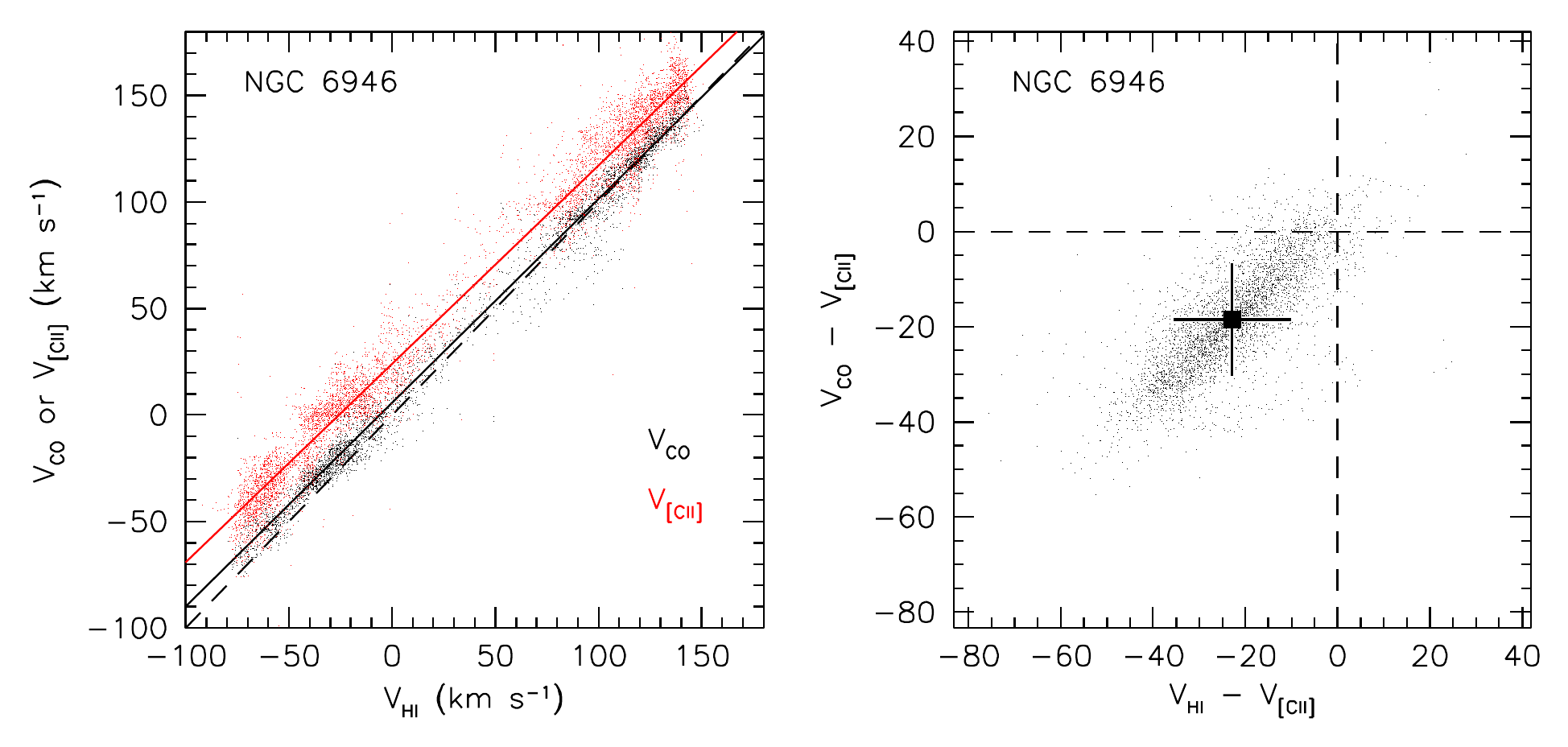}
\caption{As Fig.\ \ref{fig:vels_app1} but for NGC 6946.\label{fig:vels_app9}}
\end{figure*}

{\bf Velocity comparison:} We compare the \HI, CO and \cii velocities
as derived from the respective velocity fields in
Figs.\ \ref{fig:vels_app1} to \ref{fig:vels_app9}. See
Fig.\ \ref{fig:vels} in the main text for the results for NGC 5055.

\begin{figure*}
\begin{center}
\includegraphics[width=0.4\hsize]{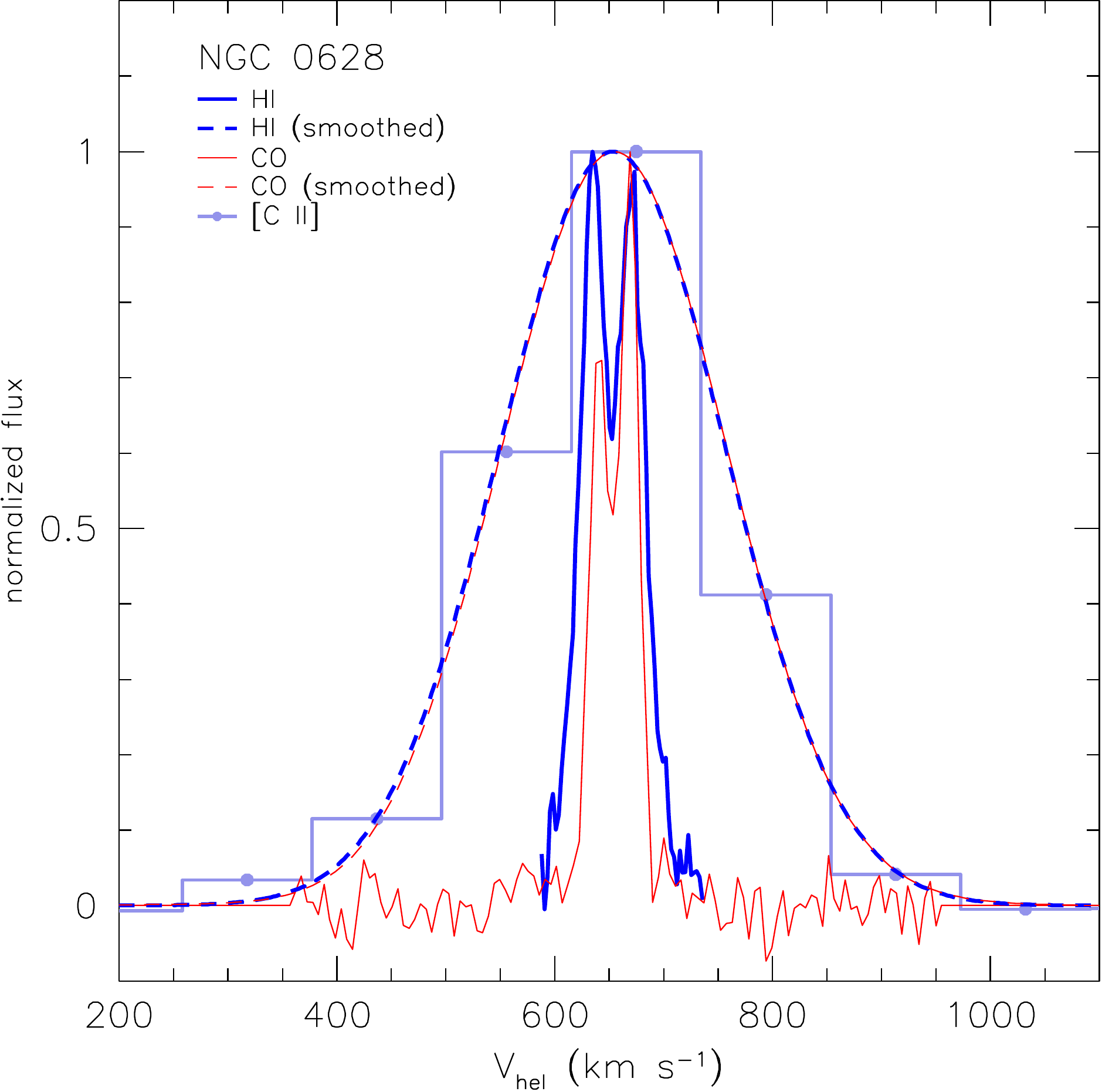}\hspace{1cm}\includegraphics[width=0.4\hsize]{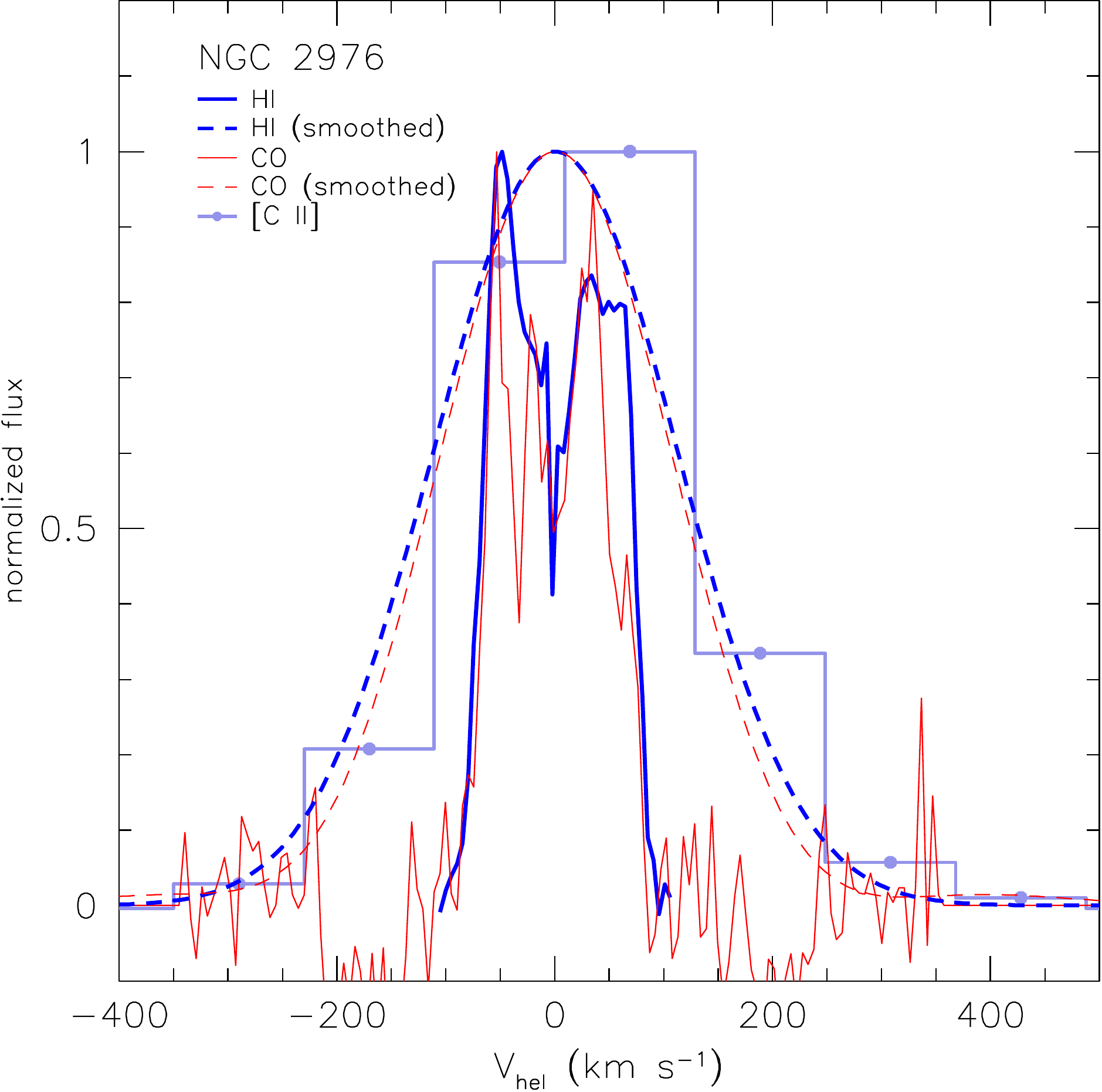}
\caption{Global profiles as measured in CO, \cii and \HI. Global
  profiles are all derived from the masked ``strip'' cubes. Thick,
  blue full line shows the \HI profile, thin red full line the CO
  profile. The light-blue histogram with filled circles superimposed shows the \cii
  global profiles. The thick blue and thin red dashed lines show the \HI
  and CO global profiles, respectively, both convolved to the velocity
  resolution of 239 \kms of the \cii data. Left panel shows NGC 628, right panel
  shows NGC 2976.
\label{fig:globprofs_apps1}}
\end{center}
\end{figure*}

\begin{figure*}
\begin{center}
\includegraphics[width=0.4\hsize]{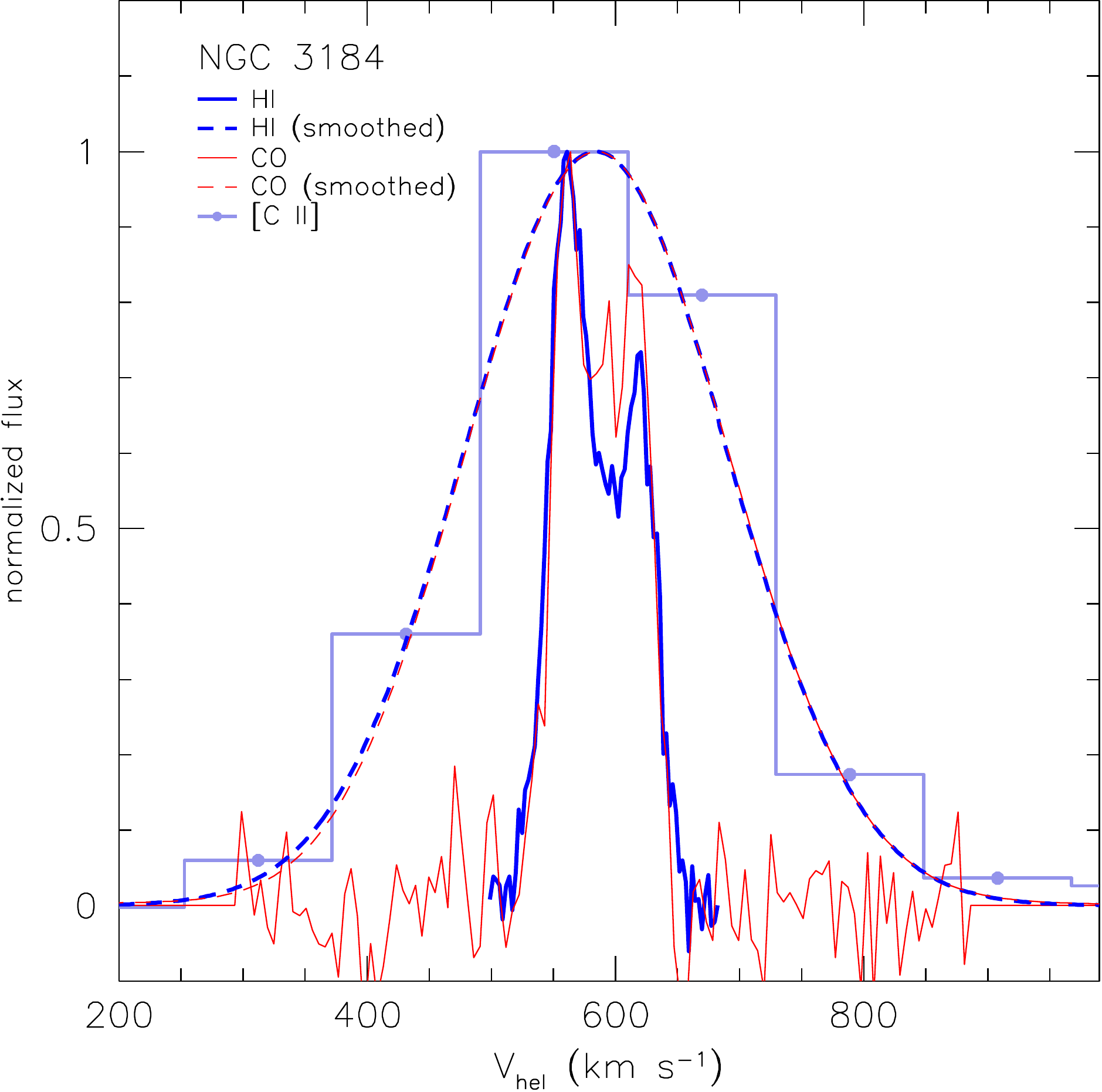}\hspace{1cm}\includegraphics[width=0.4\hsize]{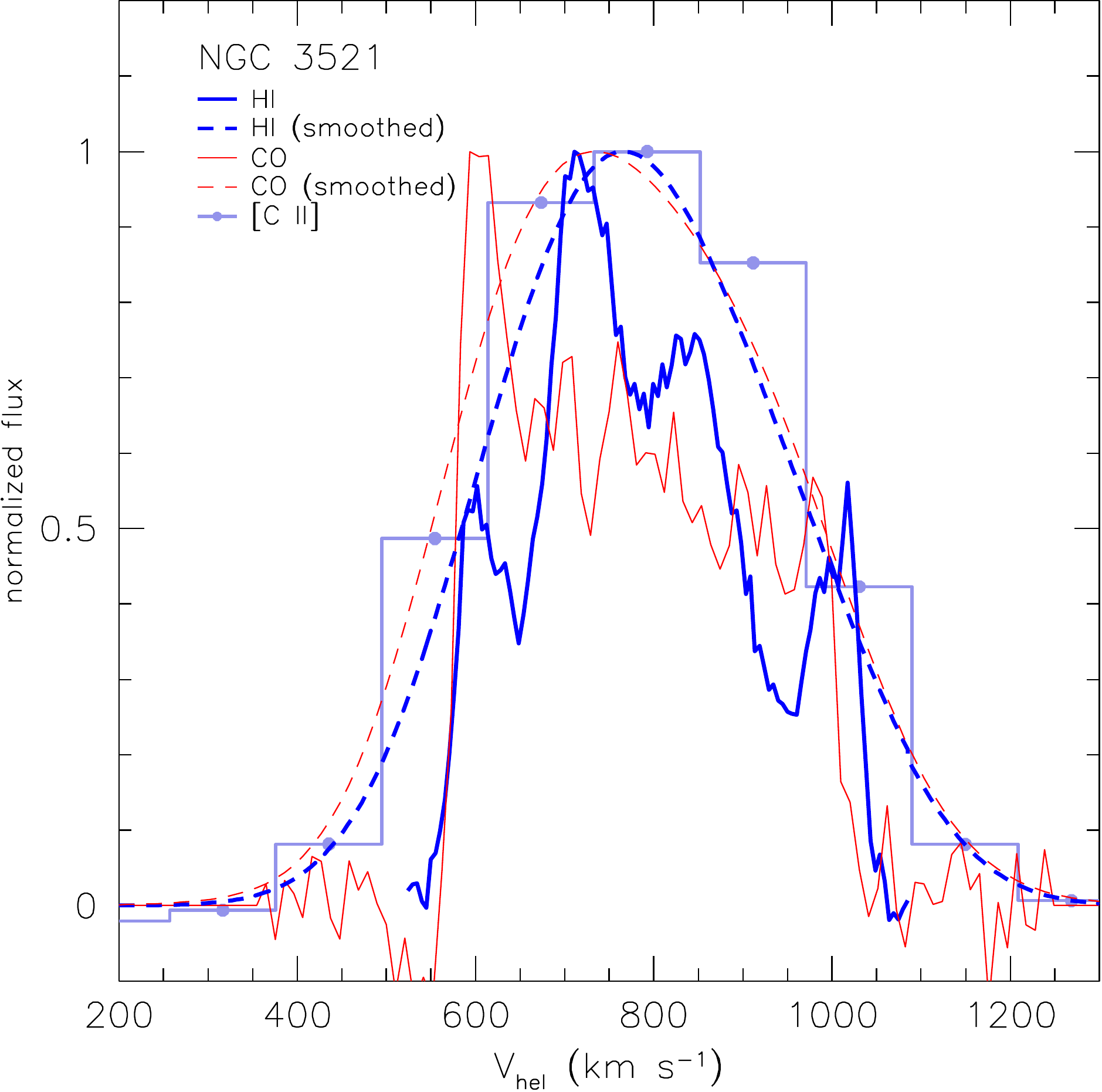}
\caption{As Fig.\ \ref{fig:globprofs_apps1}.
 Left panel shows NGC 3184, right panel
  shows NGC 3521.
\label{fig:globprofs_apps2}}
\end{center}
\end{figure*}

\begin{figure*}
\begin{center}
\includegraphics[width=0.4\hsize]{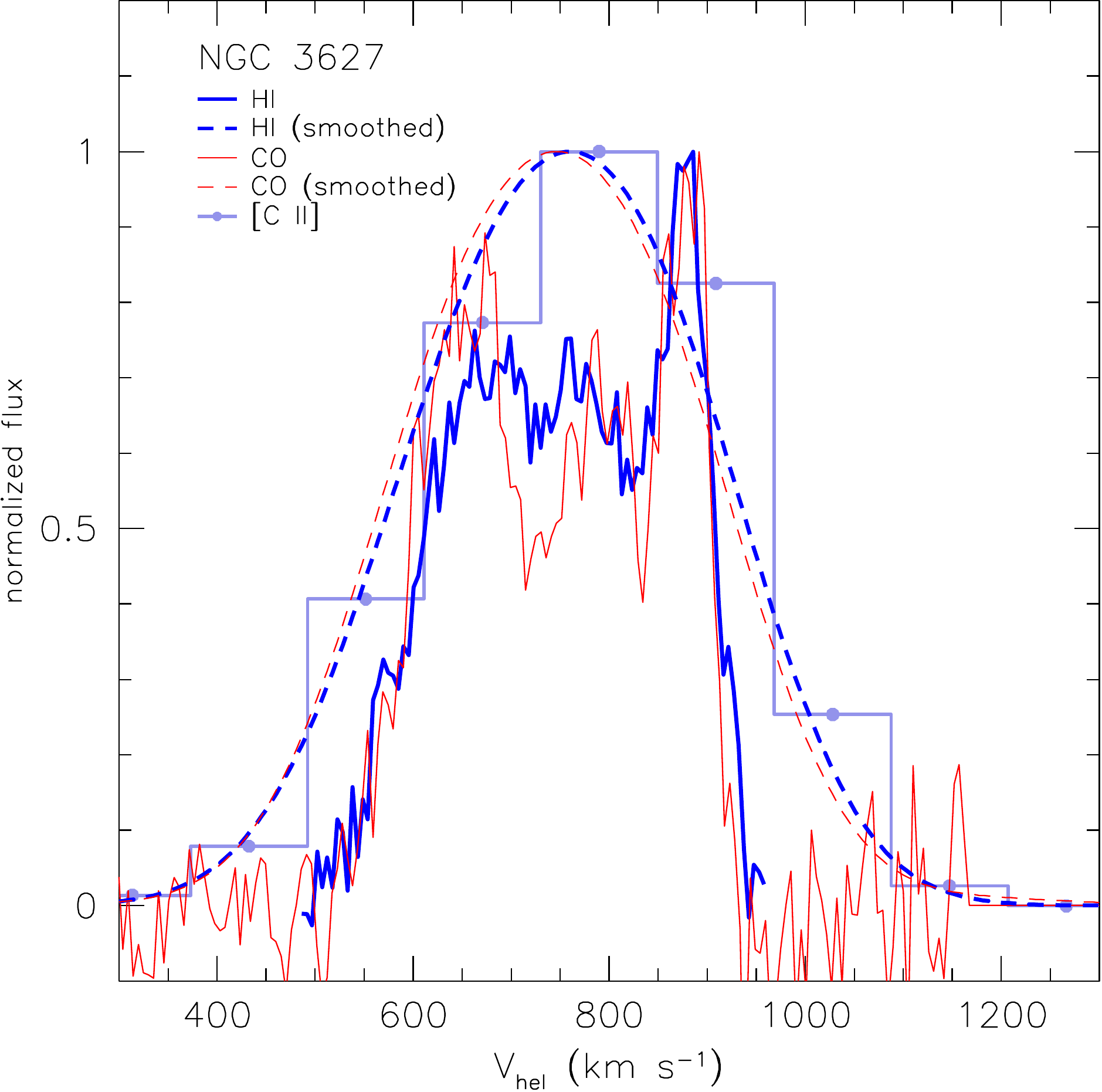}\hspace{1cm}\includegraphics[width=0.4\hsize]{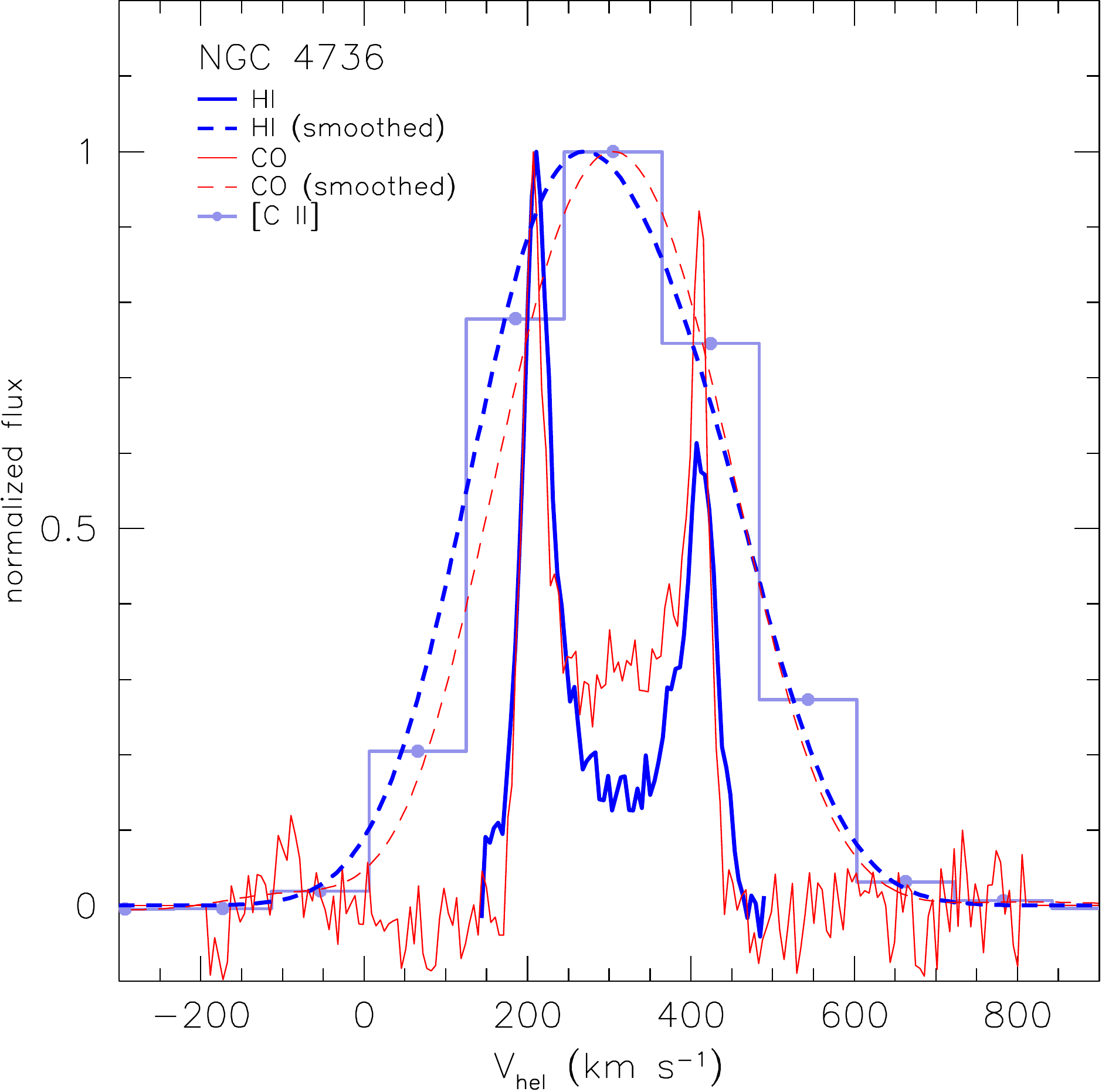}
\caption{As Fig.\ \ref{fig:globprofs_apps1}.
 Left panel shows NGC 3627, right panel
  shows NGC 4736.
\label{fig:globprofs_apps3}}
\end{center}
\end{figure*}

\begin{figure*}
\begin{center}
\includegraphics[width=0.4\hsize]{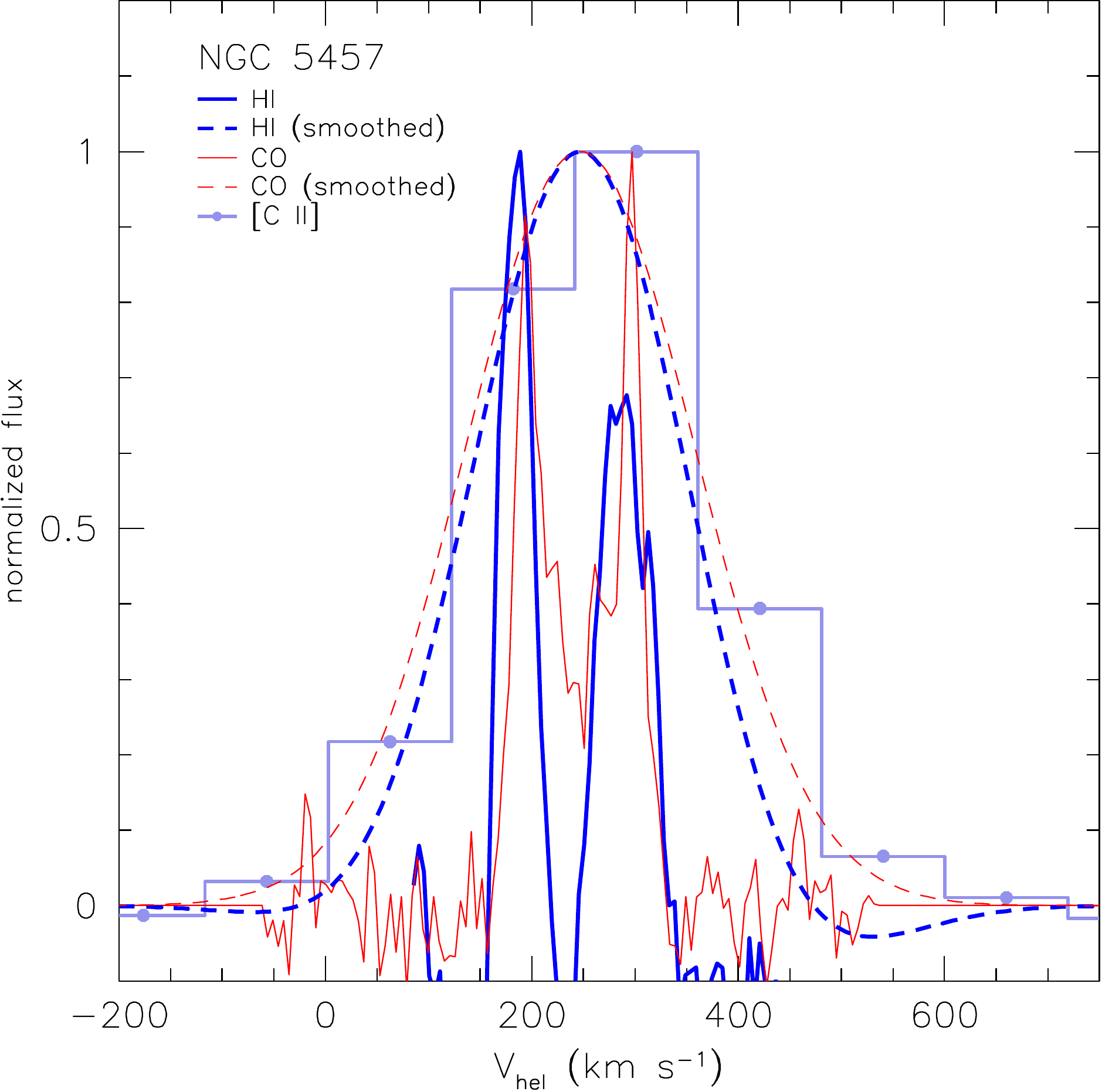}\hspace{1cm}\includegraphics[width=0.4\hsize]{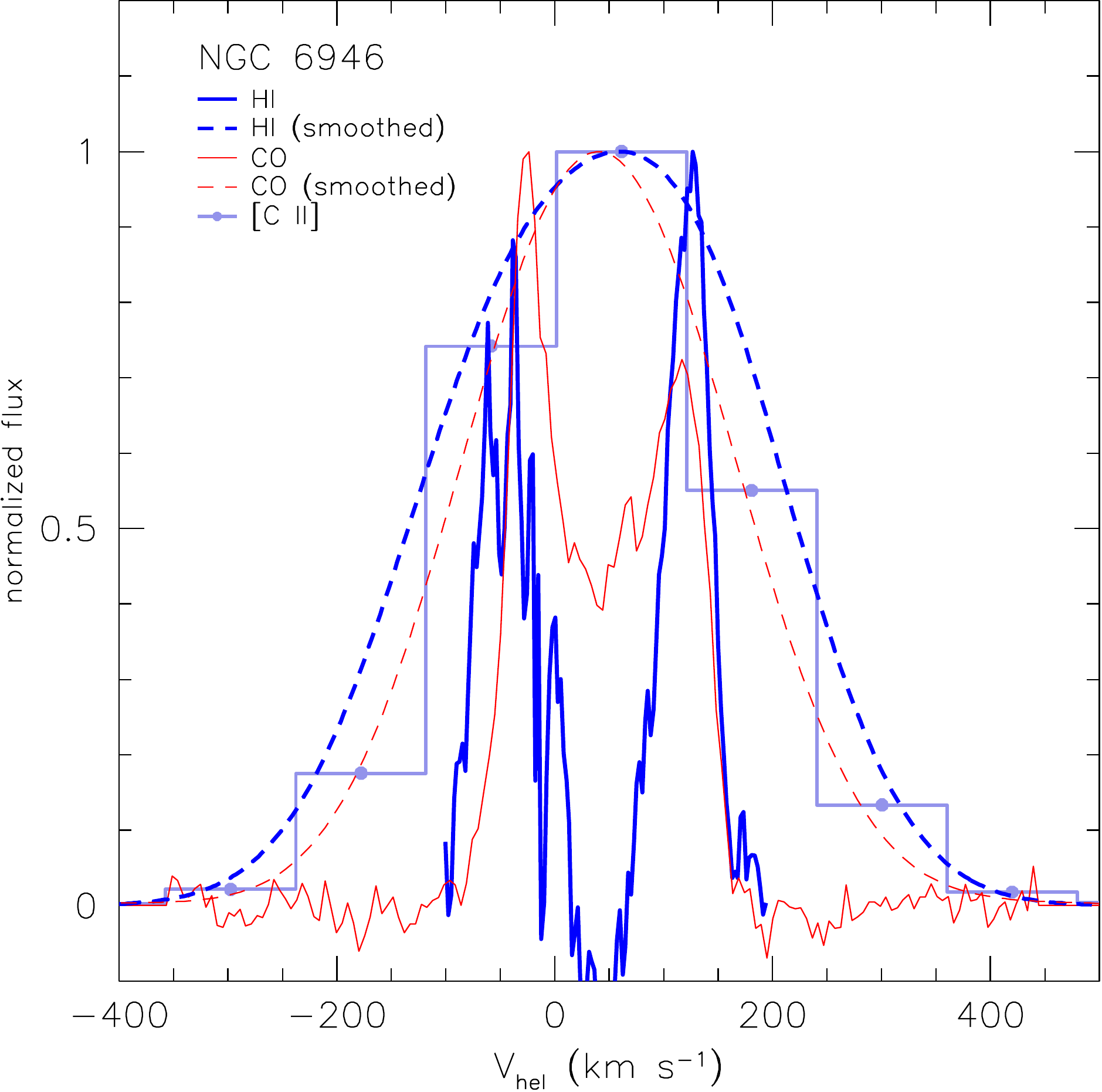}
\caption{As Fig.\ \ref{fig:globprofs_apps1}.
 Left panel shows NGC 5457, right panel
  shows NGC 6946. Note that the \HI profile of NGC 5457 suffers from missing short spacings effects.
\label{fig:globprofs_apps4}}
\end{center}
\end{figure*}

{\bf Integrated spectra (global velocity profiles):} In
Figs.\ \ref{fig:globprofs_apps1}--\ref{fig:globprofs_apps4} we compare
the integrated spectra (global velocity profiles) as measured in \HI,
CO and \cii. The results for NGC 5055 and NGC 3351 are shown in
Fig.\ \ref{fig:globprofs} in the main text.

\end{document}